# AI for Manufacturing and Healthcare:

## a chemistry and engineering perspective

Jihua Chen[1], Yue Yuan[1], Amir Koushyar Ziabari[2], Xuan Xu[3], Honghai Zhang[1], Panagiotis Christakopoulos[1], Peter V. Bonnesen[1],Ilia N. Ivanov[1], Panchapakesan Ganesh[1], Chen Wang[4], Karen Patino Jaimes[5], Guang Yang[6], Rajeev Kumar[1], Bobby G. Sumpter[1], Rigoberto Advincula[1,5]

[1]Center for Nanophase Materials Sciences, Oak Ridge National Laboratory
[2]Energy and Electrification Infrastructure Division, Oak Ridge National Laboratory
[3]1Data, K-State Olathe, Kansas State University
[4]National Institute for Occupational Safety and Health, Centers for Disease Control and Prevention
[5]Chemical Engineering, University of Tennessee at Knoxville
[6]Chemical Sciences Division, Oak Ridge National Laboratory

This manuscript has been authored by UT-Battelle, LLC, under Contract No. DEAC05-00OR22725 with the U.S. Department of Energy. The United States Government and the publisher, by accepting the article for publication, acknowledges that the United States Government retains a nonexclusive, paid-up, irrevocable, world-wide license to publish or reproduce the published form of this manuscript, or allow others to do so, for United States Government purposes. DOE will provide public access to these results of federally sponsored research in accordance with the DOE Public Access Plan (http://energy.gov/downloads/doe-public-access-plan).

# Abstract

Artificial Intelligence (AI) approaches are increasingly being applied to more and more domains of Science, Engineering, Chemistry, and Industries to not only improve efficiencies and enhance productivity, but also enable new capabilities. The new opportunities range from automated molecule design and screening, properties prediction, gaining insights of chemical reactions, to computer-aided design, predictive maintenance of systems, robotics, and autonomous vehicles. This review focuses on the new applications of AI in manufacturing and healthcare. For the Manufacturing Industries, we focus on AI and algorithms for (1) Battery, (2) Flow Chemistry, (3) Additive Manufacturing, (4) Sensors, and (5) Machine Vision. For Healthcare applications, we focus on: (1) Medical Vision (2) Diagnosis, (3) Protein Design, and (4) Drug Discovery. In the end, related topics are discussed, including physics integrated machine learning, model explainability, security, and governance during model deployment.

# Table Of Content

# 1. Overview

Artificial Intelligence (AI) and machine learning (ML) [1–4] are revolutionizing many aspects of our life including Chemistry, Engineering, and Industries. The algorithms [1–4] developed through decades of research range from tree models, unsupervised clustering, dimensional reduction, classification, regression, to natural language processing (NLP), computer vision (CV), various types of deep learning, and reinforcement learning, facilitating a series of progress and opportunities for future technologies in manufacturing [5–7] and healthcare.[8,9] On the chemistry side,[10–16] examples include molecule design, properties prediction, mechanistic insights on chemical reactions. On the engineering side,[5,6,8,9,17–20] impacts can be reached through risk assessment, mechanism optimization, pattern recognition, self-automation of additive manufacturing, sensors and optoelectronics, battery safety and personalized healthcare.

This review focuses on some new and important cases of AI applications in manufacturing and healthcare. In manufacturing, well-known AI technologies include predictive maintenance, defect inspection and quality control, anomaly detection, and robotics. Herein, we focus on AI and algorithms for (1) Battery State of Health, (2) Flow Chemistry, (3) Additive Manufacturing, (4) Sensors, and (5) Machine Vision. In Healthcare, AI is applied widely for automated analysis of medical scans, early detection of diseases, personalized medicine, robotic surgery, and virtual assistants. Our focus of AI's healthcare applications in this review will be on (1) Medical Vision (2) Diagnosis, (3) Protein design, and (4) Discovery of small molecule drugs. Overall, AI has significant potential to transform Chemistry, Engineering and Industries through increasing levels of automation, data-driven insights, and intelligent decision making. Continued progress in AI will lead to accelerated discovery and innovation, increased efficiency and productivity, improved quality, and reliability - and ultimately a reimagining of entire industries. But there are also risks and challenges posed by AI that must be addressed through guidelines and policies to ensure the responsible development and application of AI technology. In the end of the review, related issues are discussed, on physics integrated machine learning, model explainability, security and governance during model deployment.

Navigating the vast chemistry and engineering space can be considered as an optimization problem for manufacturing and health industries,[14] considering that the synthesized substances to this date in human history is on the order of $10^8$ and the number of potential candidates can be many orders of magnitude higher (for example, the number of drug-like molecules is up to $10^{60}$).[14] **Figure 1** shows a way[14] to tackle this problem with recent advances of machine learning, using a Simplified Molecular-Input Line-Entry System (SMILES) representation of molecules and probabilistic generative models. Instead of using discrete methods, a continuous fixed-length-vector representation of the SMILES strings (or latent representation) enables a two-component autoencoder[14] framework, with an encoder network to convert a SMILES string into a vector and a decoder network to revert a vector into a SMILES string. For continuous optimization in the space of latent representation (or latent space) to be effective, invalid areas representing invalid SMILES strings are mitigated with a variational autoencoder (VAE).[21] An open source cheminformatics toolkit RDKit is used to filter validated output molecules.[22] An additional multilayer perceptron (MLP) is used for subsequent property prediction.[14] **Figure 2** showcased a new optimization method based on autoencoder, with joint evaluations of synthetic accessibility score (SAS), Qualitative Estimate of Drug-likeness (QED) ranging from 0 to 1, and objective score percentile.[14] It is noted that the objective of choice in this work is O = 5QED - SAS, representing an estimate of locating the likely drug molecule that can be easily synthesized. For algorithm trained with the ZINC[23] data set, the objective may include the water−octanol partition coefficient (logP), QED, SAS, while for algorithms trained on the QM9 data set [24–26], the objective can be highest occupied molecular orbital or HOMO energies, lowest unoccupied molecular orbital or LUMO energies, and the electronic spatial extent $R^2$. In **Figure 2a**, VAE based optimization with Gaussian Process (GP) consistently outperforms the baseline methods (random Gaussian search and genetic algorithm). In principal component analysis (PCA) representation (**Figure 2b**), the path of optimization is given. In **Figure 2c**, the optimization path is decoded and shown in their molecular formula, with a three-number index of their SAS, QED, and objective score percentile.

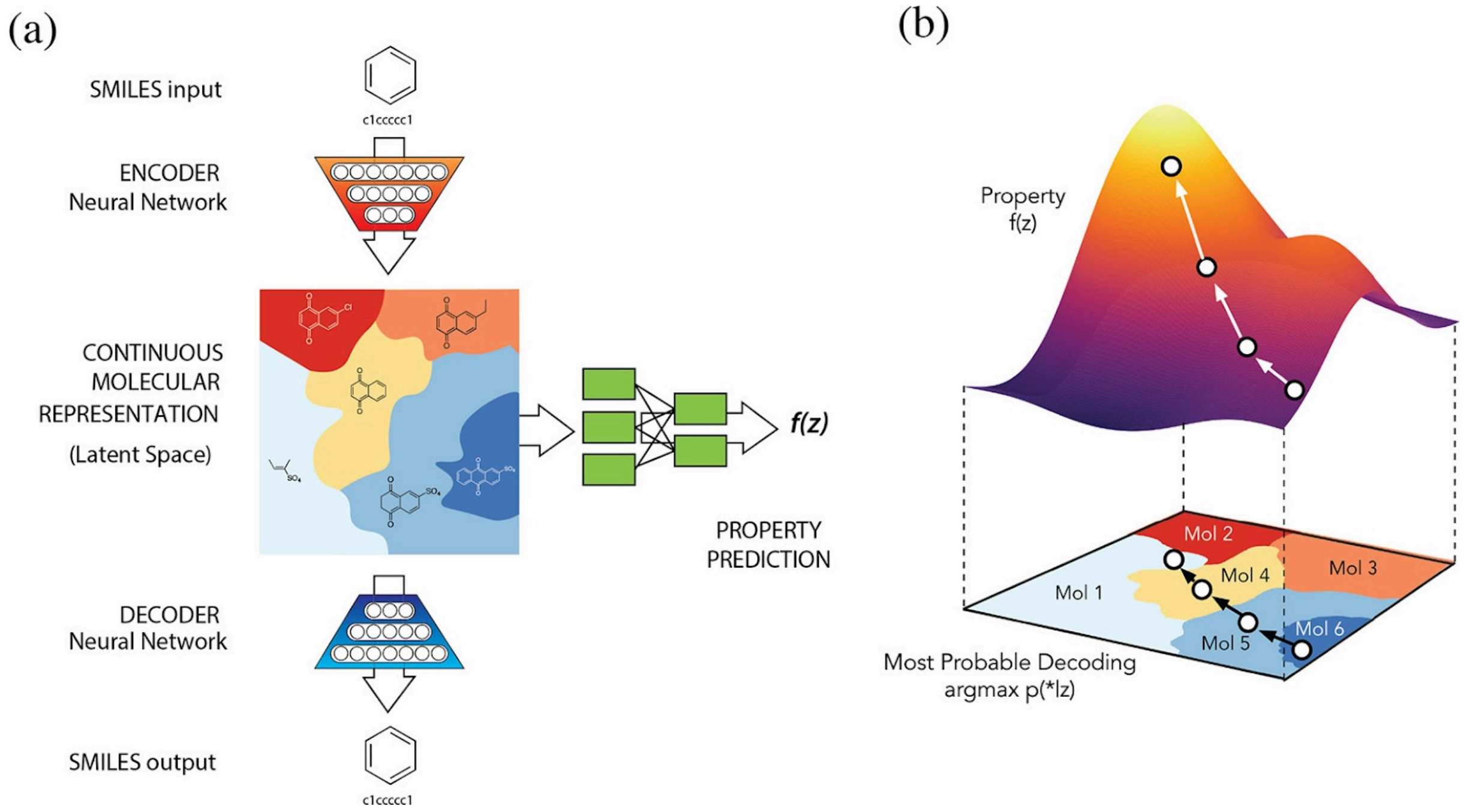

**Figure 1**. (a) A diagram of the autoencoder used for molecular design, including the joint property prediction model. Starting from a discrete molecular representation, such as a SMILES string, the encoder network converts each molecule into a vector in the latent space, which is effectively a continuous molecular representation. Given a point in the latent space, the decoder network produces a corresponding SMILES string. A multilayer perceptron network estimates the value of target properties associated with each molecule. (b) Gradient-based optimization in continuous latent space. After training a surrogate model f(z) to predict the properties of molecules based on their latent representation z, we can optimize f(z) with respect to z to find new latent representations expected to have high values of desired properties. These new latent representations can then be decoded into SMILES strings, at which point their properties can be tested empirically. (Figure and Caption from reference[14], without change, under CC-BY License[27].)

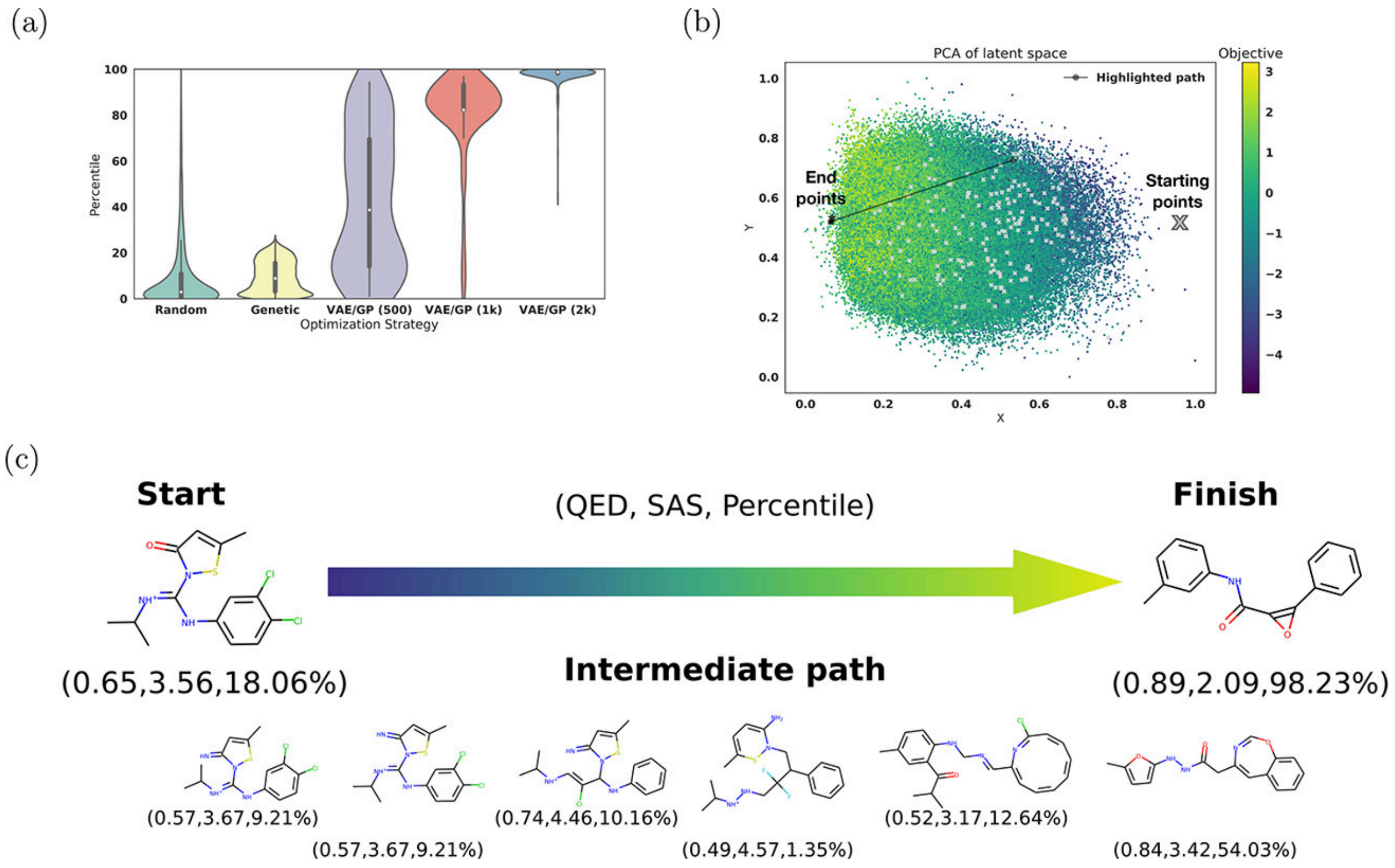

**Figure 2**. Optimization results for the jointly trained autoencoder using 5 × QED – SAS as the objective function. (a) shows a violin plot which compares the distribution of sampled molecules from normal random sampling, SMILES optimization via a common chemical transformation with a genetic algorithm, and from optimization on the trained Gaussian process model with varying amounts of training points. To offset differences in computational cost between the random search and the optimization on the Gaussian process model, the results of 400 iterations of random search were compared against the results of 200 iterations of optimization. This graph shows the combined results of four sets of trials. (b) shows the starting and ending points of several optimization runs on a PCA plot of latent space colored by the objective function. Highlighted in black is the path illustrated in part (c). (c) shows a spherical interpolation between the actual start and finish molecules using a constant step size. The QED, SAS, and percentile score are reported for each molecule. (Figure and Caption from reference[14] , without change, under CC-BY License[27].)

To fully take advantage of AI and big data analytics, an iterative process may be necessary when monitored metrics deviate from a preset range to ensure an effective AI model is in place.[3] An end-to-end data pipeline can be typically divided into 2 parts: 1) data engineering (DE) and 2) data science (DS).[3] The former involves data collection, data cleaning, as well as data extraction, transformation, and loading (ETL); the latter typically has exploratory data analysis (EDA), feature engineering, metrics selection, train/validation/test splitting, model training and evaluation, as well as Continuous Integration and Continuous Deployment (CICD).[3] Some of the common challenges in building machine learning models include missing values, outlier detection and elimination, feature engineering, metric defining, model selection and evaluation, and mitigation of overfitting.[3] For AI in manufacturing and healthcare, these steps and challenges can be largely similar. **Figure 3a** outlined a workflow for supervised learning towards molecular properties prediction. **Figure 3b** gives an example of virtual screening that starts from a large candidate space and ends with a few experimental tests to verify.

a.

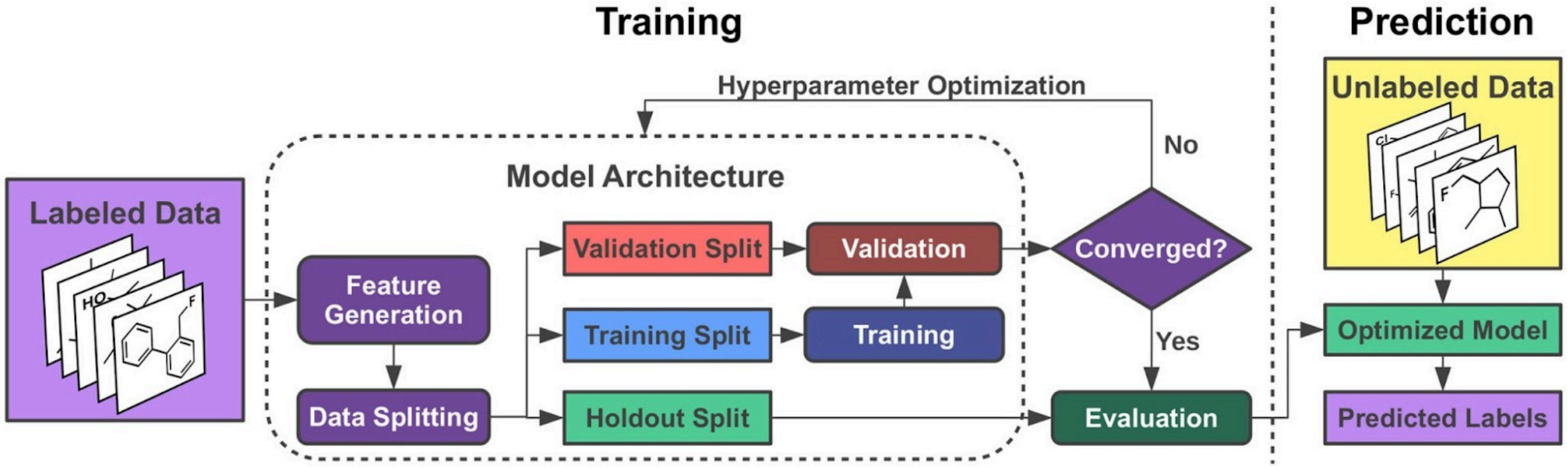

b.

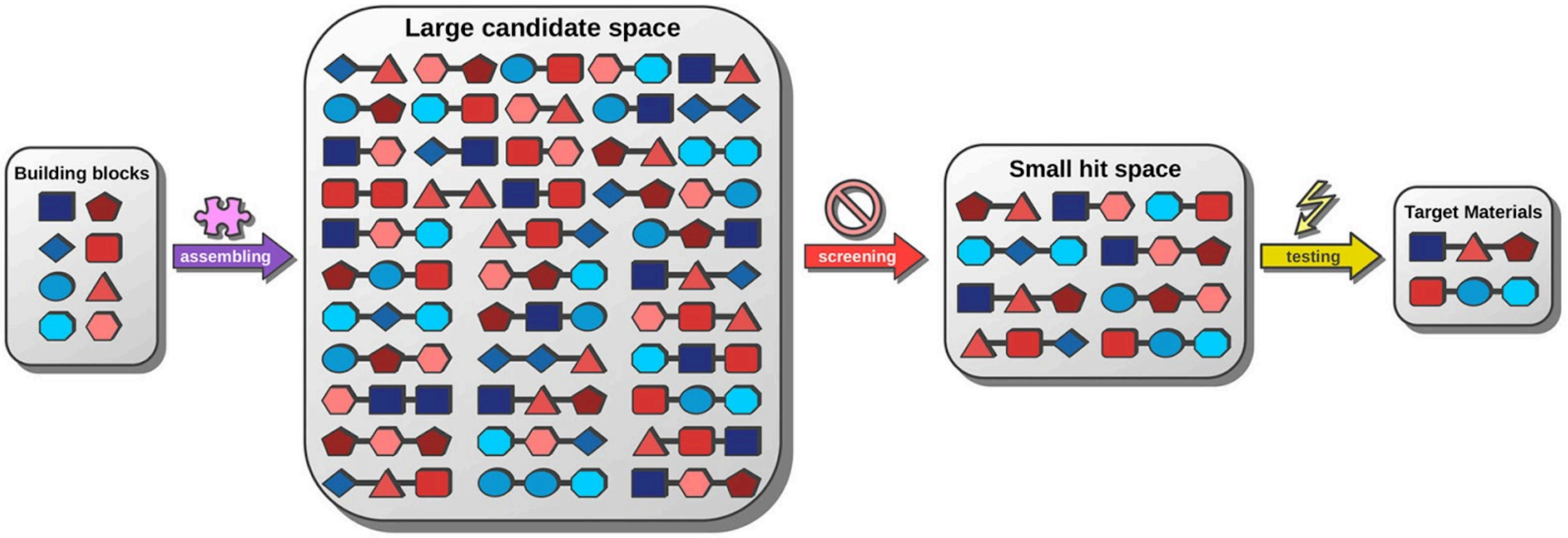

**Figure 3 a**. Workflow for supervised learning of molecular properties. A known (labeled) data set is used to optimize a model, which is subsequently used to estimate molecular properties for an unknown (unlabeled) data set. (Figure and Caption from reference[10], without change, under CC-BY License.[27]) **b.** High-throughput virtual screening starts from a large space of candidates (e.g., generated combinatorically, as illustrated). Using virtual screening, most candidates are eliminated, such that fewer (more expensive and time-consuming) experimental tests can be performed. (Figure and Caption from reference[10], without change, under CC-BY License.[27])

To zoom out from the specific example given in **Figure 1** and **Figure 2**, a substance-category perspective of AI utilization is given in **Figure 4**.[13] Journal articles and patent trends in substance classes from Chemical Abstracts Service (CAS) suggest a distinct increase of AI adoption from 2000 to 2020.[13] Many inorganic class labels are straightforward, such as Alloy, Inorganic Compound, and Element. It is noted that the class with a label "Manual Registration" requires further human inputs, and is mainly made up of biomolecules, including enzymes, hormones, vaccines, as well as antibodies.[13] Ring Parents refers to molecular scaffolds that define the connectivity of a molecular ring system. It is noted in **Figure 4** that Small Molecules represent the most important substance category for journal publications, while Nucleic Acid Sequence dominates the patents.[13] Soft materials, including small molecule, polymer, ring parent, peptide sequence, and nucleic acid sequence, represented the majority of the AI publications and in most cases determine the trend of the AI development. Ionic compounds such as coordination compounds and salts represent another important substance class of interest. Based on this observation from the survey[13] of publication, soft materials and ionics will also be the focus of this review whenever possible. We will discuss some of the key roles that soft materials and ionics are playing in revolutionizing manufacturing and healthcare industries at later sections.

Furthermore, **Figure 5a and 5c** gives the numbers of (journal or patent) publications and substances as a function of different types of substances.[13] Challenges in AI implementation can include substance representation, data availability, as well as modeling and evaluation. The number of substances within each substance category (indexed by CAS) can be thus correlated with research success in overcoming such challenges. The large volume of Small Molecules publication is likely enabled by the simplicity of their structures and representation as compared to other classes such as Polymer. **Figure 5b and 5d** shows the year-to-year change in the number of AI-relevant (journal or patent) publications as a function of the substance class.[13] Nucleic Acid Sequences and Peptide Sequences are the largest category for patent (**Figure 5c**).

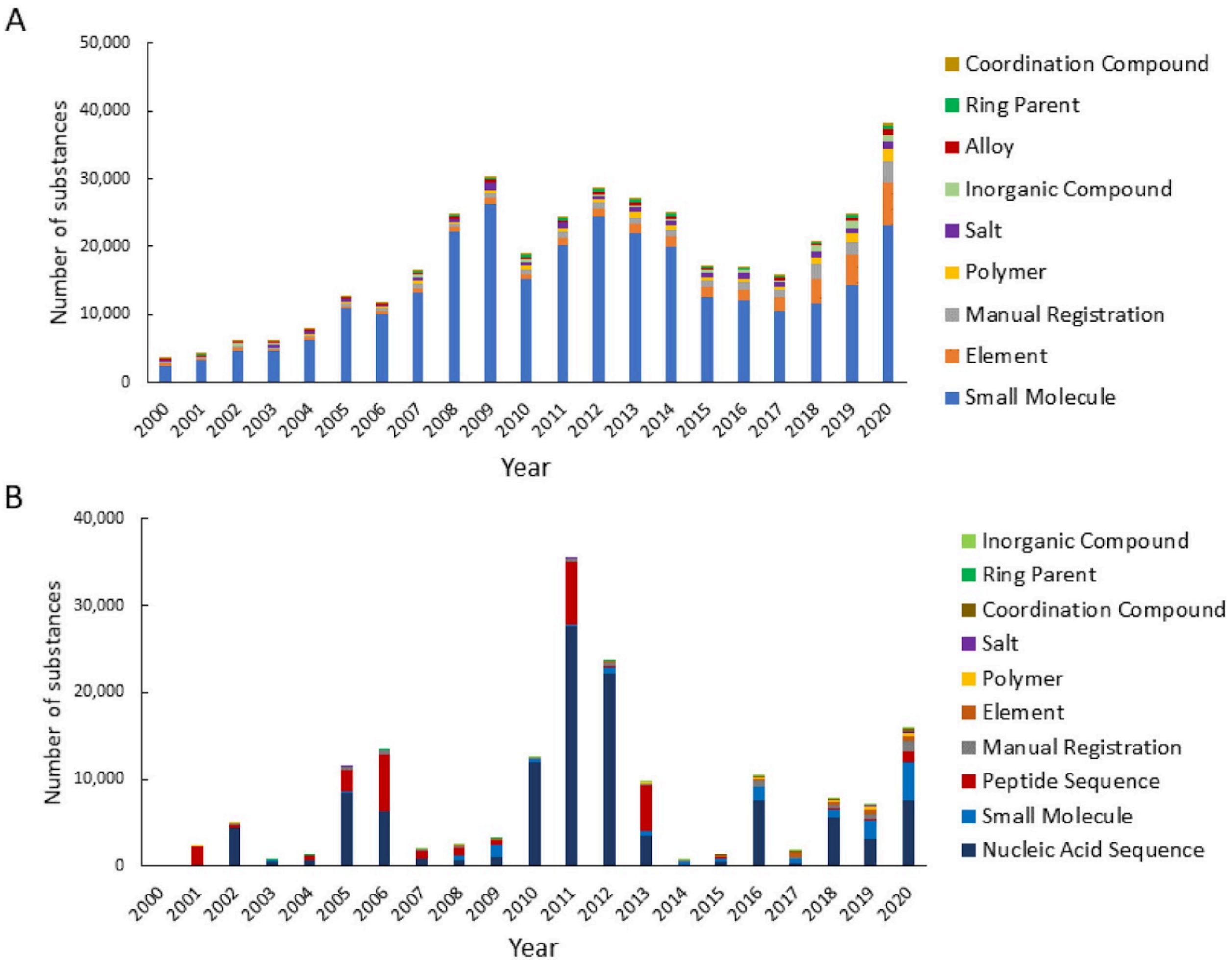

**Figure 4.** Trends of substance classes in AI-related chemistry publications from 2000 to 2020: (A) journal publications and (B) patent publications. (Figure and Caption from reference,[13] without change, under CC-BY 4.0 License.[28])

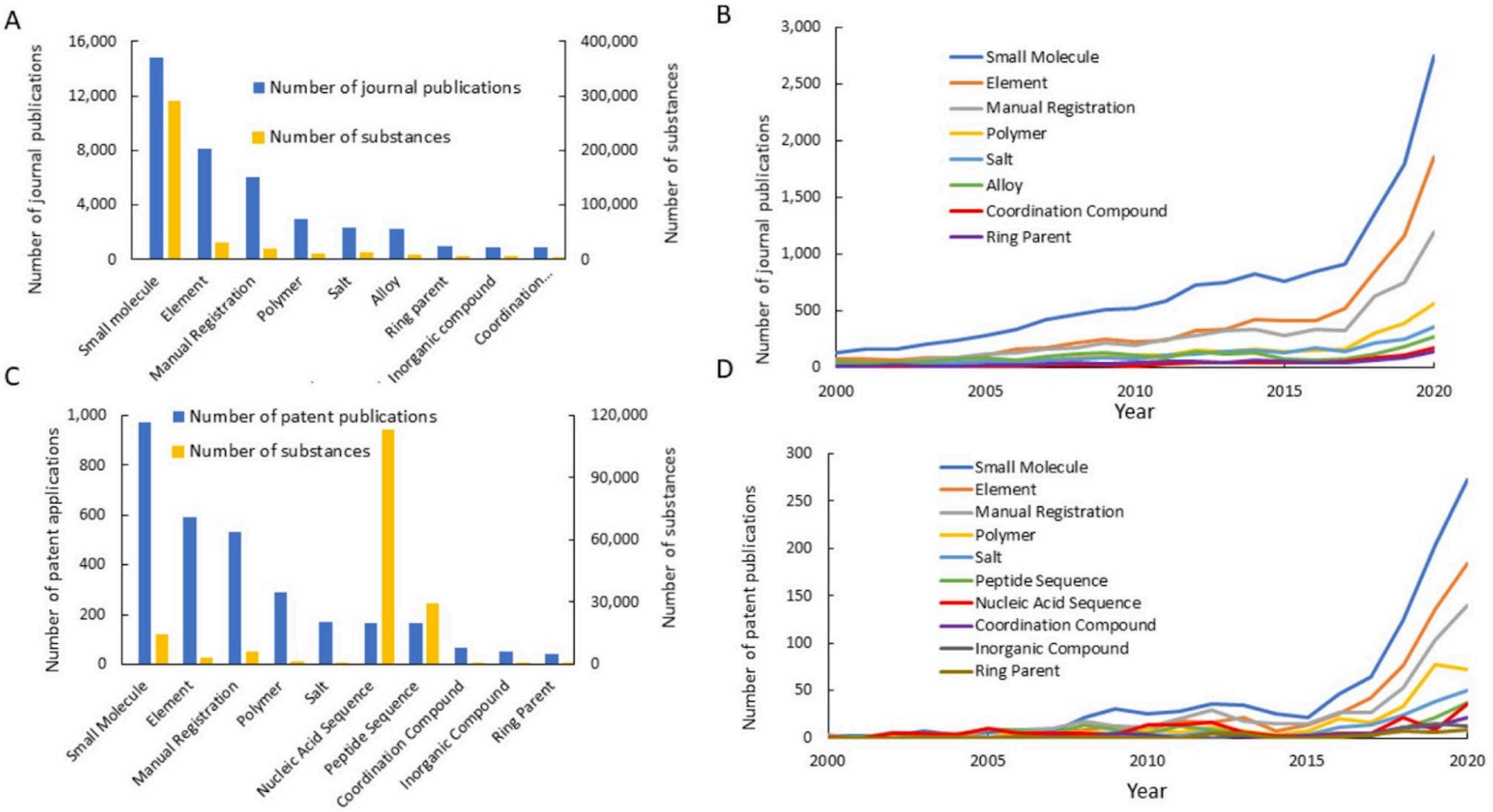

**Figure 5**. Publications in AI-related chemistry associated with substance class from 2000 to 2020. (A) Number of AI-related journal publications and number of substances associated with each class. (B) Trends of AI-related journal publications associated with each substance class. (C) Number of AI-related patent publications and number of substances associated with each class. (D) Trends of AI-related patent publications associated with each substance class. (Figure and Caption from reference,[13] without change, under CC-BY 4.0 License[28].)

The foundation of any research in data science and AI lies in the quality and availability of data sets. Although it is possible to build a data set from scratch, or search a specific dataset from Kaggle (now part of Google), Github (now part of Microsoft), and Google Dataset Search Engine[29], we are listing some important dataset for manufacturing and healthcare industries in **Table 1**. Many of these curated dataset will be the data source for training relevant algorithms that are discussed in later sections of the review. Soft matter and ionics are the focus, as indicated by **Figure 4 and 5**.

**Table 1**. Some important OPEN datasets that are curated for data science in manufacturing and healthcare. Soft matter and ionics are the focus of this review, with their importance indicated in **Figure 3 and 4**. Databases not curated for AI (such as ChEMBL, PubChem) are not included.

| Name | Link | Description |
| --- | --- | --- |
| ANI-1 | https://figshare.com/collections/_/3846712 | DFT calculations of 20 million off-equilibrium conformation of over 50k organic molecules [30] |
| van der Waals Radii | https://data.4tu.nl/articles/_/12688418/1 | All intermolecular contacts below 7 Å, out of over 200k Cambridge Structural Database (CSD) entries [31] |
| CKBIT | https://data.mendeley.com/datasets/829ph4nkny | Chemical Kinetics Bayesian Inference Toolbox[32] |
| ZINC | https://figshare.com/articles/dataset/ZINC_250K_data_sets/17122427<br>https://paperswithcode.com/dataset/zinc<br>https://www.kaggle.com/datasets/basu369victor/zinc250k | 750 million commercially available compounds for virtual screening [33,34] |
| Quantum Machine (QM) | http://quantum-machine.org/datasets/ | Quantum chemical properties for small |
| | | organic molecules. QM7 is composed of 7211 molecules, and QM9 from 133,885 molecules.[26] |
| USPTO | https://figshare.com/articles/dataset/Chemical_reactions_from_US_patents_1976-Sep2016_/5104873 | Chemical reactions in US patents |
| Materials Project | https://materialsproject.org/ | Calculations of 146k known or predicted materials |
| AFLOW | https://www.aflowlib.org/ | 3.5 million material with over 734 million calculated properties |
| NOMAD | https://nomad-lab.eu/nomad-lab/ | Shared data for almost 3 million materials |
| Polymer Genome | https://www.polymergenome.org/ | Polymer property prediction |
| Battery Data (Cavendish Lab) | https://www.materialsforbatteries.org/data/ | Battery materials properties and component materials |
| NREL Battery | https://www.nrel.gov/transportation/machine-learning-for-advanced-batteries.html | Battery microstructure, failure, and capacity |
| Battery EIS | https://data.mendeley.com/datasets/zdsgxwksn5/1 | Lithium Ion Battery data: impedance for varying State of Charge[35] |
| Panasonic NCR18650PF | https://data.mendeley.com/datasets/wykht8y7tg/1 | Lithium Ion Battery data |
| LG 18650HG2 | https://data.mendeley.com/datasets/cp3473x7xv/3 | Lithium Ion Battery data for State of Charge estimator |
| A123 APR18650M1A | https://data.matr.io/1/ | Dataset for 124 lithium iron phosphate (LFP) /graphite cells cycled with fast charge and varying cycle lives (150 to 2,300)[36] |
| Center for Advanced Life Cycle Engineering | https://calce.umd.edu/battery-data | Lithium Ion Battery data from University of Maryland, which includes impedance, continuous full |
| | | /partial cycling, storage, open-circuit voltage measurements, and dynamic driving profiles. |
| Sensors of a Smart Home | https://data.mendeley.com/datasets/t9n68ykfk3/1 | Multi-sensor information of a single user's activities at home |
| The Cancer Genome Atlas - Lung Adenocarcinoma Data (TCGA-LUAD) | https://wiki.cancerimagingarchive.net/pages/viewpage.action?pageId=6881474 | A dataset to connect phenotypes to genotypes of cancers by clinical images |
| Human Connectome Projects (HCP) | https://www.humanconnectome.org/study/hcp-young-adult/data-releases | Neuroimaging and behavioral data from healthy young adults |
| Alzheimer's Disease Neuroimaging Initiative (ADNI) | https://adni.loni.usc.edu/ | A global research resource that supports the investigation and treatments of Alzheimer's |
| IXI | https://brain-development.org/ixi-dataset/ | 600 magnetic resonance images of healthy subjects |
| Mindboggle | https://mindboggle.info/ | A large manually labeled collection of human brain images |
| Predicting Molecular Properties | https://www.kaggle.com/competitions/champs-scalar-coupling | A competition to predict the scalar coupling constant or the magnetic interaction between two atoms of a molecule |
| Time Series data for M Competitions | https://forecasters.org/resources/time-series-data/ | Uncertainty Quantification of time series forecasting[37–39] for Makridakis Competitions |
| Protein Data Bank | https://www.wwpdb.org/ | 3D structure database curated for complex assemblies, nucleic acids, and proteins. |

Champion models for AI applications in manufacturing and healthcare will be discussed in later sections when specific tasks are analyzed. A typical AI and machine learning (ML) model can be considered as 1) unsupervised learning, 2) supervised learning, or 3) reinforcement learning (**Figure 6**).[1–3,40] Unsupervised learning refers to data-driven methods taught using unlabeled data (i.e. without human inputs) to look for patterns in the supplied dataset, whereas supervised learning is a task-driven strategy with data being labeled ("the ground truth") during training. When compared to supervised learning and unsupervised learning, reinforcement learning is significantly different. Reinforcement learning focuses on how agents (for example, a Chess player) interact with their surroundings (i.e. a chessboard).[1,3,40] The model effectiveness of supervised learning and unsupervised learning can be assessed and monitored by minimizing a loss function (or objective function). On the other hand, the goal of reinforcement learning, in contrast, is to maximize a cumulative reward. In addition, semi-supervised learning is a subset of machine learning techniques. Because it uses both labeled and unlabeled data during training, it falls between supervised and unsupervised learning.

Classical ML algorithms include Linear regression (LIR or LR), support vector regression (SVR) or support vector machine, random forest, logistic regression, decision tree, gradient boost, while deep learning algorithms range from artificial neural networks (ANNs), feedforward neural networks (FFNNs) or multilayer perceptron (MLP), convolutional neural networks (CNNs), recurrent neural networks (RNNs), long short-term memory (LSTM), gated recurrent unit (GRU), to generative models such as generative adversarial networks (GANs), and variational autoencoder (VAE).[40] ResNet stands for Residual Network and is a specific type of CNN,[41] in which the weight layers learn from residual functions in reference to the layer inputs and enable deep learning models to train easily with the identity skip connections or "residual connections". In addition, ML algorithms may be used to evaluate and enhance the performance of other ML models.[40] These include Bayesian learning for parameter estimation and probability comparison, Gaussian process regression for uncertainty measurements. Furthermore, graph neural networks (GNNs) operate on graphs consisting of nodes and edges, with recent progress on semi-supervised classification.[40]

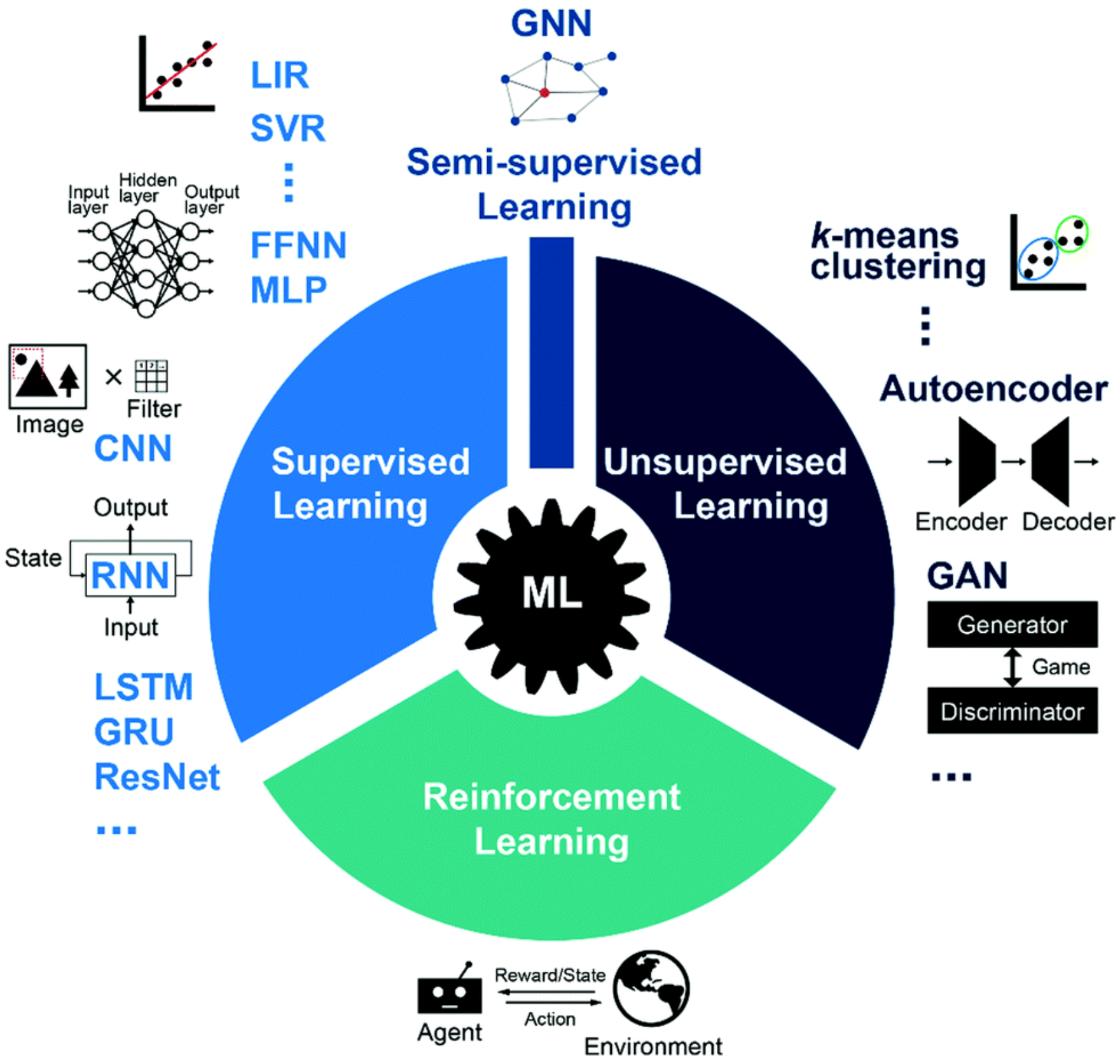

**Figure 6.** A brief overview of ML approaches, including three major categories known as supervised learning, unsupervised learning and reinforcement learning. ML approaches such as linear regression (LIR), support vector regression (SVR), feedforward neural networks (FFNNs), multilayer perceptron (MLP), convolutional neural networks (CNNs) and recurrent neural networks (RNNs) are generally used for supervised learning. Typical approaches to unsupervised learning include k-means clustering, autoencoder and generative adversarial networks (GANs). Reinforcement learning follows a general interactive loop between the agent and the environment. The difference between supervised and unsupervised learning is determined by whether training data is labeled or unlabeled, and there is a category of tasks between them called semi-supervised learning, which combines labeled and unlabeled data (generally mostly unlabeled) during training. It is worth pointing out that some of the aforementioned ML methods are not merely limited to the tasks illustrated in this schematic. For instance, graph neural networks (GNNs) have been widely used for semi-supervised learning tasks, but they are also applicable to supervised and unsupervised learning tasks involving graph representation. (Figure and Caption from reference[40], without change, under CC-BY NC 3.0 License.[42])

In **Table 2**, we highlight some of the notable programs that are beneficial to scientific literature mining, enhancing research for AI in manufacturing, healthcare, and beyond. Several open-source programs are highlighted here. These include SciBERT,[43] BioBERT,[44] BatteryBERT,[45] BatteryDataExtractor,[46] HuggingChat, Carrot2, and Gephi. It is noted that acquiring large-scale, high quality, labeled data and performing training for natural language processing (NLP) tasks is challenging and can be expensive, especially for models that require specialized large training data such as a pretrained model with enhanced performance in a field of scientific literature.[43,44] BERT means Bidirectional Encoder Representations from Transformers,[47] which is a simple yet powerful new language representation model introduced in 2019 that is designed to pre-train data from unlabeled text and can be fine-tuned later with an additional output layer for question answering or language inference, without significant architecture modifications for various NLP tasks.[47] Modern computational organic chemistry is becoming increasingly data-driven. Transformers and BERT based approaches are also attempted in areas such as functional groups, toxicity or solubility prediction , drug discovery (with drug-likeness), and molecule synthesis accessibility, based on string representations of molecules.[48]

**Table 2**. Some important programs related to scientific literature mining

| Name | Description |
|---|---|
| SciBERT | A pre-trained model[43] from Semantic Scholar corpus (1 million papers in full text) |
| BioBERT | A pre-trained model[44] for biomedical text mining |
| BERT Learns Chemistry | A transformer-based model on datasets of molecular string representations with analysis on its attention heads[48] |
| BatteryBERT | Pretrained models fine-tuned on battery paper classification, device component classification[45] |
| BatteryData Extractor | A open text-mining software with BERT models for automated battery data extraction and classification [46] |
| BERT for BioCreative VII | ALBERT-based system for chemical Identification and multiclass labeling of full-text Pubmed article[49] |
| Bard | Chatbot from Google |
| Bing AI | Chatbot type search engine from Microsoft |
| ChatGPT | Chatbot from OpenAI |
| Claude | Chatbot from Anthropic |
| HuggingChat | Open Source Chatbot: https://huggingface.co/chat/ |
| Carrot2 | Open source text clustering tools for PubMed and others: https://search.carrot2.org/#/search/web |
| Gephi | Open Source tool[50] for network knowledge graph |
| Neo4J | Knowledge graph and network analysis: https://neo4j.com/ |

# 2. Manufacturing

Artificial intelligence is revolutionizing the way we design, make, and inspect our industrial products in the energy, transportation, information technologies, communications, and other industries. For example, it is enabling fuel-efficient engines with less weight, and highly efficient turbines with less manufacturing time and less repair.[51] On one hand, AI is currently used to facilitate the parts design in subtractive manufacturability. On the other hand, it optimizes process parameters of additive manufacturing. In both cases, it can be used in generative design, tool life modeling, robotic control and automation, as well as product inspection.[52] Furthermore, highly accurate material characterization and measurement are made possible by AI-driven picture analysis, assuring the quality of various product inspection or materials examination.[51] For example, deep learning and augmented reality are used to speed up estimates of surface-stress distribution to evaluate the performance of materials in additively built components, and to automate the complicated inspection of industrial parts.[51] **Figure 7** highlighted some of the areas of AI application in manufacturing, showing that supervised, unsupervised, and reinforcement learning reach out to downstream tasks such as 1.) Predictive maintenance, 2) Energy consumption, 3) Industrial automation, 4) Quality control, 5) Process optimization, 6) Human-machine interaction, and 7) Physical and cyber security, via champion algorithms like regression, classification, anomaly detection, generative models, CV, and NLP.

**Figure 8 and 9** showcased an example to use AI to automate chemical and materials research towards self-optimized experimental design and execution.[5] In this example, a self-driving laboratory named “Ada” has two robots (N9 and UR5) with their work envelopes overlapping with each other to both synthesize and characterize a series of thin film samples. The robot N9 has a 4-axis robotic arm, and is used to mix, cast, and anneal towards target thin-film samples. Imaging and 4- probe conductance measurements are performed subsequently within the same system. The other robot UR5 is a 6-axis arm system, and is used to move samples and check palladium content under an XRF microscope. After a sample is made and characterized, a machine learning algorithm will predict and decide the synthesis parameters of the next sample to be created.[5] In **Figure 9**, the synthesis conditions on the Pareto front are plotted with colors, with synthesis conditions away from the Pareto front in gray. In this reaction that aims to maximize the resultant thin-film conductivity, $x$ determines the fuel blend, and φ refers to the fuel-to-oxidizer ratio.[5] **Figure 9b** shows an empirical Pareto front from each of the four experimental campaigns to exhibit a trade-off between combustion temperature and conductivity.[5]

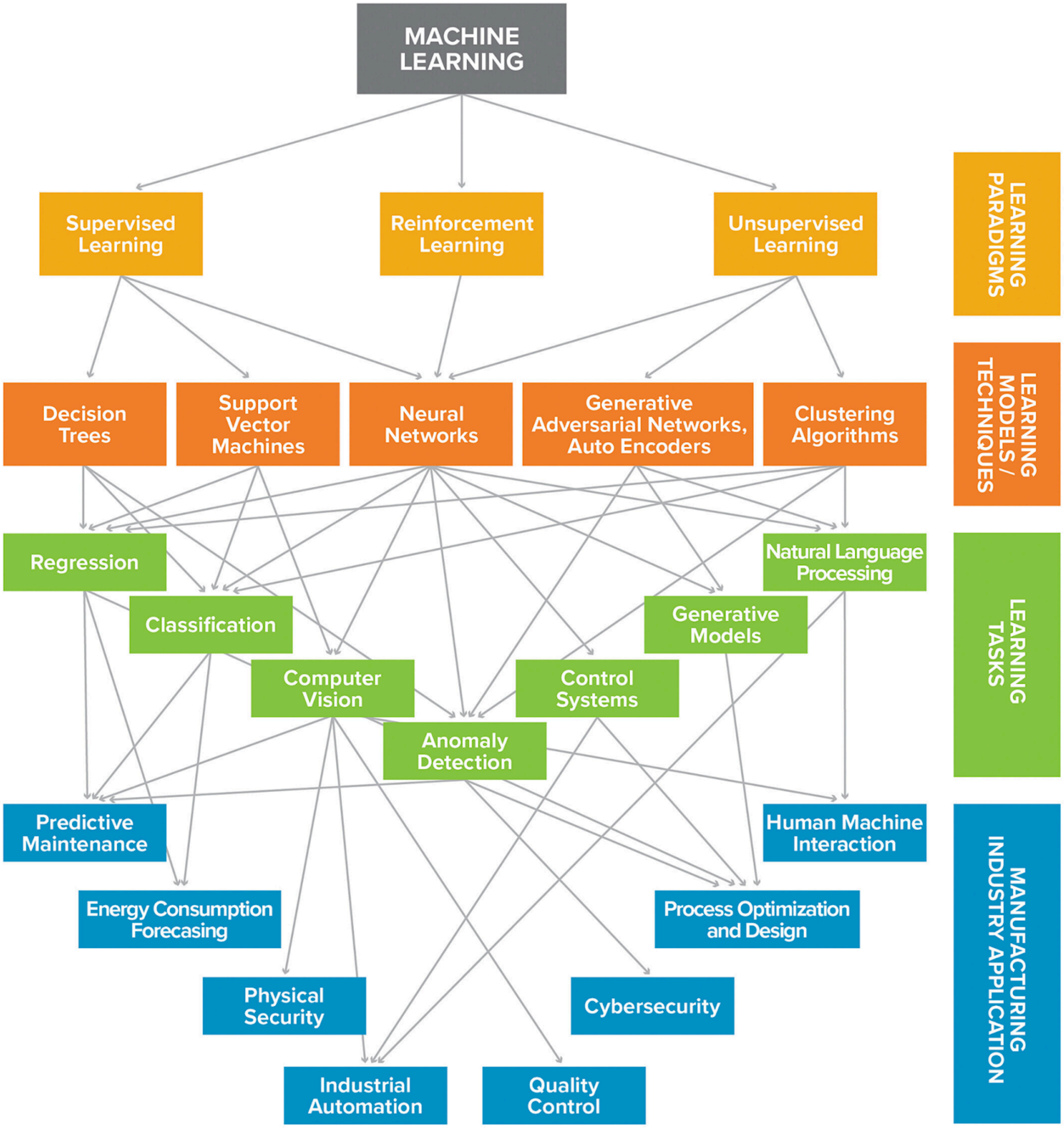

**Figure 7.** Common categories for various aspects of machine learning, grouped into paradigms, techniques, tasks, and relevant manufacturing industry applications. (Figure and Caption from reference,[53] without change, under CC-BY 4.0 License.[28])

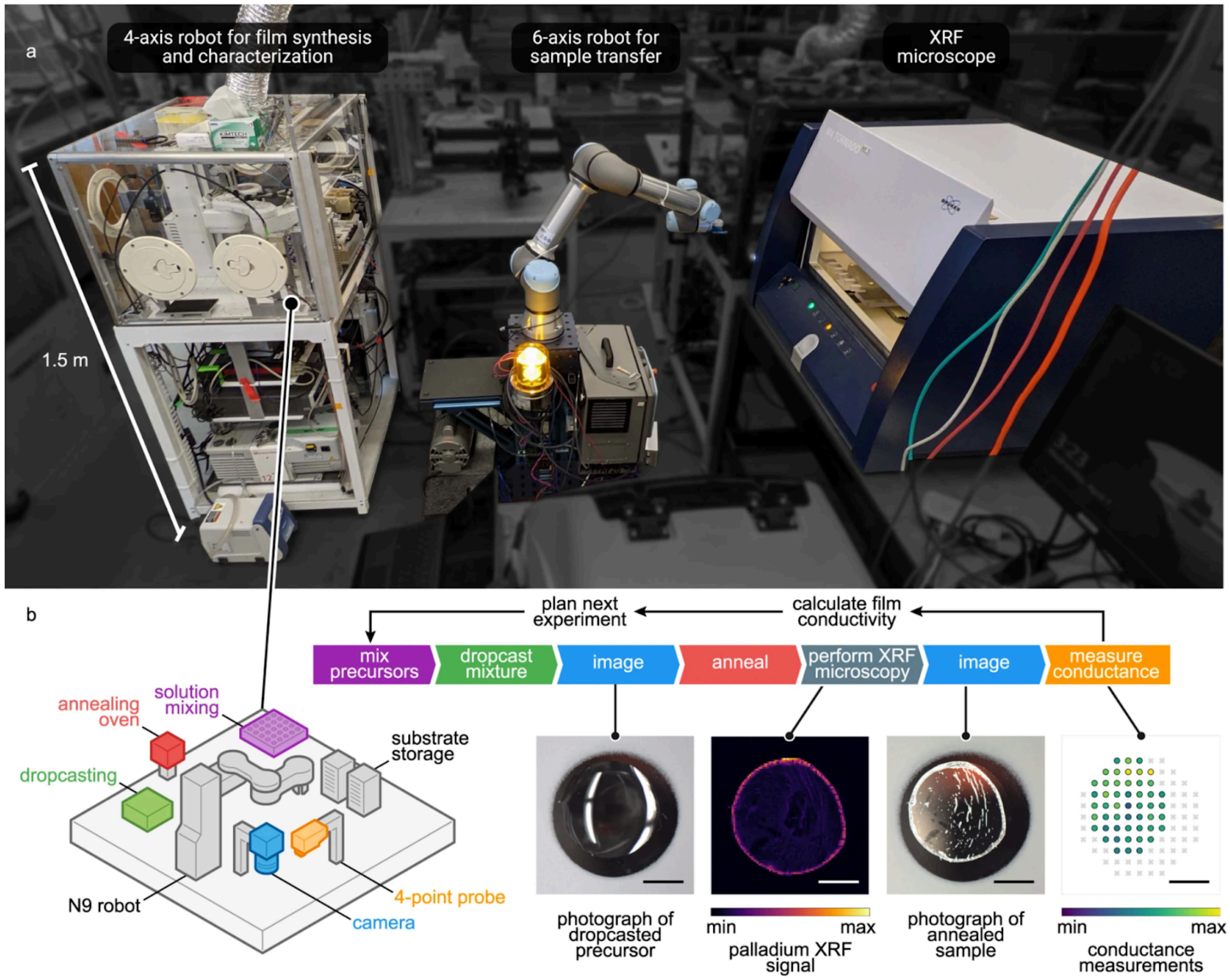

**Figure 8 a.** Schematic of the Ada self-driving laboratory. Ada consists of two robots (N9 and UR5e) with overlapping work envelopes. These robots work together to synthesize and characterize thin film samples. The N9 robot is a 4-axis arm equipped to mix, drop cast, and anneal precursors to create thin-film samples. The N9 also performs imaging and 4-point probe conductance measurements on the films it creates. The UR5 robot is a larger 6-axis arm equipped to transport samples to additional modules, including an XRF microscope. **b**. Steps in the automated experimental workflow. Each iteration of the experiment produces a single, drop-cast, thin-film sample; images of the sample before and after annealing; an XRF map of the quantity of palladium in the film; and a map of the film conductance measured by the 4-point-probe at different locations on the sample. After the sample is characterized, the film conductivity is calculated and the qEHVI algorithm is used to autonomously plan the next experiment. All scale bars are 5 mm. (Figure and Caption from reference[5], without change, under Creative Commons Attribution 4.0 International License[28].)

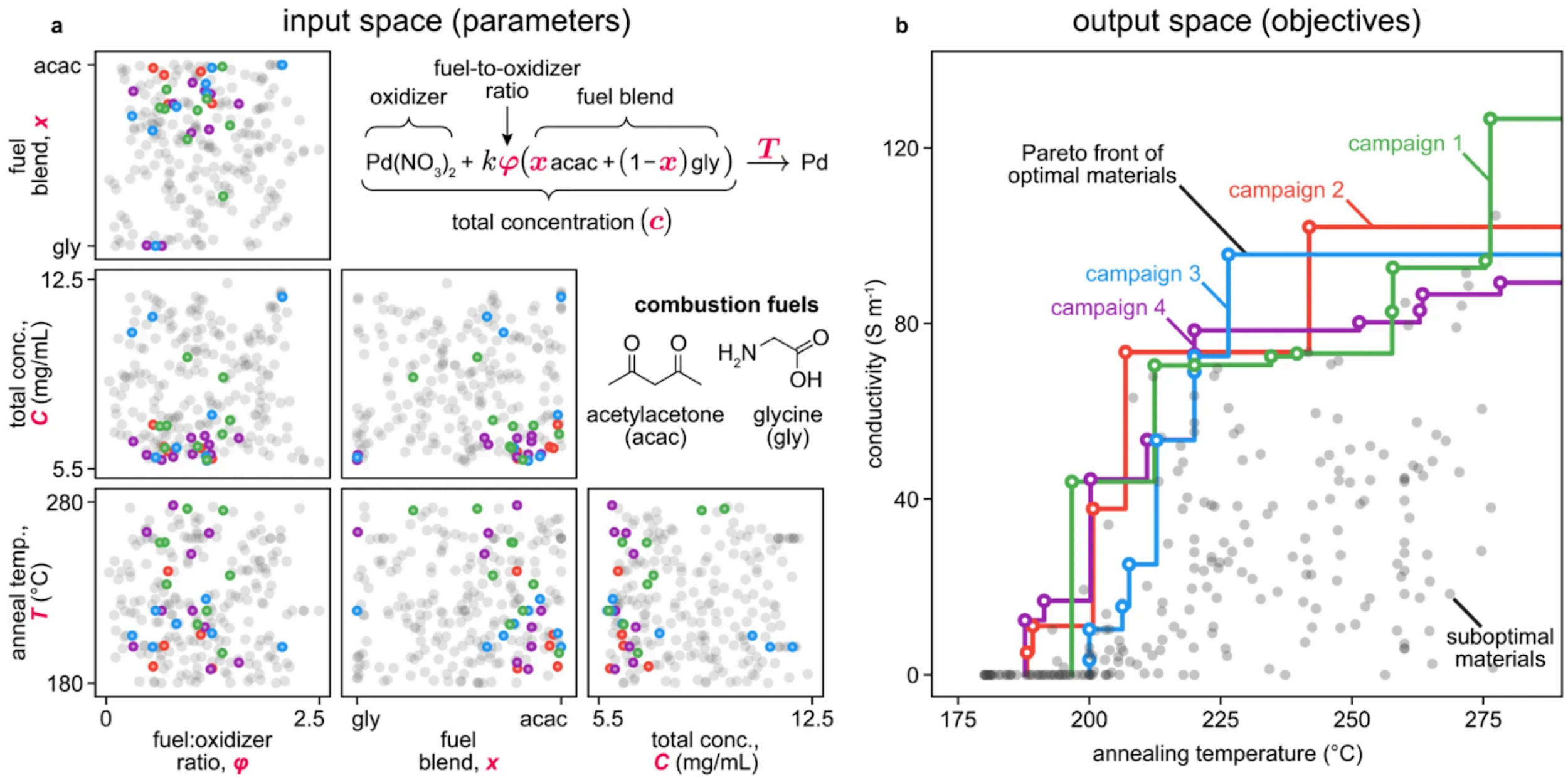

**Figure 9. a** Maps of the combustion synthesis conditions required to obtain experimental outcomes on the Pareto front. Sampled points not on the Pareto front are shown in gray. The combustion synthesis reaction, with parameters manipulated during the optimization highlighted, is shown. In this reaction $x$ controls the fuel blend and φ is the fuel-to-oxidizer ratio, as calculated using Jain's method (see Supplementary Methods). The fuel-to-oxidizer ratio used is scaled by $k = 36/5(8 - 5x)$ to account for the differing reducing valences of the acetylacetone and glycine fuels. **b** The empirical Pareto fronts from each of the four campaigns (solid lines) reveal the observed trade-off between temperature and conductivity. The experimental points which define the fronts are shown with open markers. Sampled points not on the Pareto front are shown in gray. (Figure and Caption from reference[5], without change, under Creative Commons Attribution 4.0 International License[28].)

A few review papers [53–55] in recent years highlighted on-going progress and challenges of machine learning and deep learning on manufacturing. For example, Kim, *et al.*, reviewed AI advances in manufacturing with regard to i) product enhancement and ii) process enhancement.[54] The former covers a) autonomous driving, b) battery management, c) robotics, and d) renewable energy generation, while the later addresses scenarios such as e) steel production, f) semiconductor processing and wafer fault detection. Xu, *et al.*, reviewed AI advances in manufacturing with a mathematical-oriented approach, analyzing challenges and solutions for data quality, data security, and model reliability.[55] Plathottam, *et al.*,[53] discussed some of the popular machine learning algorithms for i) manufacturing operations and ii) designs. The former includes a) real-time operations in terms of process optimization and safety, as well as b) planning: predictive maintenance, quality assurance, energy consumption estimate, and supply chain, while the latter includes c)generative design and d) automated experimentation. In addition, a number of works are dedicated to applying machine learning in characterizing and designing soft matter.[56–61]

Herein we are adopting a chemistry- and engineering- oriented perspective, and breaking it down into the following specific areas of manufacturing, to discuss AI of smart and scalable manufacturing in 1) Battery, 2) Flow Chemistry, 3) Additive Manufacturing, 4) Electronics, and 5) Machine Vision.

## 2.1 Battery

The demand for improved energy storage devices, particularly batteries, has increased significantly due to their crucial role in various technologies we use daily.[62] While lithium-ion batteries (LIBs) have been around since the 1970s and have proven effective, there is still a large room for improvement in terms of life, performance, cost, and safety.[63–72] In this context, we elaborate on two subtopics. (1) **Battery Management Systems (BMS)** and **Battery Life Model**s: one of the significant concerns in the LIB industry is battery failure modes, such as internal shorts, which pose safety risks. An accurate estimate of battery lifetime can be achieved through data-driven approaches to partially replace costly accelerated aging tests.[73–77] Additionally, efficient heat and operating management are crucial for high-power devices to prevent damage and extend battery lifespan.[62,78] (2) **Battery Design**: Machine-learning-guided performance prediction, as well as cell and material design, are also important aspects of battery AI.[79–82]

Battery life and management are of intense interest, because the battery is a complicated electrochemical system with a combination of mechanical, thermal, spatial and time dependent behaviors, for widespread applications such as personal electronics and electrical vehicles (EV). Consequently, BMS is an electronic system designed to monitor and control the operation of a battery, with its primary purpose to ensure the battery's optimal performance, safety, and longevity.[62] To maximize battery performance and lifespan, it is crucial for a BMS to monitor and optimize battery parameters such as state of health (SOH), state of charge (SOC), and state of power (SOP).[62] BMS also manages temperature levels, detects faults, and provides diagnostics. SOH reflects the battery's current capacity compared to its initial state. SOC indicates the battery's current energy level or capacity in relation to its maximum capacity. SOP measures the ratio of peak power to nominal power, signifying the battery's power delivery capacity.[62] AI techniques are well suited for predicting SOH and SOC by learning from historical data to capture the complex behavior of batteries.[83,84] Predictive AI/ML algorithms play a crucial role in extracting complex and nonlinear patterns from training datasets, translating the underlying metadata into statistical models and enabling accurate state estimation and monitoring of the battery system.[85]

A machine learning pipeline for SOH estimation is shown in **Figure 10**, which comprises 4 steps: 1) data acquisition, 2) data processing, 3) model training, and 4) output (SOH estimation).[85] **Step 1**) includes collecting datasets on voltage, current, temperature, and SOH reference. SOH estimation has direct and indirect approaches (**Figure 11)**. Indirect SOH estimations include model-based, data-driven, and hybrid approaches, while direct-measurement methods are based on capacity, impedance, and internal resistance readings.[85] It is noted that the direct measurement methods are highly demanding for resources and may not be practical outside a laboratory setting, while the data based methods can establish a correlation between SOH and real-time battery signals to provide accurate SOH estimations efficiently. Compared to model-based methods that require precise but costly models (i.e. equivalent circuits models, electrochemical models), data-driven methods rely on high quality training data and machine learning algorithms. **Step 2**) covers feature extraction (incremental capacity analysis or ICA, differential voltage analysis or DVA, and time interval), correlation study, and normalization. Feature engineering represents a critical step to implement ML algorithms, and model performance closely depends on the selected features. Four types of features are shown in **Figure12,** including incremental calculation, time, envelope area, as well as model parameter.[84] **Step 3**) Typical machine learning algorithms includes back propagation neural network BPNN, support vector machine SVM, long-short-term memory LSTM network, Gaussian process regression GPR, and ensembles. **Step 4**) Finally, model evaluation can be based on Mean Absolute Error (MAE), Root Mean Squared Error (RMSE), confidential interval (CI), and computation cost.

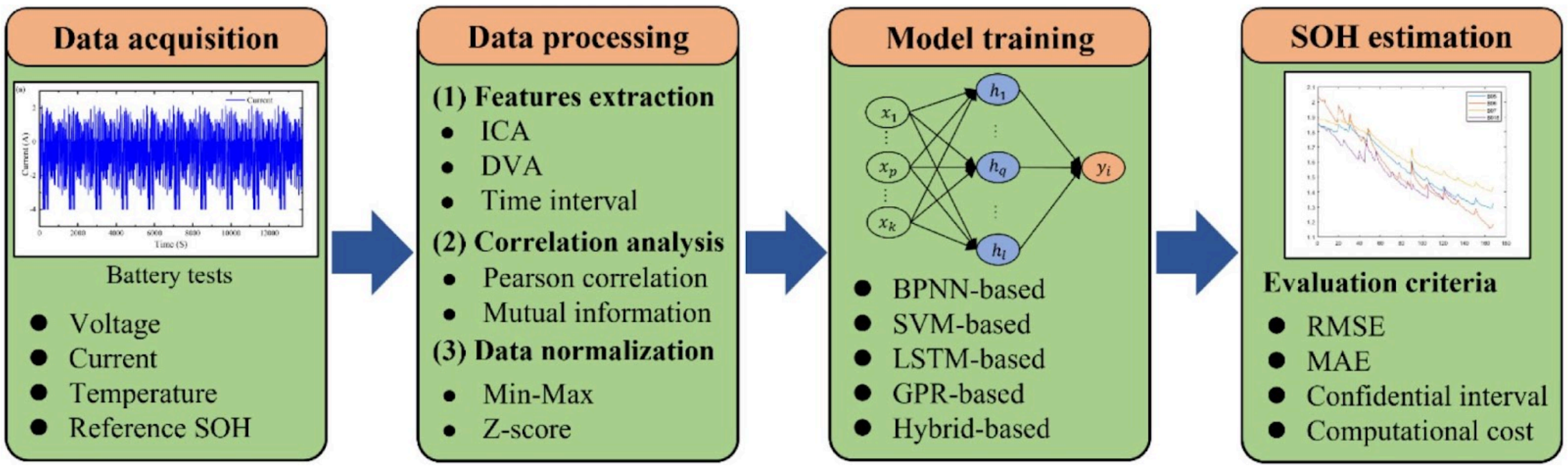

**Figure 10**. The procedures for developing ML SOH estimation algorithms. (Figure and Caption from reference[85] without change, under CC-BY License[28].) Abbreviations: incremental capacity analysis ICA, differential voltage analysis DVA, back propagation neural network BPNN, support vector machine SVM, long-short-term memory LSTM network, Gaussian process regression GPR, Mean Absolute Error MAE, Root Mean Squared Error RMSE.

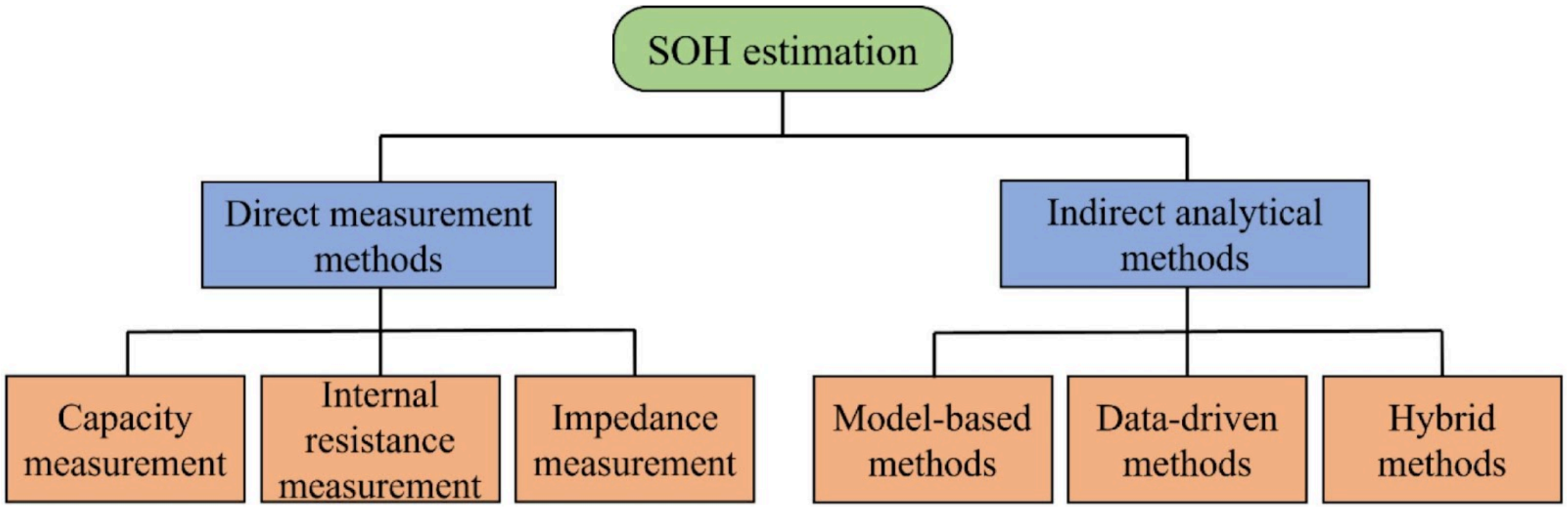

**Figure 11**. Classification of SOH estimation methods. (Figure and Caption from reference[85] without change, under CC-BY License[28].)

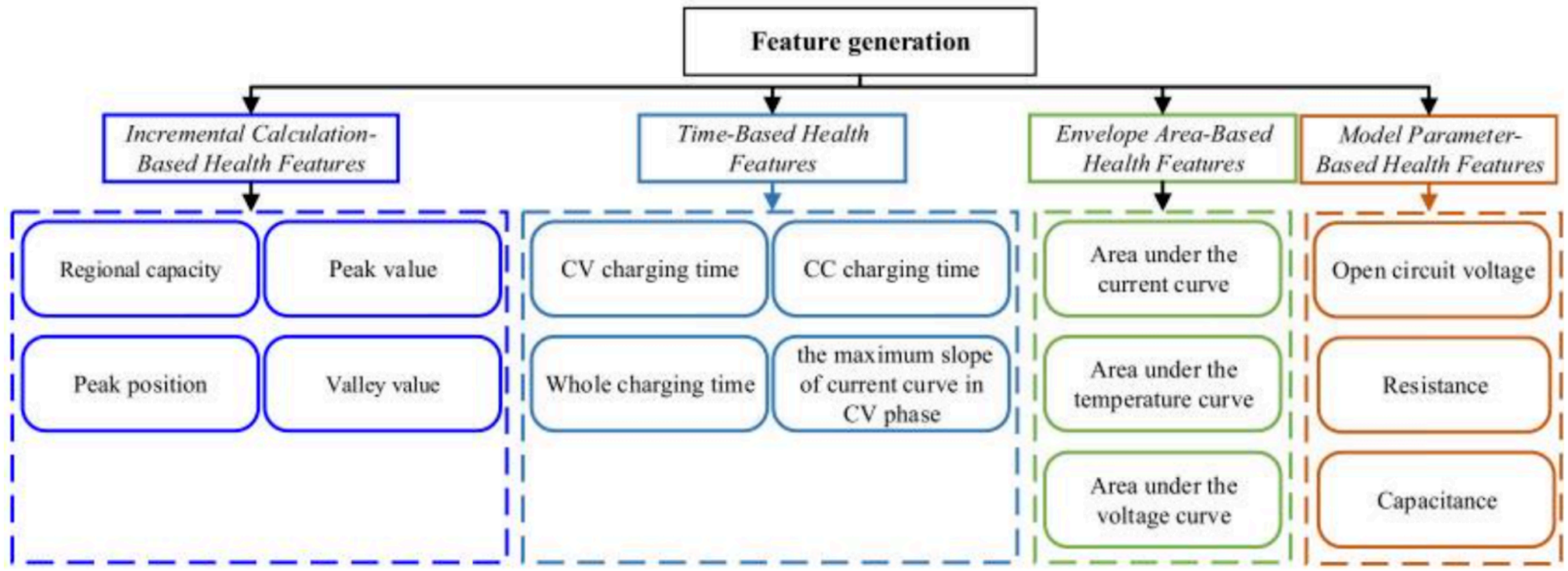

**Figure 12.** Summary of different health features (Figure and Caption from reference[84] without change, under CC-BY License[28].)

Unlike SOH estimation, there are no direct ways to measure SOC.[85] A machine learning pipeline for SOC estimation is shown in **Figure 13**,[85] which is in general similar to the four steps described earlier for SOH estimation. SOC estimation methods include 1) look-up table methods, 2) Ampere-hour integral methods, 3) filter-based methods, 4) observer-based methods, as well as 5) data-driven methods (**Figure 14)**.[85] Methods 1) and 2) are straightforward, but suffer from low accuracy and vulnerability to sensor errors. Methods 3) and 4) can be highly accurate and robust to noises, but require resources to build a well-parameterized model beforehand. In contrast, data or AI based methods require little pre-knowledge of the battery system, but the quality of the training data and the parameters of machine learning algorithms can be the key to fight against overfitting (and sometimes underfitting).

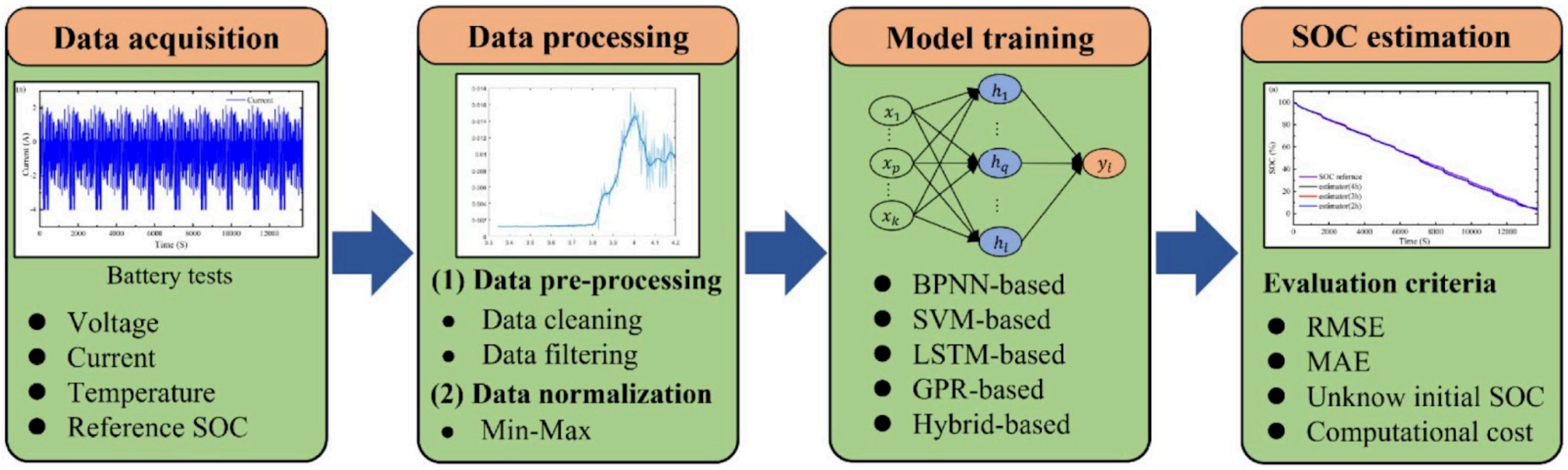

**Figure 13**. The procedures for developing ML SOC estimation algorithms. (Figure and Caption from reference[85] without change, under CC-BY License[28].) Some abbreviations are shared with **Figure 10**.

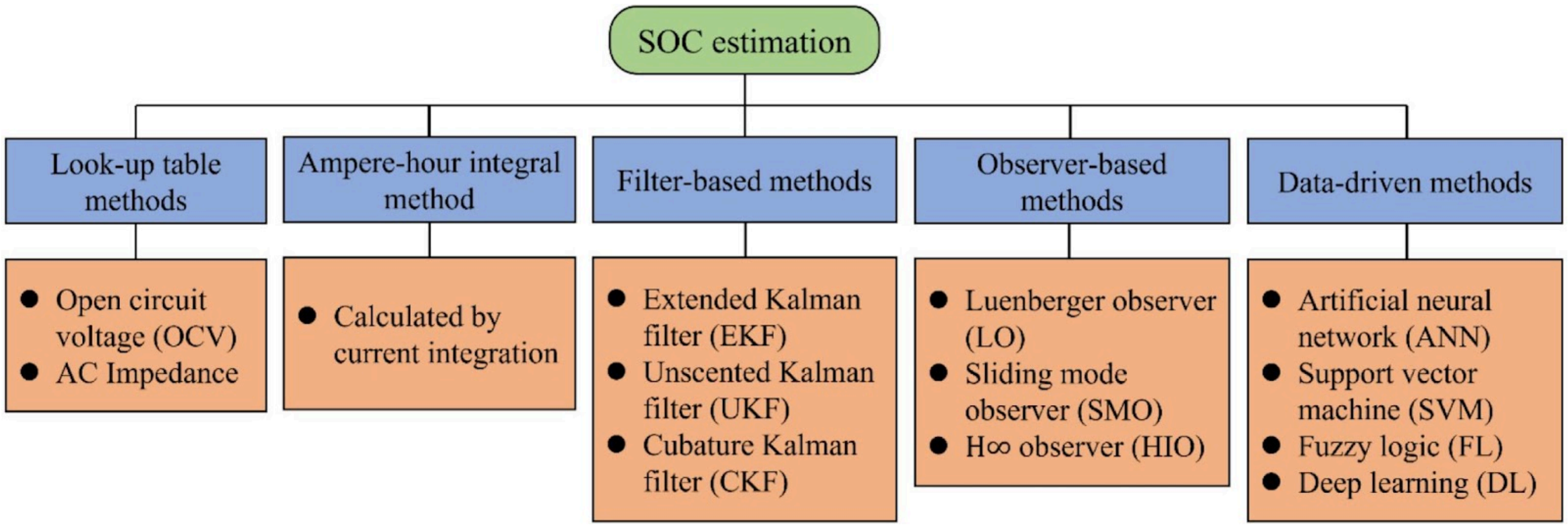

**Figure 14**. Classification of SOC estimation methods. (Figure and Caption from reference[85] without change, under CC-BY License[28].)

Battery lifetime models provides critical information on mechanical and chemical degradation, which remain challenging considering the fact that it may take years to fully characterize thousands of cycles with traditional testing methods and the aging of batteries can be highly cell-chemistry-dependent.[73,75–77,86] To tackle these challenges, machine learning has been used to 1) aid identification of battery lifetime models with significantly reduced uncertainty,[73] 2) train battery lifetime models [77] to compare against a semi-empirical model, or 3) combine with physics models in various ways to achieve enhanced results.[76] As one of the key indicators of battery lifetimes, the remaining useful lifetime (RUL) is the duration between the first arrival (or observation) of the battery and its end-of-life (EOL), which can be largely nonlinear and dependent upon history of the battery usage, affected by temperature, charge/discharge rate, side reactions, conditions of battery components (anode, cathode, electrolytes, and interfaces).[87–89] The prediction of battery RUL can be approached in three ways (**Figure 15**): 1) deterministic machine learning, 2) filter-based methods, and 3) stochastic methods.[87,88,90] It is noted that, sometimes, stochastic methods and filters may be included as machine learning methods, but here they are listed as separate classes to emphasize their common characteristics. Compared to model-based filter methods (Kalman filters and Particle filters), deep learning and other machine learning methods can have a downside because of the complex nature of the algorithms, which may contribute to overfitting and explainability issues. Nevertheless, they may offer unmatched advantages of higher accuracy, fast response, and simple input (**Table 3**).[87,91]

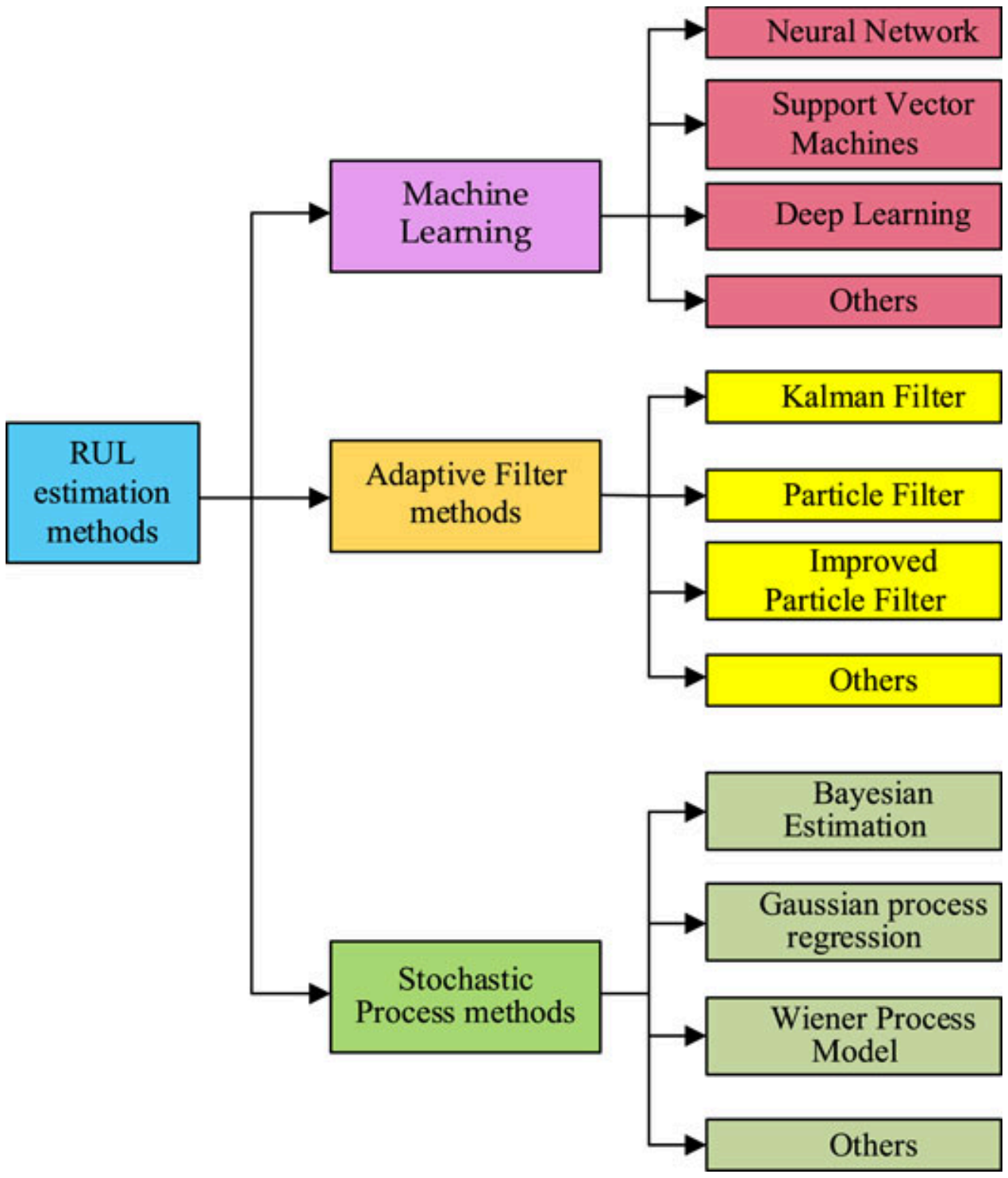

**Figure 15**. The main RUL prognosis methods. (Figure and Caption from reference[87] without change, under CC-BY License[28].) It is noted that stochastic and filter methods are sometimes included as machine learning methods, but here they are listed as separate classes to emphasize their common characteristics in contrast to deep learning and other machine learning algorithms

**Table 3.** The comparison of ML, Filter, and Stochastic process methods. (Table and Caption from reference[87] without change, under CC-BY License[28].) It is noted that stochastic and filter methods are sometimes included as machine learning methods, but here they are listed as separate classes to emphasize their common characteristics in contrast to deep learning and other machine learning algorithms.

| Method | Accuracy | Input complexity | Forecast cycle percentage | Advantage | Disadvantage |
|---|---|---|---|---|---|
| Machine learning | High | Simple | Less | High accuracy<br>Simple input<br>Short forecast period | Complex algorithm |
| Adaptive filter methods | Medium | Complex | Many | Simple algorithm | Complex input<br>Long forecast period |
| Stochastic process methods | Medium | Complex | Medium | Simple algorithm | Complex input |

In the area of battery design, machine learning is accelerating the discovery of newer and better battery components. Kim, *et al.,*[82] used machine learning models to identify energy-dense, durable, and cost-effective cathodes out of 1617 candidates in the categories of Ni-rich layered cathodes, specifically $LiNi_xCo_yMn_zO_2$ (NCM) and $LiNi_xCo_yAl_zO_2$ (NCA). Lv and coworkers[81] reviewed and summarized over 20 previous works on the application of AI in optimizing or predicting behaviors of battery cathode, anode, as well as liquid and solid electrolyte; and in a case with solid electrolytes,[92] screening of 13,000 candidates is achieved. Furthermore, a closed-loop framework was proposed for the design, discovery, property prediction, and battery management using digital twins and AI capabilities.[92] The framework comprises three modules: 1) the physical system/scenarios, 2) the digital twin, and 3) the AI engine. The physical system/scenarios module represents the actual system, in battery design and material discovery. The digital twin module encompasses digitalized models of real-world scenarios, including both rule-based models derived from physical principles and data-driven models trained using extensive historical data from the real system. These models strive to accurately replicate the real environment within the digital twin module. The AI engine module employs learning-based approaches to optimize, diagnose, and control real-world systems. The three modules interact with each other through various mechanisms. Descriptive AI is employed to learn system behaviors, prescriptive AI is adopted for cyclic training via digital twins to optimize and solve real-world problems, and predictive AI is used to forecast future system states based on given inputs.[92] The proposed framework offers several benefits, including data enrichment, efficient and safe deployment.

## 2.2 Chemistry and Automation

In recent years, there has been a significant expansion in the utilization of artificial intelligence in the field of chemistry. To illustrate the evolution of AI, Baum, *et al.*[13] studied 70k papers and 17.5k patents from AI-related Chemical Abstracts Service (CAS) content collections over the past 20 years, and conducted quantitative topic analysis in 12 categories (1. Pharmacology & Toxicology, 2. Food & Agriculture 3. Biochemistry, 4. Physical Chemistry, 5. Materials Science, 6.Analytical Chemistry, 7. Energy Technology, 8. Industrial Chemistry, 9.Inorganic Chemistry, 10. Organic Chemistry, 11.Synthetic Polymers, 12. Natural Products). Biochemistry and Analytical chemistry have the largest number of AI-related papers (18k and 12k, respectively).[13] Furthermore, 15k articles and 3k journal articles are identified as interdisciplinary in this analysis, with a determined primary and secondary disciplines for each work from one of the 12 identified categories.[13] In journal publications, strong correlations between primary and secondary research areas are shown in Category 6 and 3 (Analytical Chemistry vs. Biochemistry), Category 5 and 4 (Materials Science vs. Physical Chemistry), as well as Category 3 and 1 (Biochemistry vs. Pharmacology & Toxicology). In patent publications, a similar trend is observed, with additional strong correlations between Category 7 and 8 (Energy Technology vs. Industrial Chemistry).[13] The results in **Figure 16** demonstrate how AI unified disciplines together (shown as positive values), and how AI is falling behind within other disciplines (shown as negative values).

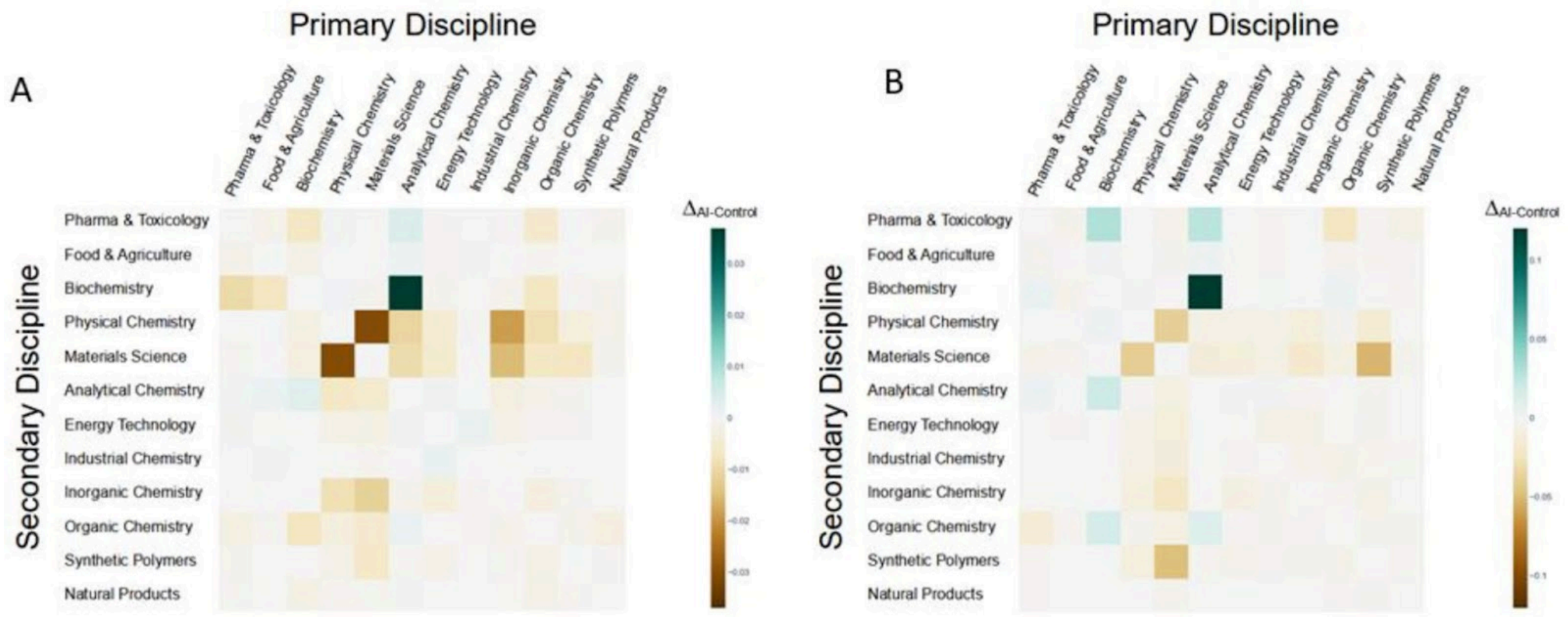

**Figure 16**. Difference in proportion of total AI-related publications and control group (non-AI related) by interdisciplinary pair: (A) journal publications and (B) patent publications. (Figure and Caption from reference[13] without change, under CC-BY 4.0 License[28].)

Network analysis of co-occurred concepts reveals trends of AI research in chemistry along the years.(**Figure 17**) The use of AI in drug discovery and quantitative structure–activity relationships (QSAR) is associated with neural networks since 2000,[13] while previously unseen research topics, such as Diagnosis, Prognosis, Peptides, and Transcription factors, are more and more connected to algorithms such as neural networks, SVM, random forest, and deep learning after 2014.

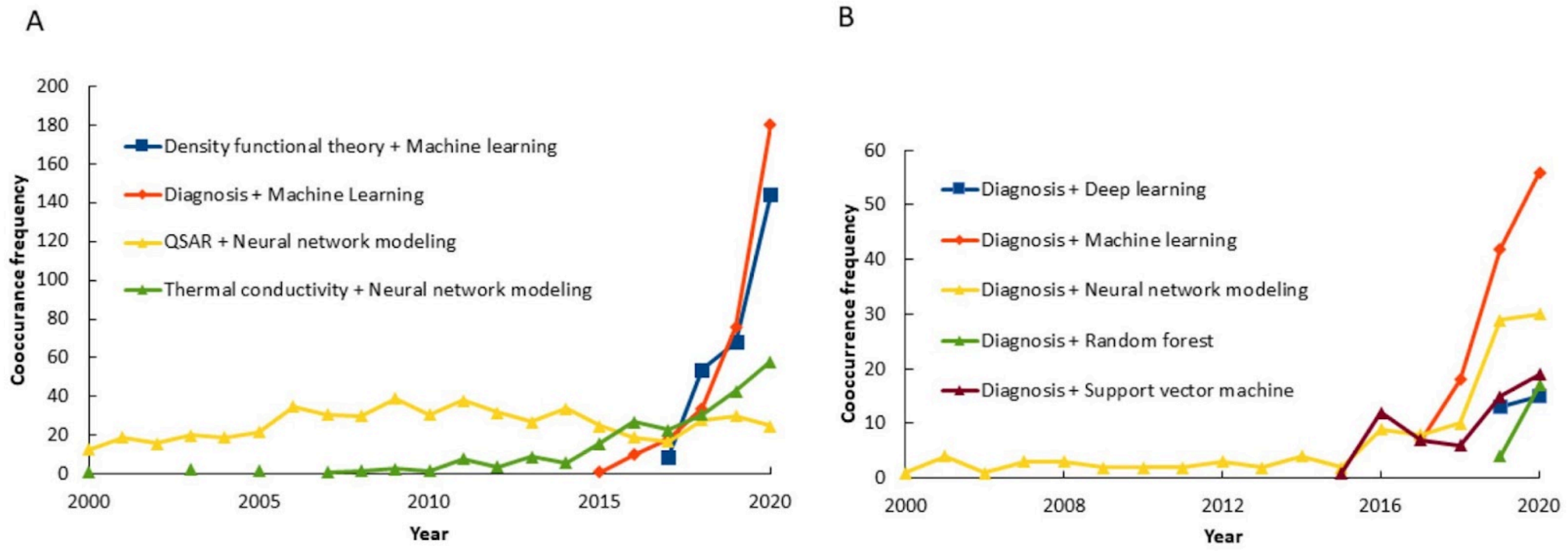

**Figure 17**. Trends of co-occurrence in scientific publications for selected research topics and AI algorithms: (A) journal publications and (B) patent publications. (Figure and Caption from reference[13] without change, under CC-BY 4.0 License[28].)

Flow chemistry refers to chemical synthesis conducted in a continuous stream rather than using conventional batch processing.[11,12,93–97] Instead of placing reactants in a flask, flow chemistry system pumps the reactants through an in-stream optimized reactor to enable continuous reactions with improved control, efficiency, safety, and sustainability. Some key benefits of flow chemistry include **1)**. Fast and efficient reactions, improved mixing and heat transfer, the ability to conduct reactions safely at higher temperatures and pressures, **2)**. automated optimization and simplified scale-up. [11,12,93–97] Although flow chemistry requires higher initial capital investment, improved process efficiency can provide rapid payback. With its many advantages, flow chemistry is seeing greatly expanded use in drug development, active pharmaceutical ingredient (API) production, specialty and fine chemical synthesis, petrochemical processing, and other applications.[11,12,93–97]

For flow reactors, single objective optimization can be achieved with algorithms such as Simplex[98,99] and Stable Noisy Optimization by Branch and Fit (SNOBFIT).[100,101] SNOBFIT aims to efficiently find the global optimum of the objective function by iteratively exploring the search space and adaptively refining the sampling points. An example showcasing the application of SNOBFIT algorithm in flow chemistry can be found in a study by Holmes, *et al.*[101] in which "self-optimizing" flow reactors are integrated with online high-performance liquid chromatography (HPLC) to make AZD92921, an irreversible epidermal growth factor receptor kinase inhibitor developed by AstraZeneca. An impressive 89% yield of AZD9291 was achieved after a second round of optimization.

In the context of flow chemistry, BO efficiently explores the high-dimensional search space with limited data, and converges towards favorable solutions while minimizing the overall number of required experiments. Xie, *et al.*[102] utilized a platform that integrates a synthesis robot with the Bayesian Optimization (BO) algorithm to improve the crystallinity of synthesized Metal-organic frameworks (MOFs) ZIF-67 in highly complex and non-linear chemical and reaction spaces. BO was chosen for its simplicity of implementation and its excellent ability to process sparse datasets. Bayesian methods utilize probabilistic models to strike a balance between exploration and exploitation, making them well-suited for handling noisy and computationally expensive objective functions. Kwon, *et al.*, reported a BO powered automated microreactor to explore the synthesis of thioquinazolinone.[96]

Multi-objective optimizations are common for flow chemistry and other chemical engineering processes.[95,97,103,104] In addition to genetic algorithms[103] and Pareto Efficient Global Optimization (ParEGO),[105] Lapkin and coworkers adapted BO into Thompson Sampling Efficient Multi-Objective (TS-EMO) to tackle multi-objective optimization in automated flow-chemistry processes.[11,12] TS-EMO uses Gaussian processes as surrogates, and use Thompson sampling with the hypervolume quality indicator and Non-Dominated Sorting Genetic Algorithm II (NSGA-II) to decide the next evaluation point after each iteration. The hypervolume of a Pareto front or a set of optimal solutions in a multi-objective optimization, is the volume or area of the criterion space defined by the Pareto curve. The hypervolume indicator is a performance metric that evaluates the quality of a non-dominated approximation set. It considers the proximity of the points to the Pareto front, diversity, and spread. TS-EMO has the advantages of 1) no requirement of *a priori* knowledge, 2) reduced hypervolume calculations, 3) robust to noise, and 4) the capability of batch-sequential usage.[106] **Figure 18a** provides a flow-chemistry case study in which the condensation of benzaldehyde (1) and acetone (2) are catalyzed by sodium hydroxide (3) to yield benzylideneacetone (4). The formation of dibenzylideneacetone (5) and acetone polymerization (highlighted in red) are to be avoided or minimized by a self-optimizing algorithm.[11] **Figure 18b** is a TS-EMO self-optimization flow reactor built from commercial equipment including R2 and R4 modules from Vapourtec for solution flows and reactor temperatures control.[11] In this case, a MATLAB user interface was used to control the self-optimizing flow reaction, which includes controlling physical equipment through a software interface, generating training data, reading HPLC signals, evaluating optimization objectives, and performing further iterations. The four variables optimized are 1) the molar ratio of acetone and benzaldehyde, 2) the molar ratio of sodium hydroxide and benzaldehyde, 3) temperature (<70 °C), and 4) residence time. This self-optimizing flow reactor can efficiently identify optimal reaction parameters and Pareto front trade-offs between multiple objectives such as yield, cost, space-time yield (STY reactor productivity,or product mass divided by reactor volume times residence time), and E-factor (waste mass divided by product mass).

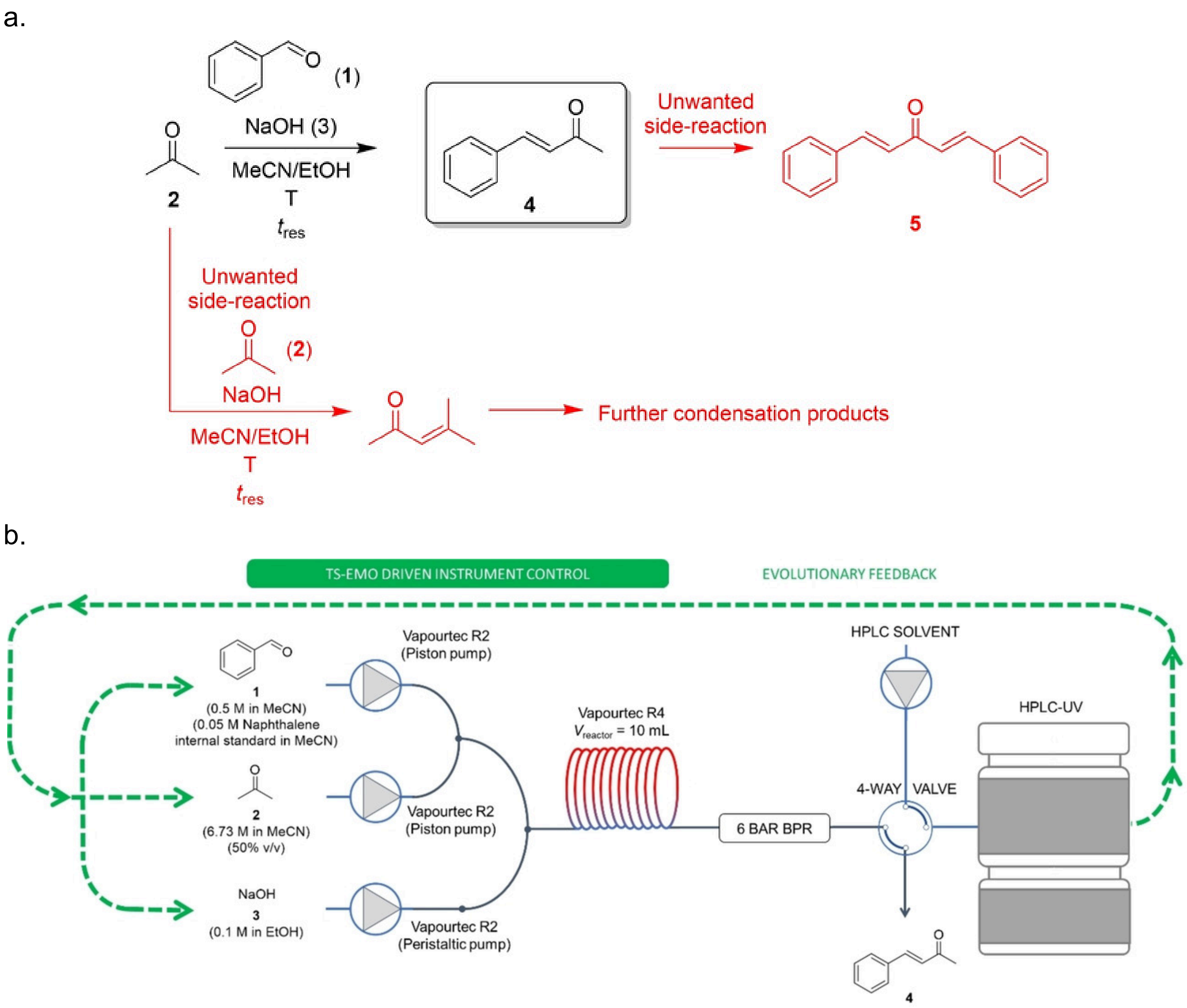

**Figure 18 a.** Reaction scheme for the sodium hydroxide (3) catalyzed Aldol condensation case study between benzaldehyde (1) and acetone (2) to produce benzylideneacetone (4) at reactor temperature, *T*, with residence time, *t*res. **b**. Schematic of the self-optimization systems containing a Vapourtec flow chemistry pumps and reactor, 4-way sample injector, HPLC-UV analysis, and algorithmic reaction optimization, controlled using a MATLAB based environment. BPR: back pressure regulator. (Figures and Captions from reference[11] without change, under Creative Commons Attribution 4.0 International License.[28])

Reinforcement learning (RL) offers additional valuable and powerful opportunities in recent competition with human experts in activities such as the Game of GO.[107] Unlike traditional black-box methods that focus on observing outputs corresponding to given inputs, RL monitors the system state and maps actions to the corresponding state response, which is well-suited for multi-step processes such as automated flow chemistry. RL breaks down decisions into isolated steps, predicting the future effects of those smaller steps.[107] RL algorithms then learn through iterative interactions within a designated environment and discover optimal strategies for these multi-step processes. However, it is noted that RL may require extensive exploration to identify optimal policies, which can be computationally expensive.[107] Volk *et al*,[94] showcased the capabilities of reinforcement learning systems in tackling challenges related to multi-step nanoparticle chemistry. In this work, RL based AlphaFlow is used as a self-driven fluidic laboratory to autonomously explore and discover core-shell semiconductor particles, performing reactions of variable sequence, phase separation, cleaning, as well as spectral monitoring. The RL agent in AlphaFlow assesses the reactor's state and response, and makes informed decisions on the next best action to navigate through a high-dimensional space efficiently, all without prior knowledge of conventional synthesis parameters.

Furthermore, Neural Networks (NN) algorithms and a closed loop system were used by Caramelli *et al*[108] to discover new organic compounds and chemistries. The system had three main parts: a liquid handling robot responsible for executing and analyzing reactions, an online analytics system for interpretation of NMR and Mass Spectra, and an algorithm that established correlations between reaction outcomes and a decision on next moves. The software was designed to run up to 36 reactions/day, each with a reaction time of 3.3 h, and it was able to navigate a chemical space made of six simple molecules mixed in binary and tertiary random combinations.[108] H NMR spectra data from 440 reactions was collected and used to train the convolutional NN (CNN) which emulates the reactivity assignments made by a human experimenter.[108] The model's performance was validated on 1018 reactions between 15 different starting reagents. To facilitate exploration of the chemical space, a junction tree variational autoencoder algorithm was employed to convert the molecular structures of the reagents into fixed-length fingerprint vectors.[108] The NN powered robotic platform was able to autonomously explore a large number of potential reactions and assess the reactivity of mixtures.[108] As another example, Orimoto, *et al.*[109] used an NN-based algorithm to predict synthesized nanoparticle properties with high accuracy. This approach is ideal to extract condition–property relations from increasing large combinatorial chemistry data, even for new reaction parameters and unknown properties.

## 2.3 Additive Manufacturing

Additive manufacturing (AM) refers to a process that can build three-dimensional objects one layer at a time, which can involve jetting, energy deposition, lamination, extrusion, powder bed fusion, layer-by-layer deposition, as well as photopolymerization.[110,111] This definition outlines AM's characteristics of less reliance on tooling, suitability for high-value or low-volume production, connecting to issues in complex geometries, mass customization, and supply chains disruption for on-site production. Machine learning can be effectively incorporated with the automation involved with additive manufacturing.[112,113]An example of combining autonomous experimentation (**Figure 19**) and Bayesian Optimization (**Figure 20**) to improve additive manufacturing is given in **Figure 19-20,** with a reported 60-fold reduction in number of experiments needed to achieve high-toughness structures.

A good number of review papers are available on machine learning applications in additive manufacturing.[112,114–128] AI has significant applications for 3D printing in design automation, customization, generative design, and defect detection, for soft materials [129–133] as well as multi-material printing.[134–145] The combination of AI and 3D printing technologies [112,114–121] will not only automate the decisions on energy and system planning, parameter optimization and in-line monitoring, but also facilitate human-centered product developments (such as prostheses, therapeutic helmets, orthoses, and finger splints). AI aided AM poses well to transform digital manufacturing, supply chain logistics and product development across many industries. The involved algorithms include but are not limited to deep learning, SVM, ANN, Bayesian networks, computer vision, and NLP.[112,114–121]

Further progress in transfer learning, light-weight models, and cloud learning is still needed to realize advanced future applications in nanoscale printing, bioprinting, and 4D printed smart materials. Towards the full AI potential of manufacturing in Industry 4.0, the possible challenges are uncertainty quantification, data fusion, limitations in training datasets, explainability, wide range of spatial and temporal scales, as well as delays and latency in response.[112,114–121]

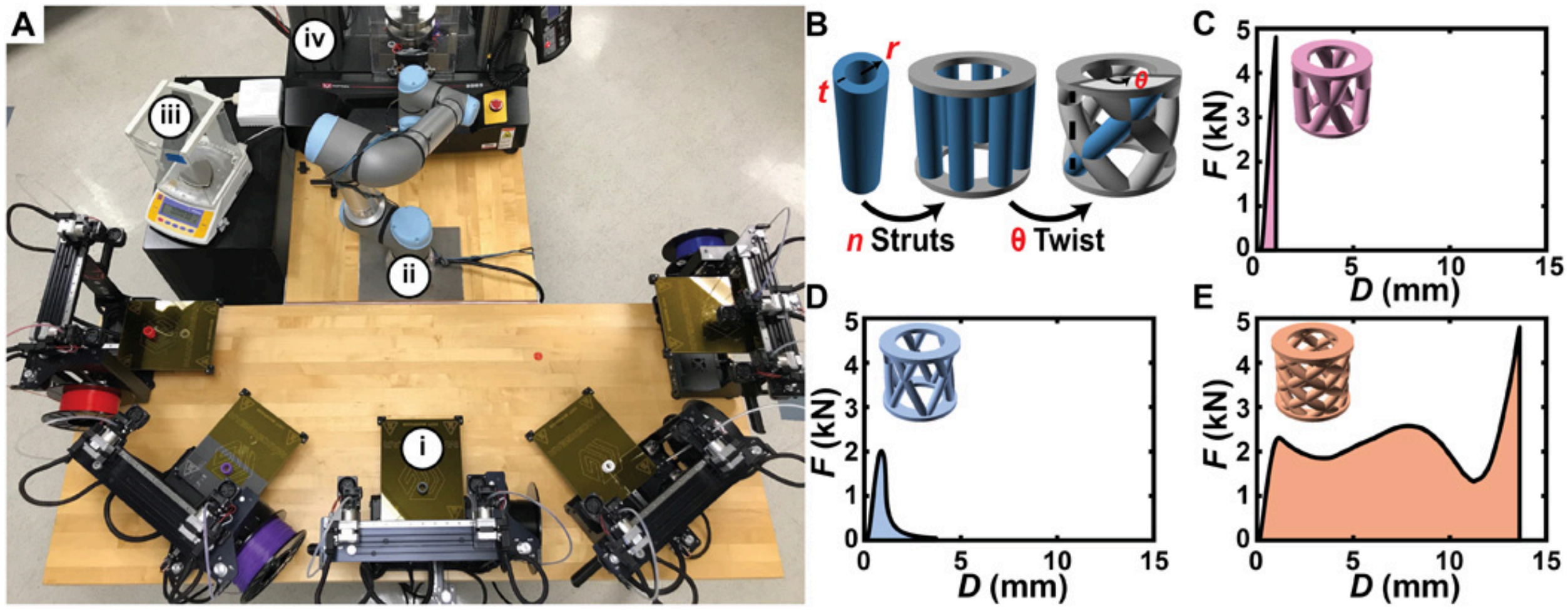

**Figure 19. BEAR for studying the mechanics of additively manufactured components.** (**A**) Experimental system composed of (i) five dual extruder fused deposition modeling (FDM) printers (M3, MakerGear), (ii) a six-axis robotic arm (UR5e, Universal Robotics), (iii) a scale (CP225D, Sartorius), and (iv) a universal testing machine (5965, Instron Inc.). (Photo credit: Aldair E. Gongora and Bowen Xu, Boston University). (**B**) Model "crossed barrel" family of parametric structures with two circular platforms that are held apart by a series of *n* hollow columns of outer radius *r* and thickness *t* and that are twisted with an angle θ. Force *F* and corresponding displacement *D* from the testing of (**C**) a crossed barrel that did not yield before ~5 kN (designated too strong), (**D**) a crossed barrel that failed in a brittle manner (designated "brittle"), and (**E**) a crossed barrel that exhibited appreciable strength after an initial yield point (designated "ductile"). (Figure and Caption from reference[6] without change, under Creative Commons Attribution 4.0 International License.[28])

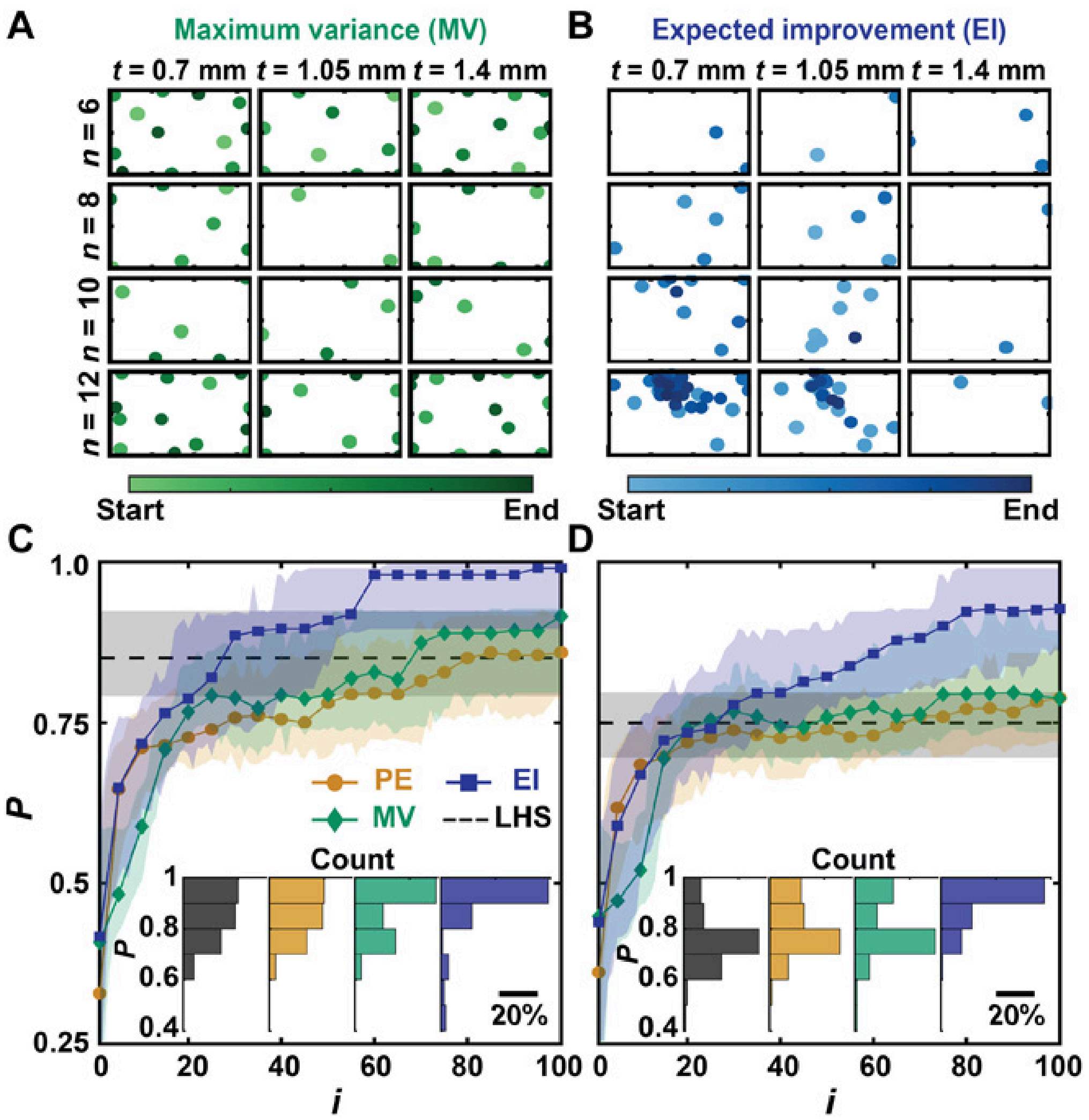

**Figure 20. Simulated learning using BO.** Distribution of experimental points when guided using (**A**) MV and (**B**) EI decision-making policies. The color gradient indicates the start and end of the campaign. Performance $P$ versus experiment number $i$ of simulated Bayesian campaigns with noise added to each simulated measurement drawn from a zero-mean Gaussian with (**C**) SD $\sigma = 0.1$ J and (**D**) $\sigma = 5$ J. EI- and MV-guided campaigns are benchmarked against PE and the average result of selecting 100 experiments using Latin hypercube sampling (LHS). Shaded regions correspond to the middle two quartiles of 100 simulated campaigns. The inset bar charts show the distribution in $P$ at $i = 100$.(Figure and Caption from reference[6] without change, under Creative Commons Attribution 4.0 International License.[28])

## 2.4 Sensors

The application of AI in the sensor market has been experiencing significant growth in recent years and is expected to nearly double by 2027 (compared to 2022).[146] Central part of this market growth is due to sensor fusion and exponential growth of Internet of Things (IOT) sensors. Combined with machine learning and predictive algorithms, sensors enable systems to monitor, understand and respond intelligently to their environment. As sensors and AI continue to advance, devices, vehicles and machines will gain human-like perception and sensitivity. But for life-critical applications like self-driving cars or medical diagnostics, AI must be carefully designed to handle errors or limitations in sensor data. There are several areas of AI in sensor fusion which can advance significantly: 1) Sensor fusion for Autonomous Vehicles, 2) Industrial Automation and Robotics, 3) Internet of Things (IoT), 4) Environmental Monitoring and Smart Cities, and 5) Healthcare and Wearable Sensors.

1. **Sensor fusion for Autonomous Vehicles** has gained significant traction with investment to advance technology of self-driving cars.[147–150] AI-powered sensor fusion techniques will combine data from video cameras, Light Detection and Ranging (LiDAR), radar, and other sensors, to enable real-time object detection, and improve decision-making technologies for self-driving cars.[147–150] Aside from cybersecurity, reliability, and repeatability of performance, of particular importance is the application of sensor fusion technology and new algorithm development to enable autonomous driving, especially under harsh weather, and/ or in complicated urban environments.[147–150]

Traditional sensor fusion algorithms used for autonomous vehicles include the following. **a**) *Kalman Filters (KF)*[151] *and Extended Kalman Filters(EKF)*[152] *based recursive estimation algorithms that can fuse data from multiple sensors, such as cameras, LiDAR, radar, and Inertia Measurement Units (IMUs), to estimate the state of the vehicle and surrounding objects.* Kalman filters [151] provide a mathematically optimal solution by iteratively updating the state estimate based on sensor measurements and system dynamics. The EKFs [152] are extensions of KF used to describe nonlinearities in sensor response (for instance, cameras and LiDAR) and system dynamics. Major limitations of KF and EKF come from their assumption of linear relationship between system states and measurements, assumption of Gaussian probability distribution of probabilities, sensitivity to initialization, inability to handle sensor outliers, and computational complexity. **b**) *Particle Filters (PF)/ Monte Carlo localization/ sequential Monte Carlo methods that use the posterior distribution of the vehicle's state as a set of particles*.[153] These particles are propagated and weighted based on sensor measurements, allowing for a more robust estimation of the vehicle's position, orientation, and surrounding objects.[153] The PFs are important complex environments, multiple sources of uncertainty, and are robust to outliers in sensor measurements. The limitations on the PF use include the so called particle depletion, or when the number of particles is limited, and the curse of dimensionality, where the computational complexity increases exponentially with the number of dimensions in the state space.[154] **c**) *Dempster-Shafer Theory (DST)/ evidence theory based mathematical frameworks used in sensor fusion to handle uncertain and conflicting information, allowing to combine information from different sensors with different levels of reliability or accuracy*.[155,156] DST provides a path to reason and make decisions based on the available evidence, integrating the uncertainties and mitigating conflicts in sensor measurements. Limitation of the DST is an increasing computational complexity which increases as the number of sensors (and states) increases, leading to a combinatorial explosion of computations. **d**) *Decision-Level Fusion (DLF)*. With this approach, the outputs of individual sensors are combined at a higher level to make decisions.[166] This approach can involve voting mechanisms, weighted averaging, or logical rules to determine the final decision based on the outputs of multiple sensors. Decision-level fusion is often used in autonomous vehicles to make critical decisions,[147–150] such as object detection and collision avoidance. As DLF relies on voting mechanisms, averaging or logical rules, information from individual sensors may be overlooked or lost. The DLF approach for autonomous vehicles often lacks contextual information, which has limited adaptability and may affect fused decisions.

**2. Industrial Automation and Robotics** have led to increased adoption of AI-driven sensors in manufacturing, logistics, agriculture, and other industries.[146] Bayesian Networks (BN)[157] have been used for probabilistic modeling of dependencies among variables, sensor measurements, system states and other factors, capturing complex relationships and dependencies between sensor measurements and the system's state. The BN allows for principled inference and fusion of sensor data, and handles dynamic sensor fusion by incorporating time-dependent information.[158] The advantage of BN is its capability of effectively combining and reasoning with uncertain sensor data, to allow accurate estimation of system state and decision making.[158]

**3. Internet of Things (IoT)** devices not only create a massive influx of multi sensor-generated data, but also force the adaptation of sensor fusion algorithms for extraction of valuable insights, to enable predictive maintenance and enhance overall system performance such as enhanced efficiency and reduced cost.[146] Compared to cases of highly localized sensors for autonomous vehicles and industrial automation, a much broader distribution of IoT sensors over different locations and environments requires additional efforts for data aggregating. Among IoT specific approaches, Fuzzy Logic (FL)[159,160] and Ensemble Methods (EM) [161,162] are notable, in addition to deep learning, Bayesian, DST, and filter based methods.

**4. Environmental Monitoring and Smart Cities** are expected to propel applications across areas including *Healthcare & Wearable Sensors, IoT, and others.*[146] AI algorithms can process large datasets from various environmental sensors, such as air quality sensors, weather sensors, and noise sensors, to analyze patterns, detect pollution sources, reduce negative environmental impact, and develop sustainable urban infrastructure operating at minimum energy cost.[146] Application of sensor fusion methods can integrate environmental data with Geographic Information Systems (GIS) including geographic and infrastructurally comprehensive information for visualizing environmental conditions, and providing real-time warnings.

**5. Healthcare and Wearable Sensors (H&WS):** Algorithms such as Artificial Neural Networks (ANNs)[2,163] were used to automatically learn patterns from sensor data, enabling applications like emotion recognition, sleep stage classification, and disease diagnosis. Revolutionary changes in health care are anticipated upon integration of sensor fusion technologies and data from wearable devices, such as heart rate monitors, accelerometers, and biosensors.

**Figure 21-23** provided examples of using Kalman filter and Markov Chain Monte Carlo (MCMC) based data assimilation (DA) to predict glucose dynamics from a type 2 diabetes patient.[164] **Figure 21A** has a two-day glucose measurements and meal data; **Figure 21B** provided point-wise DA glucose predictions using an ultradian model describing the impact of insulin on glucose utilization and production.[165] Glucose level predictions and uncertainty estimates based on the DA approach are given in **Figure 21C** as compared to the continuous glucose monitor results in **Figure 21D**. Combining the finger-prick measurements and physiological (ultradian) model, oscillations in glucose levels have been predicted with the DA approach. In **Figure 22**, the spline smoother is free to select curves that minimize errors, but unconstrained by physiology. The physiologic constraints in the DA approach (**Figure 23**) decrease model flexibility while increasing accuracy. Similar approaches and algorithms are found to be instrumental for applications such as pharmacodynamics, sleep monitoring, reproductive endocrinology, epidemiology, viral modeling, implantable defibrillators and pacemakers, artificial pancreas, cancer treatment, ICU monitoring, as well as general inverse physiology.[164] A detailed review of sensor fusion techniques can be found elsewhere.[166]

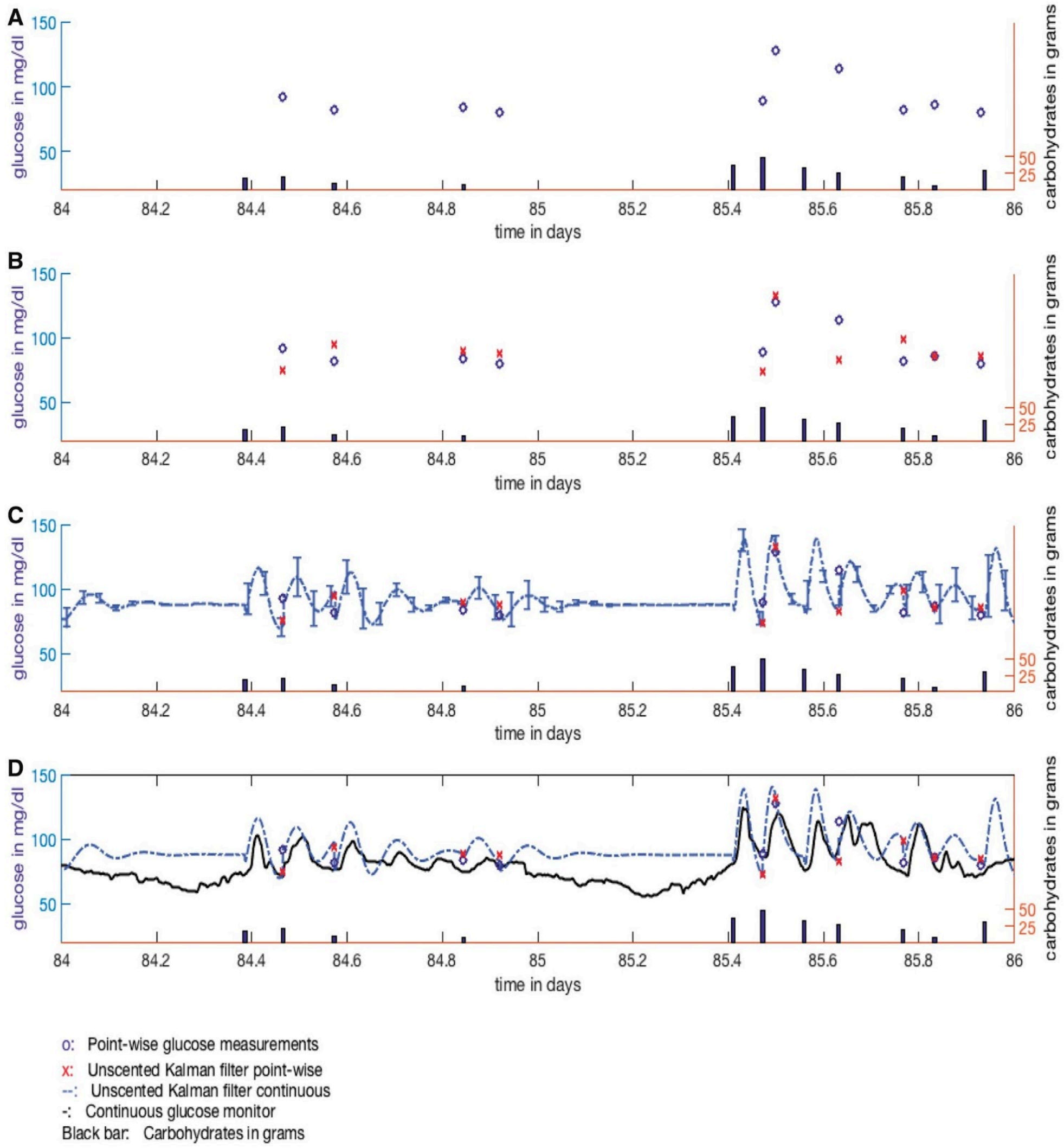

**Figure 21**. (A) Finger-prick glucose measurements and meal carbohydrates serve as the training data (one segment shown) with a goal to predict glucose in the future. (B) DA's point-wise forecasts seem reasonable but not perfect, predicting one spike but missing another. (C) Underlying continuous DA forecast with uncertainty quantification (which the point-wise forecasts are based on) appears to overfit the data with large glucose swings. (D) Continuous glucose monitoring, which was hidden from the DA, reveals striking overlap between the continuous DA predictions and the actual glucose levels. Despite insufficient information in the training set, the DA tracked glucose well based on the combination of its glucose metabolism constraints and the sparse measurements. (Figure and Caption from reference[164] without change, under CC-BY License.[28])

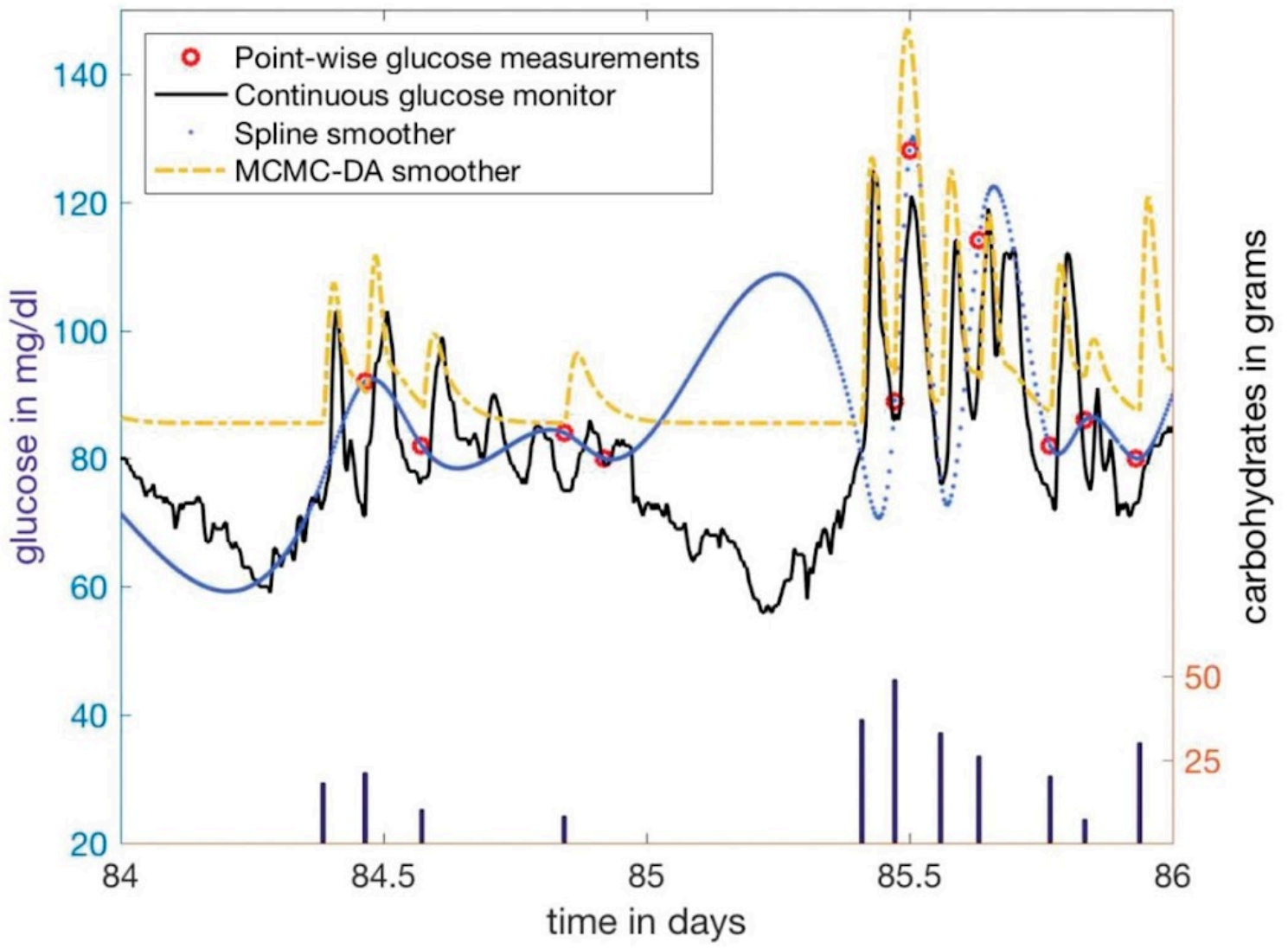

**Figure 22**. Continuous glucose monitor data, finger-prick glucose measurements, Markov chain Monte Carlo-based DA smoothing using finger-prick data, and spline smoothing using only finger-prick data. The spline, with no physiologic constraints, can deviate wildly from the real glucose measurements, while the physiologically constrained DA glucose inferences closely resemble the continuous glucose monitor data. The spline does not have enough data to infer the physiology necessary to constrain its inferences. In contrast, the DA, leveraging physiologic knowledge, is able to infer glucose dynamics that are impossible to infer according to the sampling theorem, because of the hardcoded physiologic knowledge with very little data. (Figure and Caption from reference[164] without change, under CC-BY License.[28])

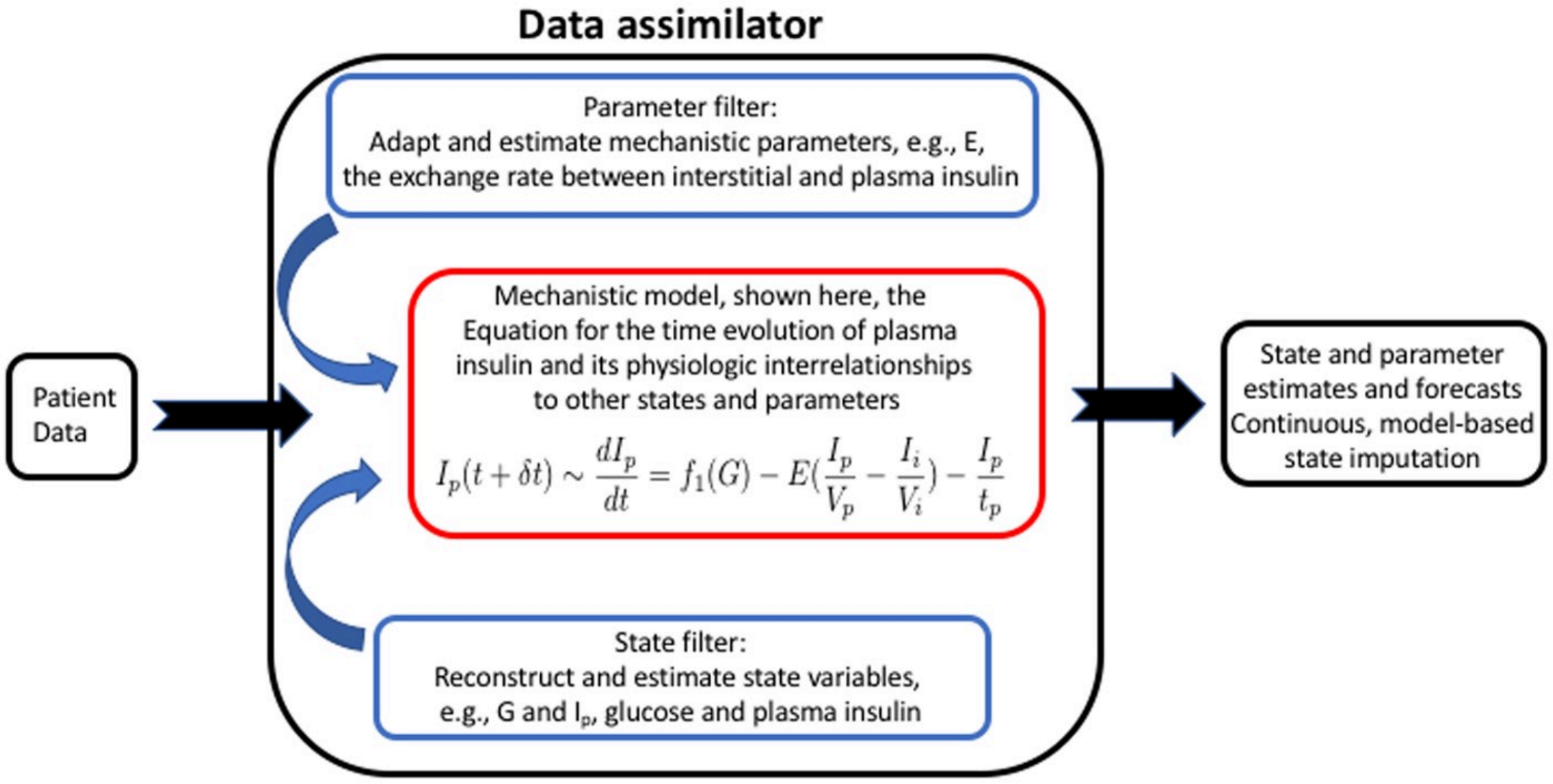

**Figure 23**. Schematic diagram of how data assimilation synchronizes a model with data by estimating parameters and states. The model equation is drawn from the ultradian glucose model. (Figure and Caption from reference[164] without change, under CC-BY License.[28])

## 2.5 Machine Vision

Deep learning[167] powers machine vision by interpreting data from hyperspectral cameras, infrared sensors, and other imaging systems.[168,169] For example, machine vision leverages data from infrared sensors, LiDAR and cameras to detect pedestrians, obstacles and traffic patterns in low light, crowded, or other challenging conditions. In manufacturing, machine vision assists robotic guidance,[170] components inspections,[171] optical metrology,[172] and lithography.[173]

.

In research labs, machine vision enables automated phase identification and crystallization monitoring via ML image recognition,[174,175] 3D or 2D grain boundary analysis,[176] microscopy,[177–181] and image reconstruction of human brain activities.[182] More specifically, in light and electron microscopy, general image datasets as well as data sets specific for microscopy or crystallography can be used for machine learning and deployment.[178] AI techniques, especially deep learning based algorithms, such as U-Net and CNN, are used for resolution enhancement, object detection and classification, image segmentation, artificial labeling, as well as tracking.[179,180]

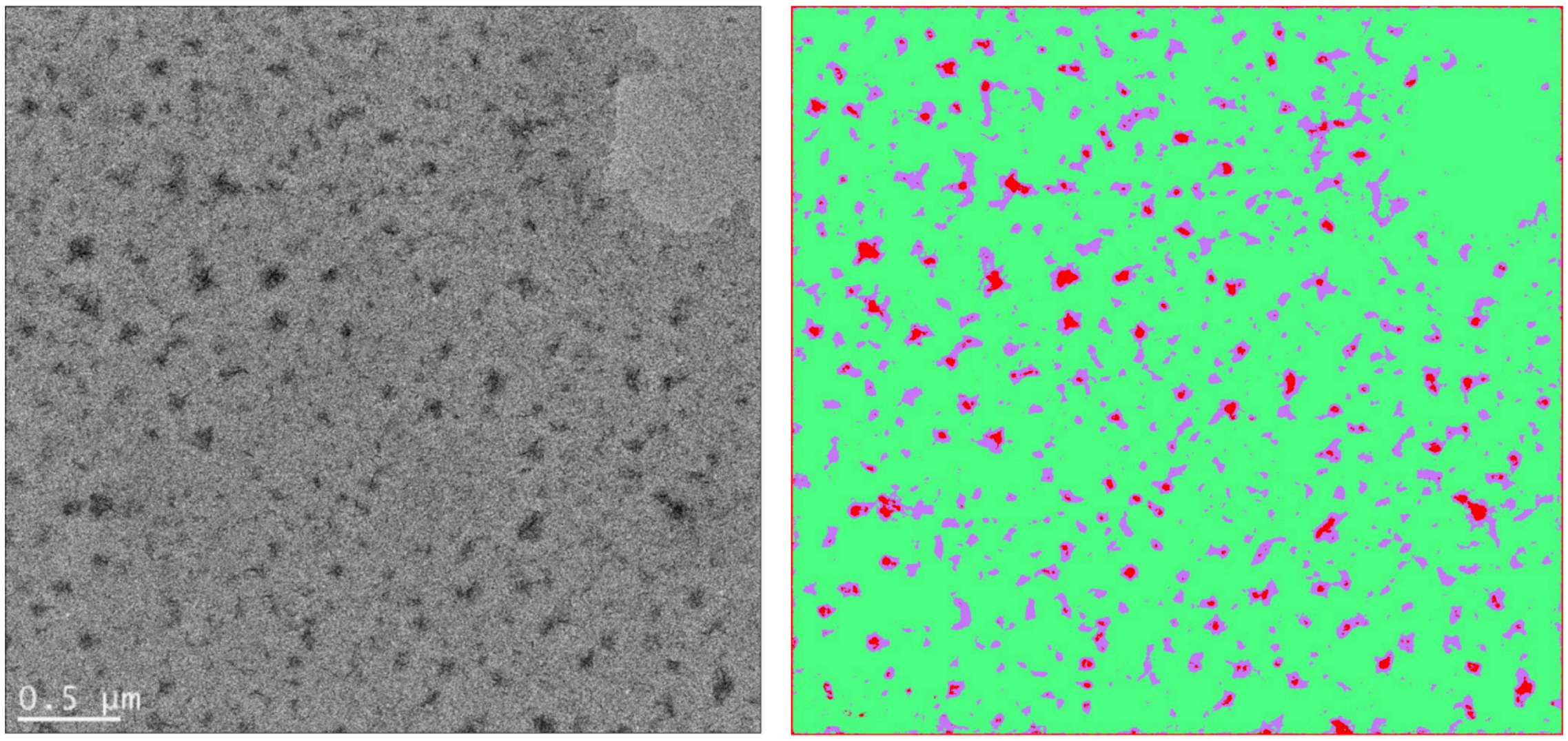

(A) (B)

**Figure 24**. (A) Bright-field TEM image of a polyvinylidene fluoride (PVDF)-block-polystyrene (PS) block copolymer film showing crystalline domains in darker regions. The volume fraction of the PVDF block is 20% in the PVDF-b-PS copolymer. (The copolymer was provided by Alex Asandei, University of Connecticut.) (B) An attempt to classify PVDF-b-PS crystalline domains (red) and amorphous domains (green), as well as their interfaces (purple), by using the Fiji package of Image J and the machine learning capability of the Weka plugin. The Weka program is based on reference[183]. The Weka program was trained by user-predefined regions and then used a chosen algorithm to automatically classify each pixel. For the case of PVDF-b-PS, a fast random forest method was used to perform image segmentation, which also yields probability maps and threshold curves for performance evaluation. (Figure and Caption from reference[184] without change, under CC-BY License.[28])

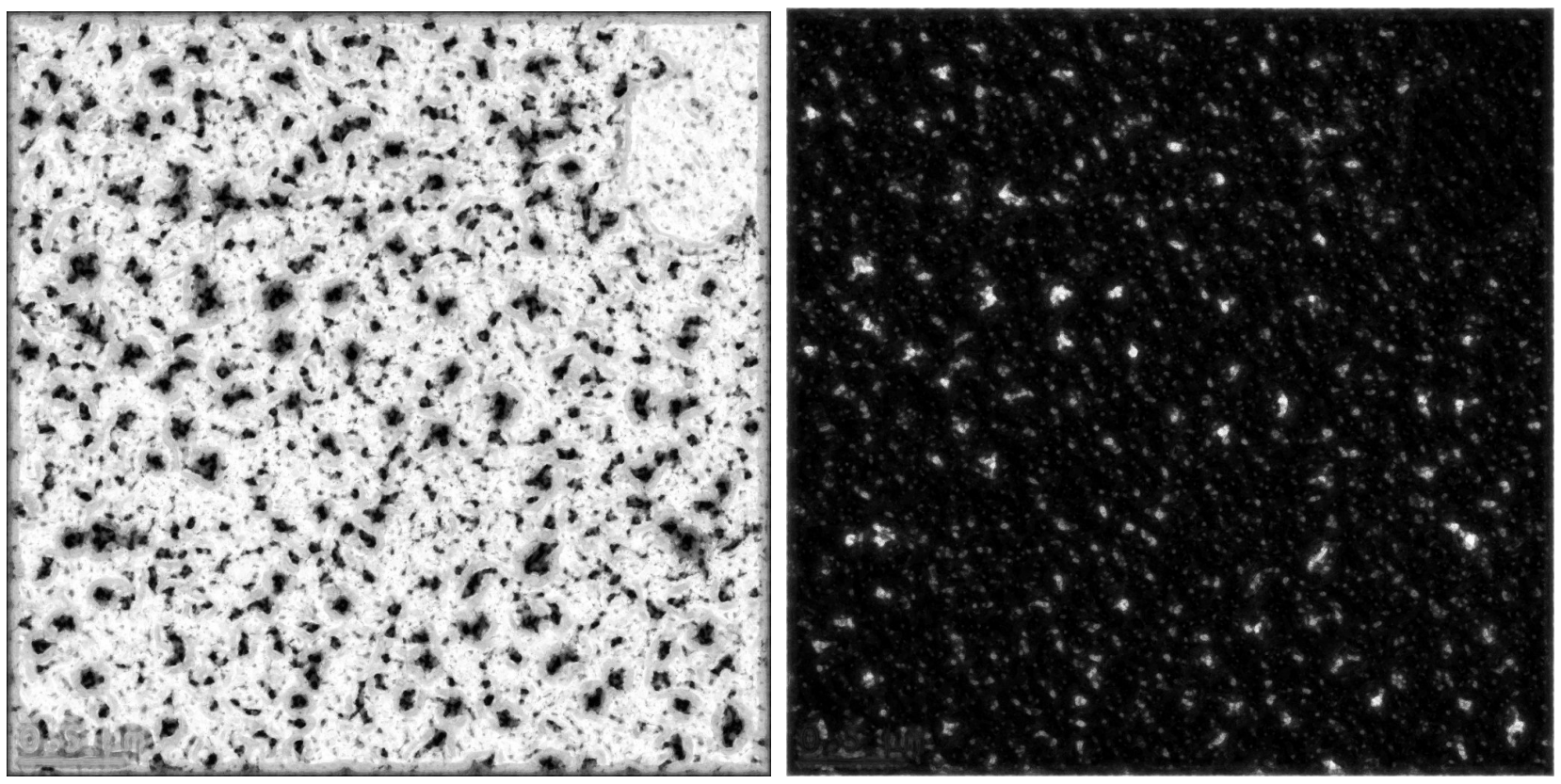

**Figure 25**. Probability Maps of Amorphous (left) and Crystalline (right) Regions for **Figure 24**. Brighter colors suggest higher probability.

Since there is a significant overlap between machine vision (this section) and medical vision (the next section) in terms of algorithms, we will only elaborate on issues unique to topics related to manufacturing such as computer vision aided robotic control and microscopy here. The latter will be used as an example to elaborate in terms of datasets, models, etc. There has been a concerted effort to apply, adapt, and design ML models specifically for scientific image data.[185,186] These models have demonstrated exceptional performance in the analysis of microscopy images in various applications for biomedical [180,187,188] and materials science [186,189] **Figure 24-25** show an example of using machine learning to semantically segment crystalline and amorphous domains of a block copolymer at the nanoscale.

## 2.5.1 Microscopy Data

Microscopy data can be generated by different imaging modalities such as optical microscopy (OM), scanning probe microscopy (SPM), scanning electron microscopy (SEM), transmission electron microscopy (TEM), and scanning transmission electron microscopy (STEM). The images obtained by employing these methods encompass features of a broad spectrum of resolution scales, ranging from atomic to mesoscale, and present spatial distributions, and morphological characteristics that are fundamentally interconnected with the functionality and performance of the biological systems or materials.[184,186,189] Microscopy data is also considered with high dimensionality and strong correlations when integrated with spectroscopy techniques, video microscopy and tomography. To attain optimal efficiency and maximum data processing capacity for the intricate informatics of these datasets, machine learning methodologies are frequently employed.

## 2.5.2. Machine Learning for Microscopy

ML has been applied with great success in various fields of microscopy with focus on improving efficiency and accuracy for human-intensive tasks such as segmentation and classification. Different ML models are chosen based on the availability of training dataset, their characteristics, and the requirements of specific tasks. Supervised learning has its wide applications in image processing and analysis, including image denoising and enhancement, segmentation, object detection and tracking, classification, and tomography reconstructions.[185,188] When labeled data is scarce or difficult to obtain, unsupervised learning is often used by creating a model that learns patterns and relationships in the existing data itself without requiring supervision or labeled examples. For example, clustering algorithms such as k-means, hierarchical clustering, and DBSCAN [190] can be used to segment images into regions with similar properties, such as cellular structures or particles based on their intrinsic features or characteristics [191]. It can incorporate generative models including Variational Autoencoders (VAEs) and Generative Adversarial Networks (GANs) for tasks such as image synthesis, inpainting, and denoising.[192] RL algorithms perform particularly well at interactive operations with information feedback, for examples, enabling mitosis detection in time-lapse phase contrast microscopy images [193] and automating operations in the STEM workflow [194].

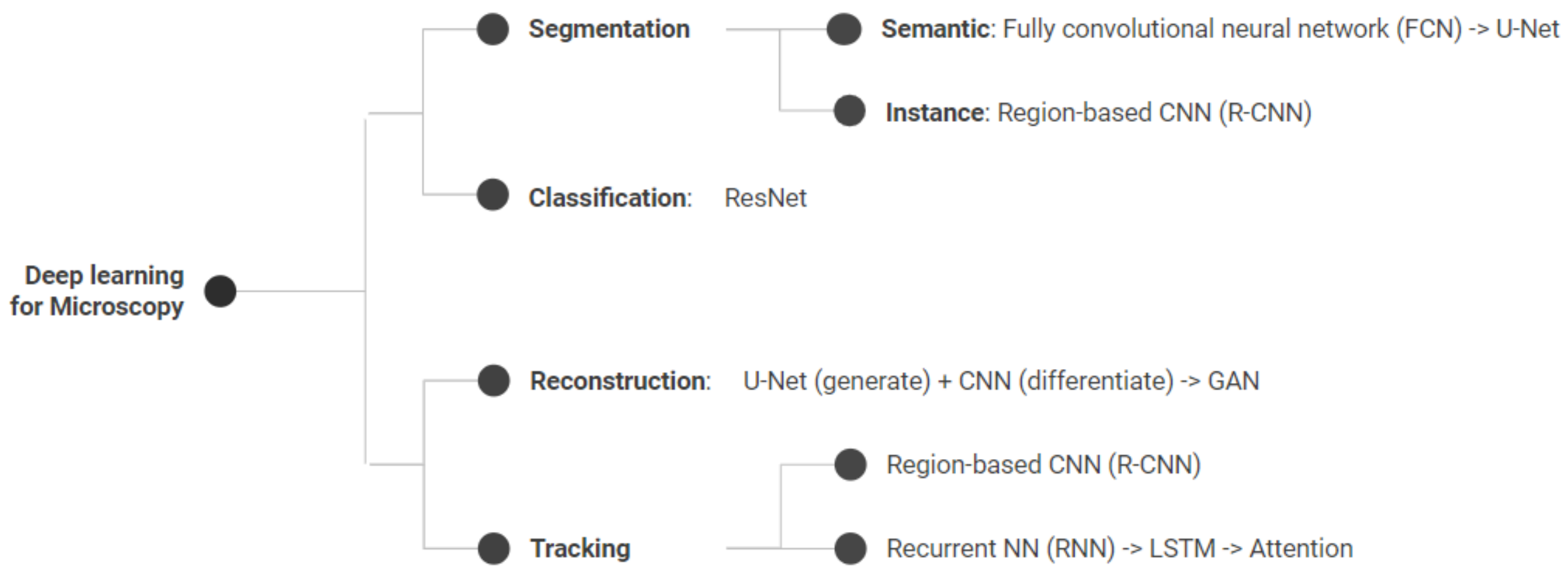

**Fig.26.** Some widely used deep learning models for microscopy.[191]

Deep learning (DL) networks, particularly convolutional neural networks (CNNs) are often chosen for microscopy image analysis. CNNs employ small matrices known as kernels to traverse across the input image, thereby facilitating the detection and classification of features at varying scales with less computational burden than using fully connected neural networks. Moreover, symmetric encoder-decoder architectures such as fully convolutional network (FCN) and U-Net [195] offer the capability of pixel-wise prediction, showing promising results in both noise reduction and semantic segmentation. Other variant of CNNs such as region-based CNNs (R-CNN), Fast R-CNN, Faster R-CNN, and Mask R-CNN can take an input image and propose a set of bounding boxes, and detect an object and its category in each region.[191] They are becoming ideal models for object detection and instance segmentation. The implementation of ResNet and U-Net in the three-dimensional domain allows for the recognition of videos and segmentation in tomography.[196] In order to leverage temporal information from microscopy videos, it is necessary to establish correlations among adjacent frames. Recurrent neural networks (RNN), such as Long Short-Term Memory (LSTM) and Attention models, possess the ability to selectively retain or discard information from preceding frames, resulting in their utility for the classification of particle trajectories and the tracking of objects. A summary of widely used DL networks is shown in **Fig. 26**.[191]

### 2.5.3. Models and Tools

With the increasing usage of deep learning approaches for image analysis, numerous pre-trained models and software tools have emerged to assist researchers and analysts. For example, MicroNet [197] offers a robust dataset consisting of more than 100,000 labeled microscopy images, which can be utilized for training network encoder architectures and enhancing segmentation when embedded with UNet and DeepLabV3+. Other models such as Deep-STORM [198] focus specifically on super-resolution microscopy, aiming for image enhancement. CryoCARE[199] is specialized to process cryo-TEM data in denoising and restoration of 2D and 3D images. For atomic-resolution STEM images, TEM ImageNet library and AtomSegNet models are specifically for tasks such as deblurring/super-resolution processing and atom detection.[200] Until recent years, the execution of deep learning processes still require substantial computational proficiency and resources. Nevertheless, more open-source DL platforms have emerged, striving to enhance the accessibility of DL-based image analysis to users with limited computational experience.[201] These applications encompasses a wide array of forms, including web applications, add-ons for the existing imaging analysis software, and interactive notebooks or cloud-based solutions.[201] One example is BioImage Model Zoo, an open-source model platform, supporting many pre-trained models for a variety of image analysis tasks.[202]

### 2.5.4. Applications

Machine learning, especially deep learning, has become the prevailing methodology for addressing a variety of image analysis challenges across different scientific and engineering fields. These methods are fundamentally transforming the ways in which we obtain, process, analyze, and interpret data. Depending on the specific application, information extraction from microscopy data may involve tasks such as image enhancement, image segmentation, object classification, and object tracking.[188]

#### Image enhancement

Image enhancement based on deep learning has proven to be an effective approach for the elimination of artifacts, improvement of signal-to-noise ratio (SNR) and resolution, and restoration of crucial information within the images. For example, CNN-based models can be trained to execute denoising and regain resolution by utilizing properly aligned pairs of low-quality and high-quality images(**Fig.27**).[181] This approach was also exemplified by Weigert et. al [203] through the content-aware image restoration (CARE) method. By implementing this pipeline for denoising, they successfully restored images with even 60-fold fewer illumination dose, and achieved near isotropic resolution of specific structures with up to tenfold under-sampling. Another example is achieving super-resolution across different fluorescence microscopy modalities based on training a GAN.[204] Using this framework, it enables transformation of diffraction-limited input images (e.g., confocal microscopy images) into super-resolved stimulated emission depletion (STED) and total internal reflection fluorescence (TIRF) microscopy images. It can rapidly produce super-resolved images, without the need for any iterations or parameter search, and has the potential to democratize super-resolution imaging.[204] Also, GAN based frameworks have been developed for image reconstructions for 3D datasets. The algorithms showed significant improvements for reconstructing missing-wedge X-ray ptychographic tomography (PXCT) dataset[205] and enabling virtual refocusing of 2D fluorescence microscopy images onto the 3D surfaces.[206] Deep learning-based and GAN-based algorithm have also found their ways in image enhancement and artifact reduction for 3D *ex situ* inspection and characterization of additively manufactured parts allowing for orders-of-magnitude speed up in the post-process inspection while enhancing the flaw detection capabilities of these systems for more consistent and accurate qualification of the printed parts.[127,137,207]

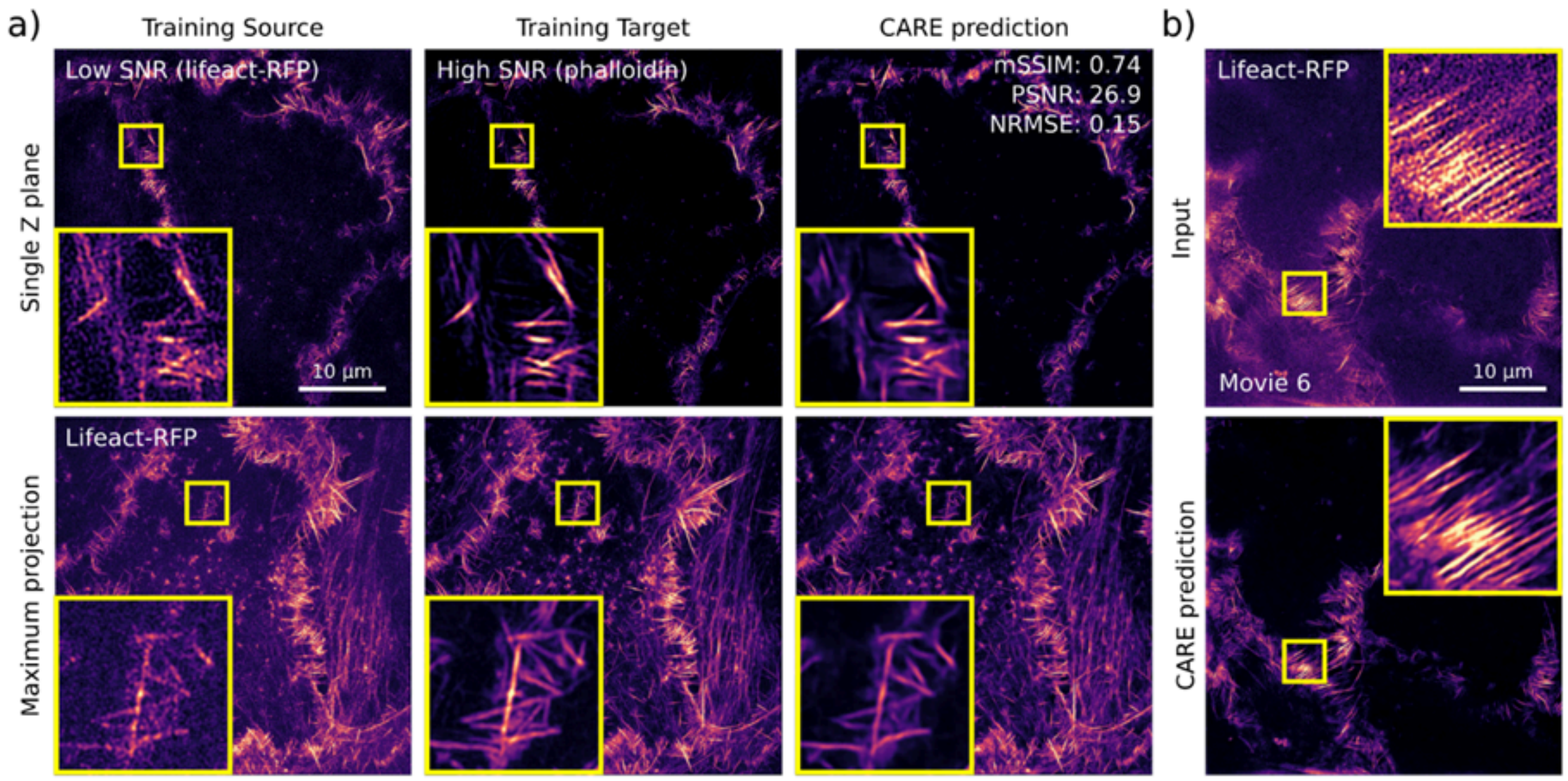

**Fig.27** Image denoising and restoration through CARE by using ZeroCostDL4Mic platform. [181] (a) A 3D CARE network was trained using structured illumination microscopy (SIM) images to denoise live-cell imaging data. (b) The low SNR image (input) and the associated CARE predictions are displayed (single plane).(Reprinted with permission from Ref [181] without change, under a Creative Commons Attribution CC-BY license[28])

## Image Segmentation

The overall goal of image segmentation is to determine the precise location of objects and their boundaries within an image based on certain characteristics. Semantic and instance segmentation are two types of segmentation for microscopy image analysis. Semantic segmentation aims to classify each pixel in an image into a specific category or class, such as background, cell, or particles. Instance segmentation, on the other hand, aims to identify and segment individual objects within an image, such as individual cells or particles. An example of semantic and instance segmentation is shown in **Fig. 28**. Both semantic and instance segmentation have various applications in biology and materials science research, including cell counting, particle tracking and classification, as well as disease diagnosis and drug discovery. [187]

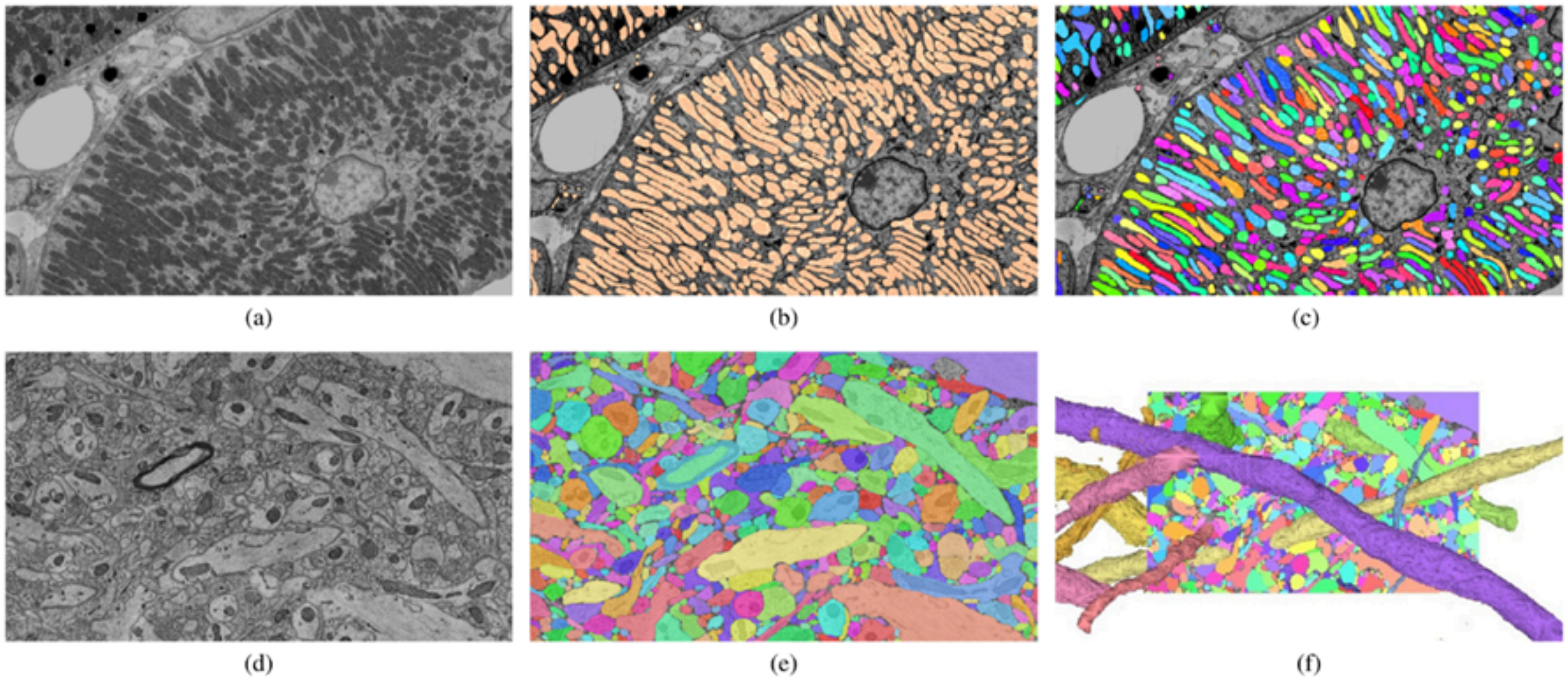

**Fig. 28.** Example of semantic and instance segmentation of a 2D EM image of mitochondria (top row) and segmentation and 3D reconstruction of selected neuronal structures that pass through the given 2D section from adjacent sections of the EM volume (bottom row).[187] (Reprinted with permission from Ref[187], without change, under a Creative Commons Attribution CC-BY license[28])

Fully convolutional neural networks (FCNs), including the popular U-Net architecture[195,208] have emerged as the preferred option for automated feature extraction and semantic segmentation of microscopy images. This U-net architecture, which is named for its U-shaped network structure, uses many convolution/pooling layers (the encoder), followed by many layers of deconvolution/upsampling (the decoder). This architecture can achieve relatively good results even with a small set of annotated images and enable precise localization and improvement of segmentation accuracy. [179,208] To date, many U-Net variants have been developed to improve the performance and meet specific requirements for segmentation.[208,209] For example, Half-UNet[210] is a simplified U-Net architecture to reduce computational parameters while maintaining the accuracy for medical image segmentation. A series of nested, dense skip pathways are used in U-Net++ [211] to reduce the semantic gap between the feature maps of the encoder and decoder sub-networks, which can improve the accuracy for various segmentation tasks. In addition, 3D-UNet[196] has become an important component in segmentation workflows for volumetric datasets. For materials science, a TEM ImageNet[200] has been used for atom segmentation and localization from atomic-resolution STEM images.

### Object Classification

One critical step in the image analysis workflow is the identification and categorization of images into pre-established classifications, which relies on the recognition of specific features. Different techniques are available for image classification, including the Bayesian classifier, geometric classifier, clustering, and neural network classifier.[191] Of all these methods, CNN-based classifiers are the most popular for classifying microscopy images. For bioimage analysis, they have demonstrated promising outcomes in the classification of cellular and subcellular structures,[212] as well as in disease diagnosis.[213]

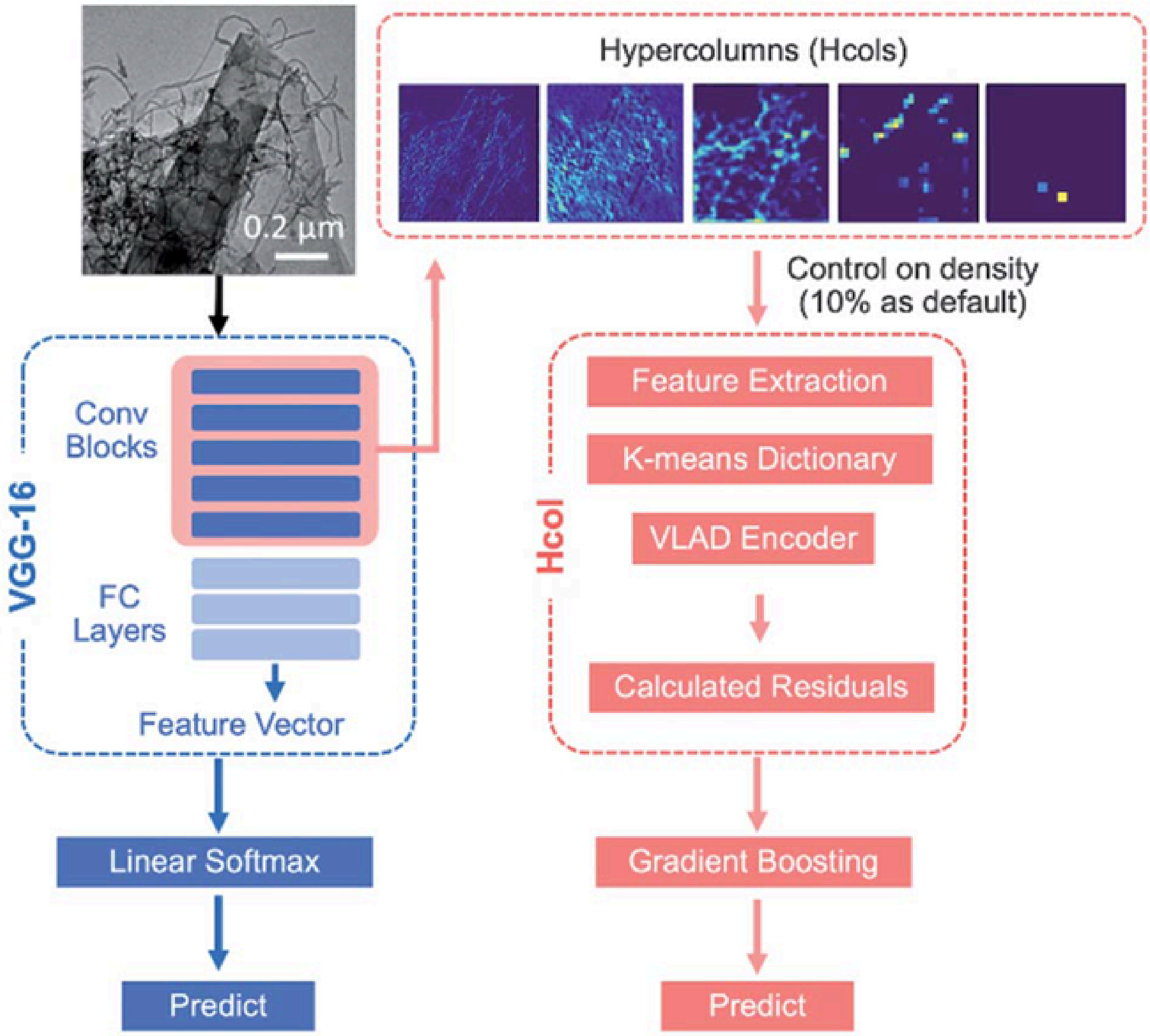

**Fig. 29.** A schematic of the classification model for CNT structures: (1) blue arrows indicate the conventional image classification pipeline with a linear softmax classifier (VGG16 + LS); (2) red arrows indicate the hypercolumn (Hcol) model followed by VLAD classifiers. (Reproduced from Ref.[214] with permission from the Royal Society of Chemistry.)

One common issue for image classification is the relative lack of annotated training datasets. Therefore, transfer learning has been implemented in the development of image classifiers. Instead of training a CNN from the ground up on a new dataset, transfer learning involves the utilization of a pre-trained CNN as a starting point and fine-tuning it on the new dataset. To illustrate, a transfer learning approach based on the CNN architecture has been employed to classify TEM images of carbon nanostructures into multiple predetermined structural categories, thus aiding in the assessment of the associated risks pertaining to the exposure to these materials.[214] In this approach, images are represented by hypercolumn (Hcol) vectors and subsequently processed by Vector of Locally Aggregated Descriptors (VLAD) classifiers (**Fig. 29**). This new method has shown significant improvement of overall classification accuracy over the conventional VGG-16 model.

## Object Tracking

As modern microscopes are now fully capable of capturing high-quality sequential images of living cells, intracellular particles, and other moving objects within the system, this calls for the development of computer vision techniques that can not only detect and segment objects but also track them over time. Deep learning has been used for both the spatial detection and segmentation and subsequent trajectory analysis in many studies. One application is for single cell tracking. Lugagne *et al.*[215] proposed an image processing tool employing a sequence of two U-Net models to segment cells and then to perform tracking and lineage reconstruction. This pipeline can accurately identify the location of cells in an image, track them over time as they grow and divide, and reconstruct their lineages. These capabilities highlight the potential applications, such as the real-time tracking of gene expression and the high throughput analysis of strain libraries at the single-cell level. Deep learning-based methods also improve the signal-to-noise ratio in low-exposure datasets. which is particularly relevant to single molecule tracking where the number of photons emitted from a single cell is limited.[216] The commonly used networks for localization and tracking are Mask R–CNNs and RNNs.They have been used to segment and track nucleus,[217] and virus structures from fluorescence microscopy image sequences.[218] Single-particle tracking can be extended to multi-particle tracking with the new tools such as DeepTrack[180] for microscopy video analysis (**Fig.30**). The ability to perform real-time imaging and analysis empowers researchers to observe and analyze dynamic processes in complex nanoscale systems, enabling quick predictions based on the acquired data.[219]

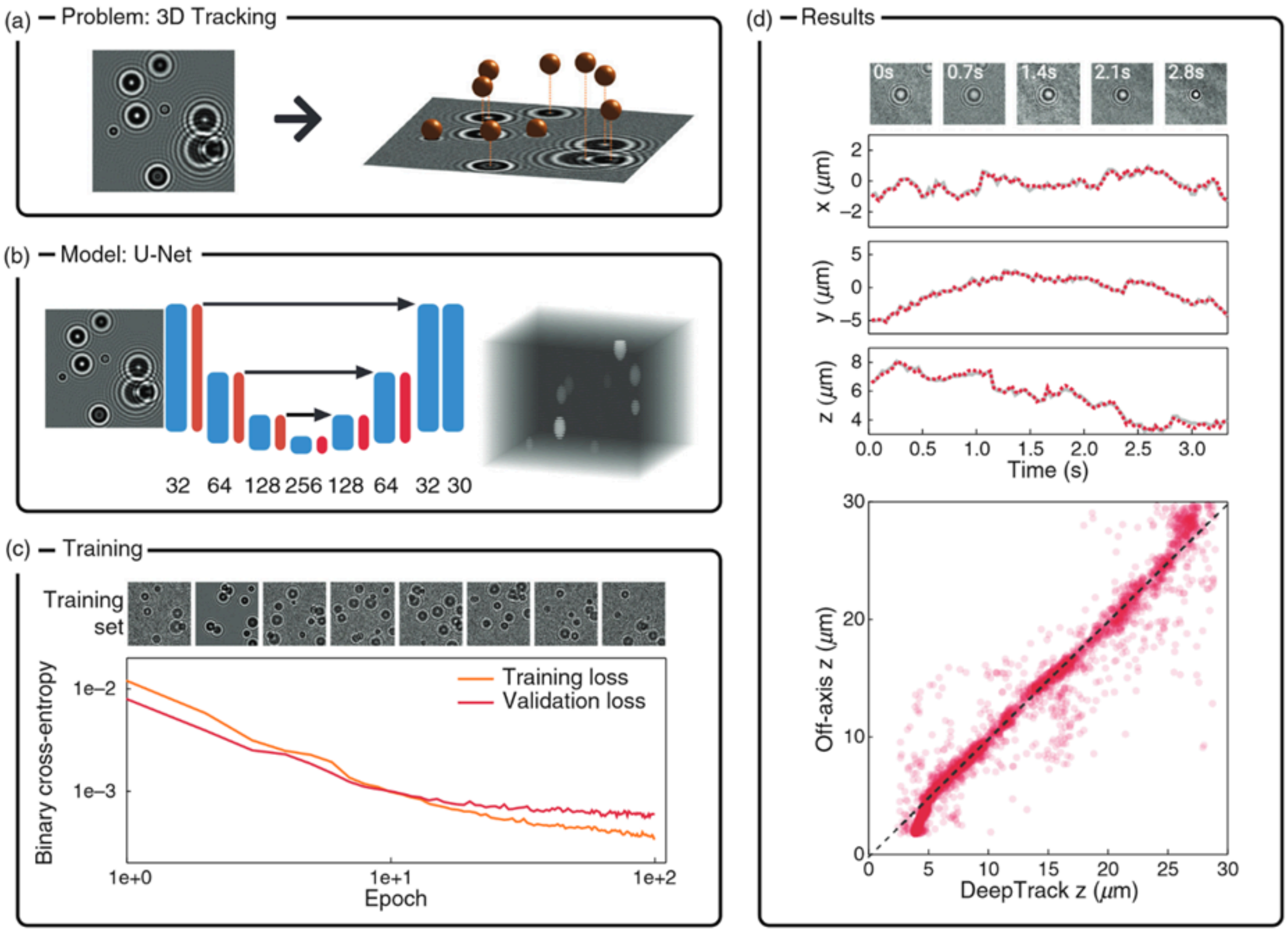

**Fig.30**. An example of using U-Net to track spherical particles in three dimensions.[180] (a) A sample network input, consisting of scattering patterns of several spherical nano particles. (b) a small U-Net architecture, with final convolutional layer outputs a volume, where each particle in the input is represented by a sphere of 1s. (c) Examples of synthetic images used in the training process, and comparison of the validation loss (magenta line) vs. training loss (orange line) after roughly 10 epochs. (d) A single particle tracked using the DeepTrack model (dotted orange line) and off-axis holography (gray line), showing the x, y, and z positions over time. (Reprinted with permission from Ref [180], without change, under a Creative Commons Attribution CC-BY license[28])

Advancements in the field of machine learning microscopy have led to significant improvements in image analysis, data interpretation, and automation of microscopy processes. Compared to traditional analysis methods, more consistent and reliable results can be delivered with less inter-observer variation. The employment of ML tools also assists the discovery of hidden patterns within images, and facilitate a better understanding of various phenomena that may not be easily discernible through manual analysis. It is no doubt that the implementation of CV and ML in microscopy analysis will lead to more accurate, efficient, and automated processes, enabling researchers to focus on the scientific insights and discoveries. Applying CV and ML in microscopy image analysis, however, also presents several challenges due to the limited annotated training data, complexity and variability of features in images, interpretability and adaptability of models and computational restrictions.[220] To overcome the challenges of available annotated training data, strategies such as data augmentation, transfer learning, and active learning can be used.[179,220] Improved feature selection, model architecture design, and training on diverse datasets can help address the complexity and variability of images.[221] Techniques like attention mechanisms, saliency maps, and combing with more interpretable ML techniques (such as decision trees or rule-based models) can provide insights into the interpretability of deep learning models.[220] Moreover, model compression, hardware acceleration, and distributed computing can address computational requirements and scalability challenges. Propelled by the expanding availability of massive datasets, breakthroughs in machine learning algorithms, development of user-friendly tools, and the evolution of microscopy techniques, the fusion of ML with other imaging techniques, real-time imaging and analysis, and data-driven discovery has become a reality.[188,189,191] The field is still undergoing an exponential development and many new data-centric approaches have the potential to reshape the future of microscopy, and lead to new discoveries and insights in various fields of biology, healthcare, and materials science.

# 3. Healthcare

The recent decades witnessed a dramatic increase in structured and unstructured healthcare data as well as a rapid evolution of machine learning algorithms. AI can play a key role to address current healthcare issues such as availability, cost, and efficiency, as well as providing novel approaches to previously unsolvable tasks through augmenting, automation, error-reduction, prediction, personalized and precision medicine.[222] To this date, AI for healthcare faces several challenges and concerns. These include (a) data quality and availability, (b) ethical, social, and privacy concerns, as well as (c) algorithm developing and implementation (such as explainability, data leakage, overfitting, and bias).[223] Nevertheless, explainable AI (XAI) can address the so-called "black box" issues of some machine learning models, towards a trustworthy AI for future healthcare.[224] Data leakage, overfitting, and bias can be minimized or eliminated with a periodically updated and carefully monitored data pipeline. Data quality and availability are expected to improve as more AI practitioners join the efforts, while ethical, safety, and social concerns may rely on governmental or organizational regulations to build up.

In terms of specific application scenarios, AI for healthcare can be used in a) basic biomedical research, b) translational research, and c) clinical practices.[225] For a) basic biomedical research, there are possible AI applications in gene function annotation, molecular dynamics simulation, and transcription factor binding site (TFBS) prediction. For b) translational research, possible AI applications include biomarker identification, drug discovery, drug-target prioritization, toxicity prediction, and genetic variant annotation. Finally, for c) clinical practices, AI could be used for disease diagnosis, risk stratification, medical vision, genome interpretation, treatment planning, patient monitoring, robotic surgery, as well as populational and chronic health management.[225–228] A review by Jiang, *et al.*, examined motivations of AI efforts in healthcare, various data types, algorithms to generate clinically meaningful results, and types of disease to tackle.[229] **Figure 31** shows a workflow from clinical data including electronic medical record (EMR), electrophysiological (EP), images, genetics information, and other sources, to natural language processing, machine learning data pipeline, and finally clinical decision making.[229] **Figure 31** also has a pie chart for the commonly used machine learning algorithms in the medical literature, generated from PubMed.[229] Support vector machine and neural networks are among the most popular algorithms identified for imaging, genetic, and electrophysiological data.[229]

A more recent review on medical AI by Bohr, *et al.,*[227] divides medical AI applications into: 1) Medical Vision for diagnosis and surgery, 2) Precision Medicine (based on patient's personal genetics, protein expression, gut microbiome, metabolics, demographics, EMR, physical or emotional activities from wearable sensors, and so on), 3) Intelligent Health Record, 4) Robotics and AI devices, and 5) Ambient assisted living. Herein, from a chemistry- and engineering-oriented perspective, we will not focus on 3) Intelligent Health Record, 4) Robotics and AI devices, and 5) Ambient assisted living. Instead, we will expand Precision Medicine into Diagnosis, Protein design, and Small Molecule Drug. Therefore the topics in this section will be 1) Medical Vision, 2) Diagnosis, 3) Designing Protein, and 4) Discovering Small Molecule Drugs.

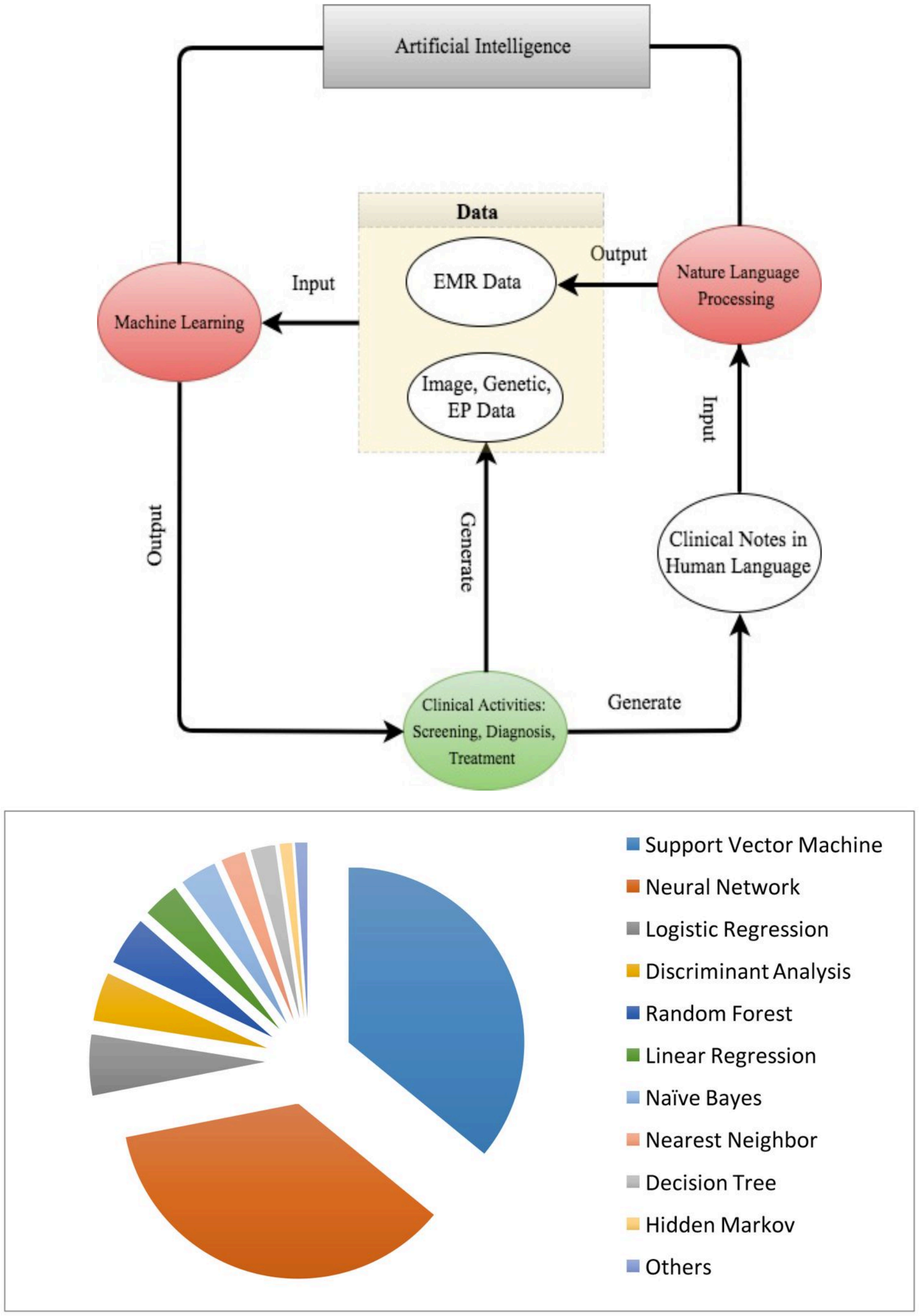

**Figure 31**. **Top**: The road map from clinical data generation to natural language processing data enrichment, to machine learning data analysis, to clinical decision making. EMR, electronic medical record; EP, electrophysiological. **Bottom**: The machine learning algorithms used in the medical literature. The data are generated through searching the machine learning algorithms within healthcare on PubMed. (Figures and Captions from reference[229] without change, under Creative Commons Attribution 4.0 International License.[28])

## 3.1 Medical Vision

AI in medical Imaging includes different medical imaging modalities (X-ray, MRI, CT, ultrasound, pathology, microscopy, etc.) for detection, diagnosis, and classification of diseases.[230–235] This can have overlaps with both machine vision (previous section) and diagnosis (next section).[236–238] While a comprehensive study of all publications in this area is prohibitive and out of the scope of this work, in this section we shed light on some of the key aspects including advantages of AI in medical imaging, as well as the challenges and limitations of incorporating AI models into medical imaging tools. As an example, **Figure 32** shows the training procedure of a machine learning tool *EasyReg* for multi-modality brain MRI registration. **Figure 33** provides the overall inference procedure using the same tool.

AI offers several advantages that have the potential to revolutionize the medical imaging field. These advantages include improving accuracy, speed, scalability, and cost-effectiveness among others. A key strength of AI, and specifically deep learning algorithms, in medical imaging lies in their accuracy, where these algorithms have the capability to discern patterns in medical images which might be too intricate even for the expert human eye. Several publications and tools [239–243] have demonstrated the superior performance of AI in medical imaging applications. For instance, in a study by Nam et al.,[240] it was shown that an automatic detection algorithm outperformed physicians in radiograph classification and nodule detection performance for malignant pulmonary nodules on chest radiographs. Interestingly, it was also systematically shown that the AI algorithm as a complementary tool can improve the physician's detection performance measured by the area under the receiver operating characteristic curve (AUROC) and jackknife alternative free-response receiver-operating characteristic (JAFROC) figure of merit (FOM). Other algorithms were used for applications such as detecting lymph node metastases of breast cancer with time constraints[241]. AI-based tools such as Automated Retinal Disease Assessment (ARDA) recently developed[244], based on original work of Gulshan *et al.*,[245] that assists in detecting diabetic retinopathy and is being developed to assist clinicians in identifying other diseases. AI and ML have also been applied in image analysis tasks for various cell microscopy modalities, ranging from pathology, transmitted light, lattice sheet and confocal microscopy among others,[137,246–252] and have overwhelmingly outperformed traditional non-ML based tools and algorithms. Specifically, they excel in tasks such as object detection, image feature extraction, classification, semantic and instance segmentation, especially in radiology, pathology, cancer biology, and immunology. The proliferation of open-source software and new neural network architectures has significantly improved the accuracy of cell detection and segmentation in microscopy images[250]. Additionally, AI techniques have improved the objectivity, sensitivity, and specificity of cervical cancer screenings using whole slide imaging, and introduced new methodologies like the AI-powered transmitted light microscopy (AIM) to visualize and analyze live cell structures and functions without the need for labeling [236,238].

The computational speed in processing medical images is another notable advantage of AI algorithms. This rapid analysis can lead to quicker diagnosis, potentially saving lives in critical situations. For instance, an AI model Chexpert developed by Stanford Machine Learning group[253] illustrated this by analyzing a chest X-ray within ten seconds, a task that took a human radiologist more than 20 minutes. Another work, recently published in the European Radiology,[254] highlighted how AI algorithms can significantly reduce the workload of radiologists and speed up triage, without lowering sensitivity while improving specificity. In the context of image analysis for cell microscopy, AI techniques facilitate rapid analysis, automating processes that would take humans several hours or even days to complete manually. Such automation, furnished by open-source software and novel neural network architectures, aids in extracting quantifiable cellular features, shedding light on cell organization in various pathologies. Consequently, these innovations streamline complex analyses and pave the way for high-throughput analysis ensuring quick turnarounds in microscopy data processing.

Furthermore, scalability of the AI systems allows them to manage vast amounts of data, and in turn simultaneous assessment of multiple image sets. The more extensive the data it's trained on, the more refined and accurate it becomes. This can be especially beneficial in large-scale screenings or in tracking disease progression over time. While presence of large, annotated datasets is scarce in medical imaging applications, recent progress in development of self-supervised models,[255–261] and models that leverage vision transformers and large language models,[262–270] along with astronomical improvement in computational power, promise development of foundation models that could learn from the large un-annotated datasets and produce high quality predictions in downstream tasks. Foundation models are novel neural networks that leverage large datasets and utilize innovative learning methodologies, eliminating the need for conventional supervised labeling. This approach grants them the unique capability to execute zero-shot learning in diverse scenarios. Notably, such models have already made significant strides in natural language processing demonstrated by OpenAI and competitors. For instance, the recently unveiled Segment Anything Model[271] stands out as a foundation model, demonstrating impressive zero-shot segmentation and generalization capabilities across several natural image databases. These models, trained on vast and varied datasets, offer unparalleled versatility, and signify a shift from task-focused models to those that adapt dynamically to a myriad of tasks. Such models, if tuned for the medical sector, could significantly reduce the need for specialized, singular-task models. Specifically, their ability to generalize across tasks can negate the need to design, train, and deploy multiple models for different tasks. Instead, a single, scalable foundation model could cater to multiple needs. This scalability doesn't just translate to efficiency in terms of computational resources but also means quicker adaptability to evolving medical scenarios and needs. The rapid development, integration, and application of such scalable models can fast-track advancements in medical AI, ushering in an era of more effective and efficient healthcare solutions.

While AI has immense potential in medical imaging, it is not without limitations. One significant challenge is the need for vast and diverse training datasets to ensure algorithm robustness. Biases in training data can lead to mis-diagnoses or reduced sensitivity in diverse patient populations. Another concern is the "black box" nature of some algorithms, where the decision-making process isn't transparent, making it hard for clinicians to trust or understand the AI's conclusions. Additionally, there's the challenge of integrating AI seamlessly into existing clinical workflows, ensuring it complements rather than disrupts the process. Lastly, regulatory and ethical considerations play a pivotal role in defining the boundaries of AI applications in this sensitive field.

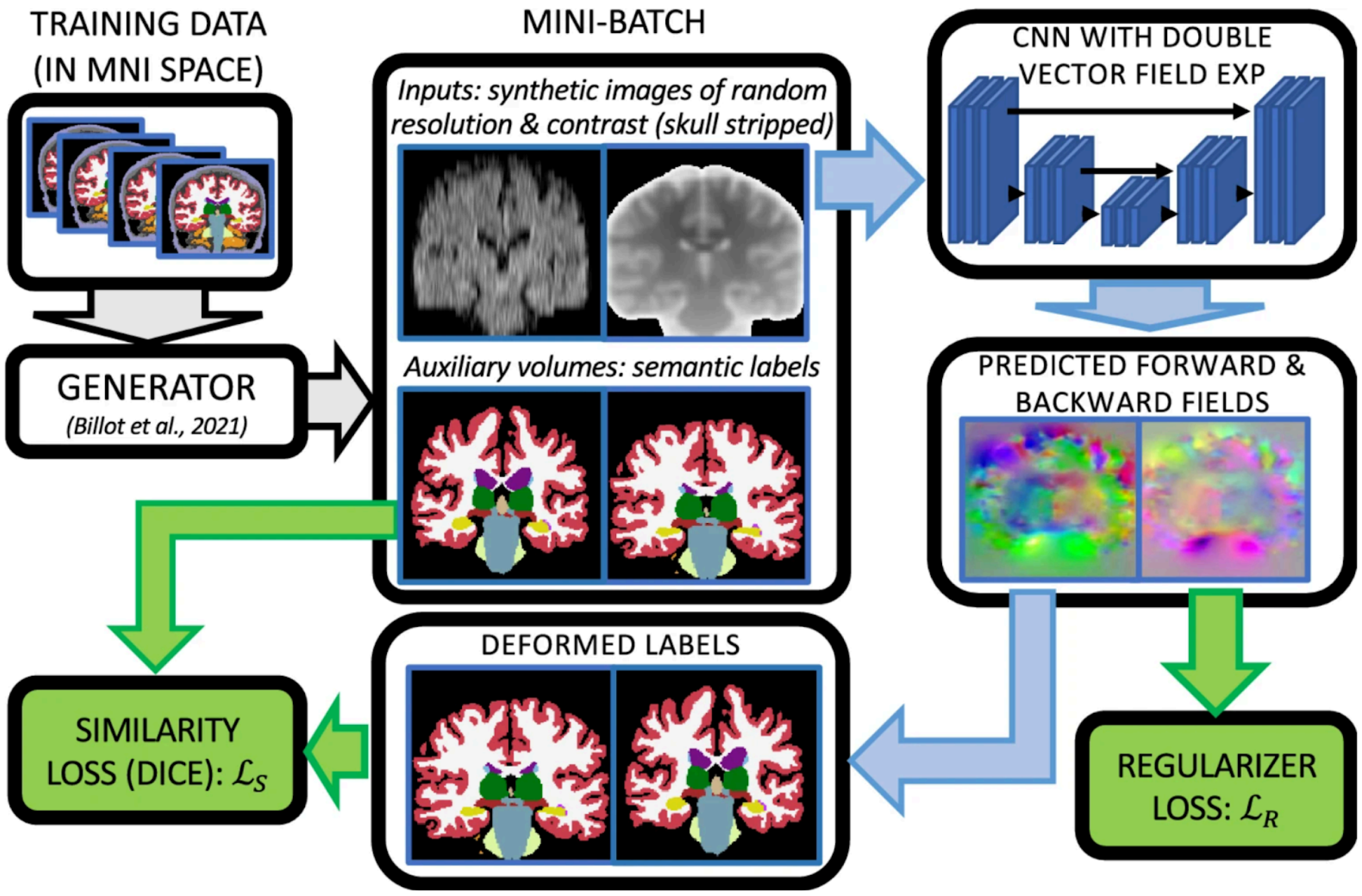

**Figure 32**. Overview of the training procedure used by *EasyReg*. The gray arrows follow the generator; the blue arrows follow the layers of the neural network; and the green blocks represent the different terms of the loss. We emphasize that the segmentations are used to compute the loss, but are not given as input to the CNN. (Figure and Caption from reference[9] without change, under Creative Commons Attribution 4.0 International License.[28])

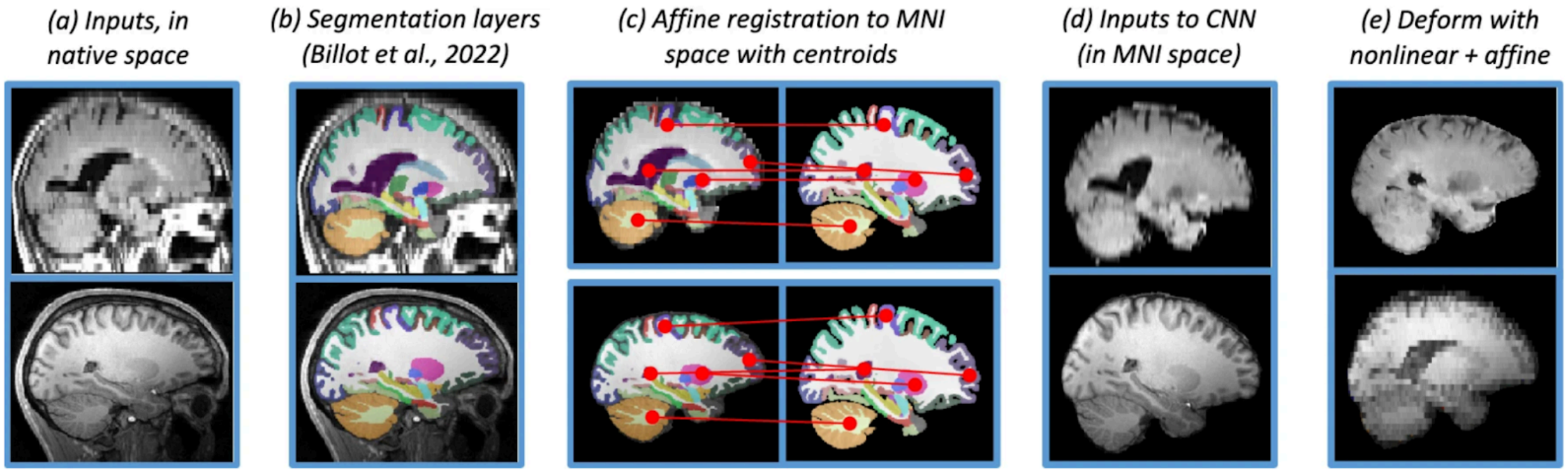

**Figure 33**. Overview of the inference procedure used by *EasyReg*, using a 5 mm axial FLAIR scan and a 1 mm isotropic MPRAGE as sample inputs. (a) Sagittal slices of the input scans. (b) Output of segmentation layers from[272], which include subcortical structures and cortical parcels. (c) The centroids of the different labels are used to compute an affine transform to MNI space using least squares regression; the centroids of the MNI atlas are precomputed. (d) The transforms are used to bring the two (skull-stripped) scans into the voxel space of the MNI reference, which is the space where the CNN was trained. (e) The CNN is used to compute the forward and backward deformation fields, which are composed with the affine transforms in order to obtain the final fields and finally deform the original images.(Figure and Caption from reference[9] without change, under Creative Commons Attribution 4.0 International License.[28])

## 3.2 Diagnosis

Diagnosis is an important part of our healthcare system, and the advances of big data and AI approaches can significantly improve the efficiency and accuracy of current diagnosis techniques.[273–275] Although further validations are required, a comprehensive review[276] based on studies from 2012 to 2019 found the performance of deep-learning based diagnostic is equivalent to that of health-care professionals in classifying diseases using medical images. In the recent fight against COVID pandemic, AI diagnosis is playing an important role and achieving impressively high accuracy.[277–282] A series of review and articles elaborate on machine-learning based disease diagnosis,[283–290] in some cases for a given disease (such as cancers,[291,292] mental disorder,[293] brain and heart diseases,[294–296] fracture,[297] and others [298,299]), or using a specific test (such as electrocardiography, radiology and imaging, flow cytometry, microbiome etc.).[232,242,300–306] Some disease diagnostics and the corresponding algorithms are shown in **Figure34.**

The fusion of the extensive utilization of electronic health records (EHRs) and implementation of machine learning approaches holds great potential for advancing investigation and improving diagnostic capabilities.[307,308] The common components in EHR data consist of recoded patient identifications, demographics, indication, medications, procedures, laboratory tests, vital signs, and utilization.[309] Despite EHR managed by healthcare systems, post-marketing surveillance of drug adverse events could constitute the groundwork for retrospectively collecting complementary EHR.[310] Given the unprecedented cumulation of EHRs, researchers have used data-driven methodologies to investigate the patterns and disclose the true signal in big data.[311] Prior to the machine learning era, data-driven approaches was applied to detect safety signals in EHR data with the emphasis on data processing and statistical methodology. Distinct methods were tailored to suit data sources exhibiting varying characteristics. Systematic approaches were applied to correct the influence of covariates and reveal the true signal when combining EHR data with multiple human health datasets.[312] Intuitively, ML/AI algorithms are supposed to be incorporated effortlessly to screen structured EHR data in hopes of safety signal detection. Deep learning and machine learning have profound influences in clinical notes, diagnoses, and medical imaging regarding the heterogeneous characteristics in EHR data.[313] The transformation of raw data into multiple layers of representations with clusters of numerous non-linear modules amplifies the hidden features via repetitively updating parameters to gain the smallest loss of a predefined task.[314] With machine learning, one could expect to improve EHR analytics with timely disease detection, efficient diagnosis for high-risk groups, precision medicine with tailored treatment plans, medical imaging analysis with enhance capabilities for otherwise undetectable abnormalities, as well as AI- and knowledge- driven patient engagement.

Patient health records could originate from diverse data sources and compose a sophisticated relationship, especially when involving genomic data. As high-throughput technologies continue to advance, substantial amounts of complex omics data have been produced and accumulated. These data sources, such as The Cancer Genome Atlas (TCGA), Uniprot, and drugbank, establish a comprehensive network interlinked via genes/proteins (biomarker).[315–318] Given the challenges in computation when involving gene expression, ML/AI can manage and mine valuable information from biomarker data confronting high-dimensional issues. Most relevant biomarkers can be identified via ML/AI models for better predictions. For example, Lewis and Kemp developed machine learning classifiers using metabolic biomarkers to investigate tumor radiation response.[317] The integration of metabolic biomarkers and other omics data sources represented metabolic features dividing tumors into radiation sensitive type and resistant type. This application showed a path of patient subgroup classification and customized metabolic biomarkers regarding radiation sensitivity. As opposed to genetic biomarkers, another study by Manak *et al.* involved machine learning algorithms in an elegant assay using dynamic live-cell biomarkers to stratify prostate and breast cancer patients with high risk of having adverse pathology before surgery.[316] They demonstrated that the implementation of machine learning could overcome the limitation in traditional diagnosis and treatment, and generate clinically relevant principles for potential adverse state after major medical intervention. With the maturity of ML/AI application in biomedical fields, researchers have incorporated statistical approaches to alleviate the non-transparency of ML/AI frameworks, increase the interpretability of feature selection.[318]

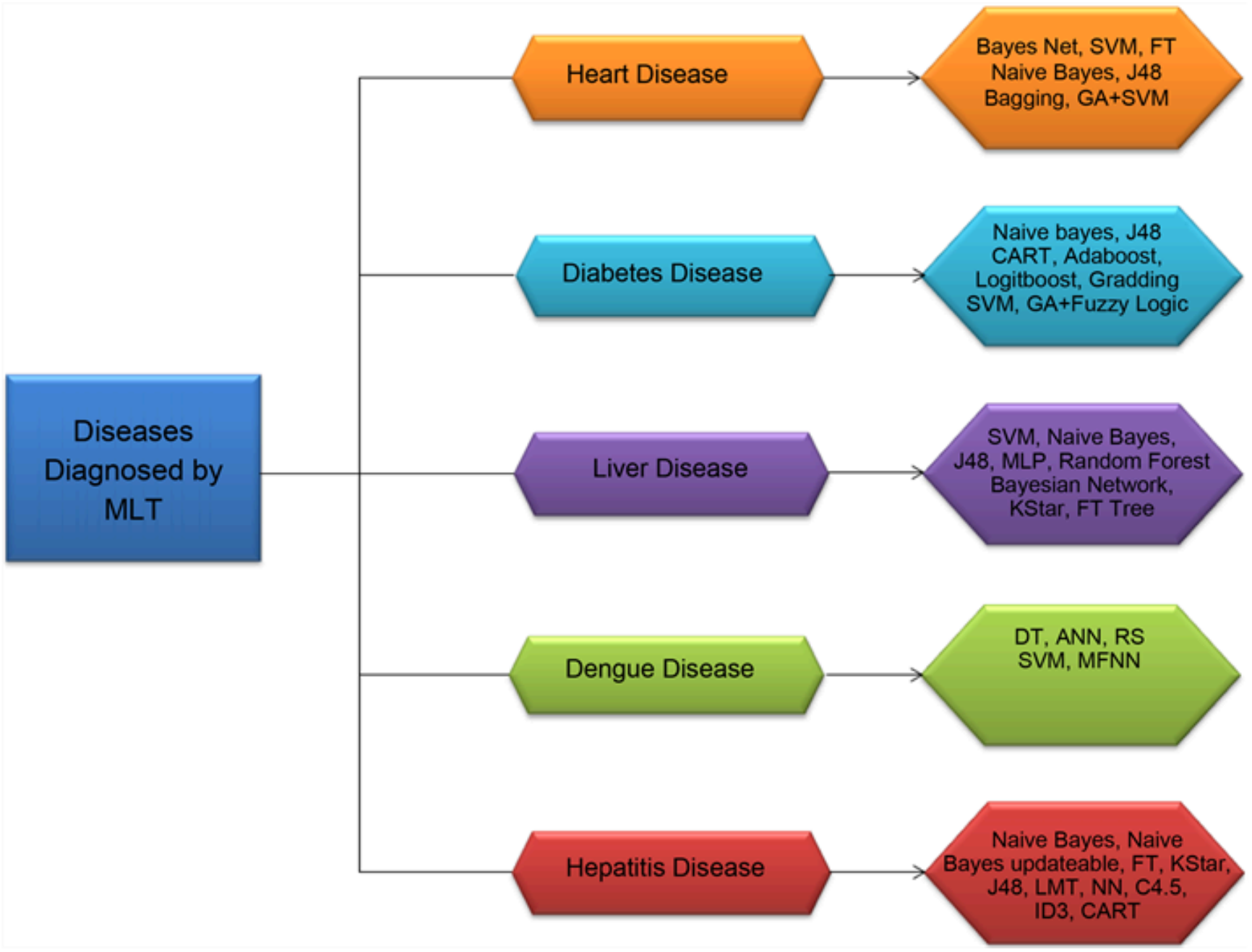

**Figure 34**. Diseases diagnosed by MLT. (Figure and Caption from reference[299] without change, under Creative Commons Attribution 4.0 International License.[28]) (MLT refers to Machine learning techniques.)

## 3.3 Proteins

Complex structures involving proteins[319] and macromolecules[320] have important applications in membranes,[321–323] energy storage and harvest,[324–327] optoelectronics,[328–332] as well as healthcare and medicine.[333–336] Proteins are macromolecules made of amino acid chains and the amino acid sequences define the structure of the proteins in order to fulfill biological functions which include catalyzing biochemical reactions (enzymes), structural support (i.e. collagen, myofiber proteins), transport (i.e. hemoglobin), signaling, regulation and defense (antibody) in the living organisms. Understanding the structure of proteins [337–339] is critical in drug design where proteins are often the target of inhibition. It will also support therapeutic approach development where proteins can be used either as therapeutic agent or carrier. By automating and optimizing the design process, AI enables the development of proteins that are personalized, robust, highly functional, and unlike anything found in nature,[340] as shown in a series of previous reviews.[341–349] Some of the challenges of machine learning workflows in protein engineering include imbalanced data and overfitting (**Figure 35**).

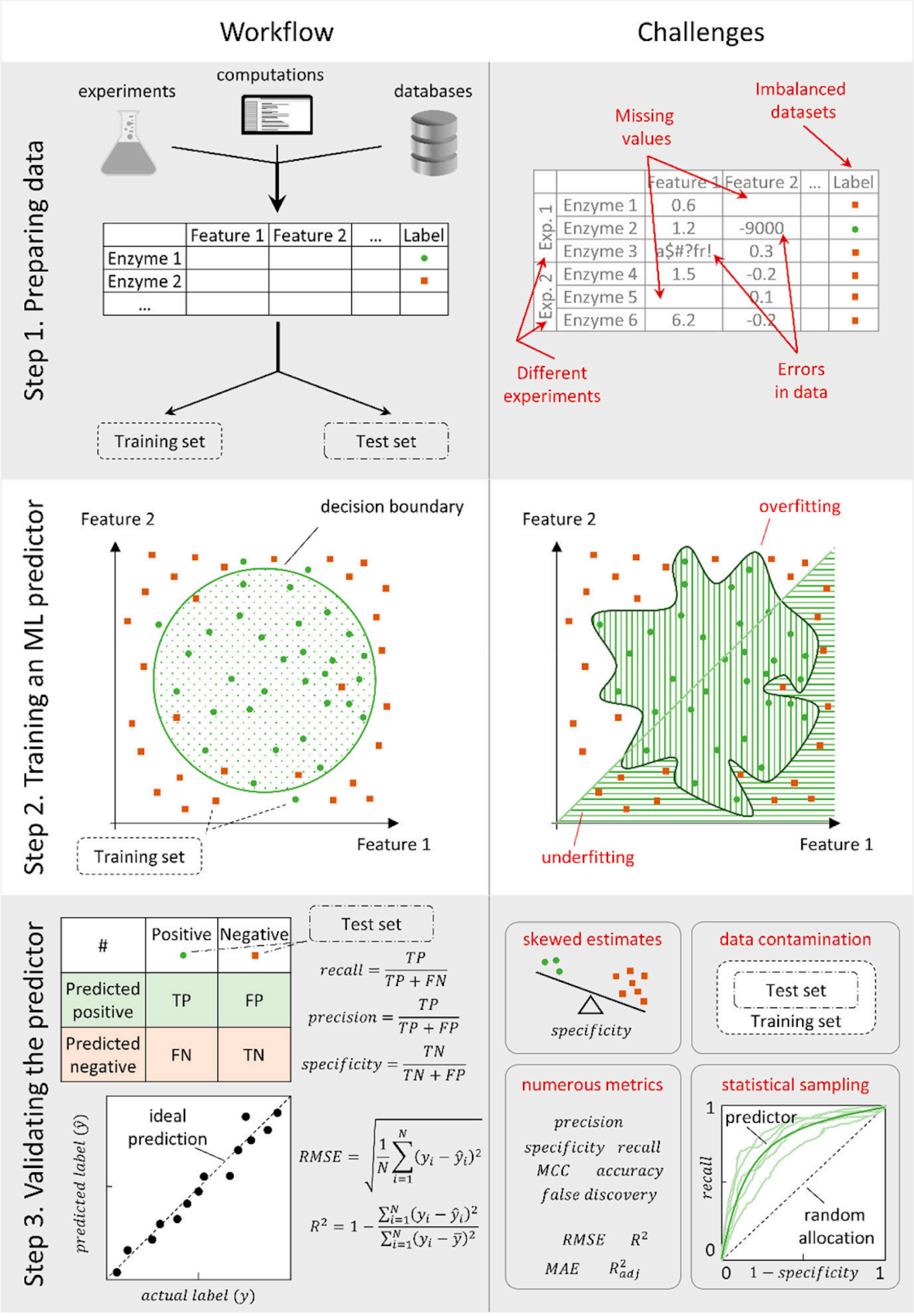

**Figure 35.** Schematic workflow of constructing an ML predictor and associated challenges. Step 1: the data are usually turned into a table format and split into the training and test parts. Any errors, biases, or imbalances will be translated to the predictor's performance and, hence, must be accounted for. Step 2: the predictor is trained on the training data set. For example, a decision boundary is derived that allows classifying future input based on whether data points are inside or outside the boundary. This is a balancing act between two extremes: explaining noise rather than fundamental dependencies (overfitting) or failure to account for complex dependencies in the data (underfitting). Step 3: the performance of the predictor is evaluated based on the test data set. For example, true and false positives and negatives and the associated measures are calculated or the root mean square error (RMSE) is calculated for continuous labels. The random nature of the initial data split as well as data imbalances might skew the evaluation,

and numerous metrics used for evaluation vary in their robustness to different data skews. Even partial inclusion of the test set at any stage of ML predictor training is called data contamination and usually invalidates the final evaluation. (Figure and Caption from reference[350] without change, under CC-BY License.[27])

Tools such as AlphaFold,[351] AlphaFold2,[352] RoseTTA fold,[353] and RFDiffusion[354] can accurately predict protein structure or assist *de novo* protein design. Machine learning is used to predict protein-protein interactions,[355,356] protein interactions with other molecules,[357,358] protein function and property,[359–361] protein stability,[362–364] as well as to advance enzyme engineering,[350,365–368] or protein engineering in general.[369–374] A approach for high-throughput collection of robust and reliable data is shown in **Figure 36**. Like any other AI/ML assisted applied fields, providing accurate training sets is essential for developing AI tools for proteins and biomacromolecules.

AI techniques with deep learning can analyze massive datasets of known protein structures to build models that predict how a new protein will fold based on its sequence. Accurate structure prediction helps researchers design proteins for applications like drugs, nanomaterials, biosensors, or industrial enzymes. Among all the structure prediction tools, AlphaFold was able to predict the most structures of more than 200 million proteins from some 1 million species, covering almost every known protein on the planet.[351] AI network shortens the cycle of solving the protein structures compared to conventional biophysics methods such as X-ray crystallography and cryo-EM, which may have lower accuracy when predicting proteins that have less defined structures, such as intrinsically disordered proteins (IDPs).[375,376] Obtaining training labels requires increased throughput and complexities of quantitative studies that combine theory, computations, and an assortment of low, medium, and high-resolution experiments.

While resolving protein structures are crucial for those protein engineering tasks, enzyme kinetic shares the equal significance in the directed engineering, as it is directly associated with the protein functionality as catalysts, or the dynamic behavior of proteins in metabolic pathways, proteome allocation and responses to reactivator and inhibitors (**Figure 37**).[377] Recent study deployed machine and deep learning methods [377–380] in predicting the $k_{cat}$ of uncharacterized enzymes, providing insights on these key indicators of proteins with known and unknown structures. Although using computational tools in prediction of enzyme kinetics is not new with reaction flux models and proteomic measurements,[381] this new approach contains numerical fingerprints that consider the complete set of substrates and products of the interested chemical reactions (**Figure 38**). Another deep learning approach, DLKcat, allows a high-throughput prediction of turnover numbers for a wide range of metabolic enzymes and substrates.[379] The genome-scale prediction on kinetics has been highlighted in multiple articles when using deep learning tools. Obtained turnover numbers will allow direct comparison of enzyme catalytic efficiency, thus adding a critical aspect in sequence generating through rapid AI-based enzyme engineering approaches.[377–380] Moreover, AI aided computational tools not only will allow us to understand the enzyme performance, or protein characteristics *in vitro*, but also will help us evaluate their *in vivo* behavior while addressing the limitations of experimental *in vivo* tests (**Figure 37**).

Towards protein design and protein engineering, AI has significant promise for revolutionizing the field if challenges around computational cost, model accuracy, and system complexity can be addressed. With proper safeguards and oversight, AI for protein design can improve lives by enabling more effective therapies, sustainable food production, eco-friendly industrial processes and beyond.

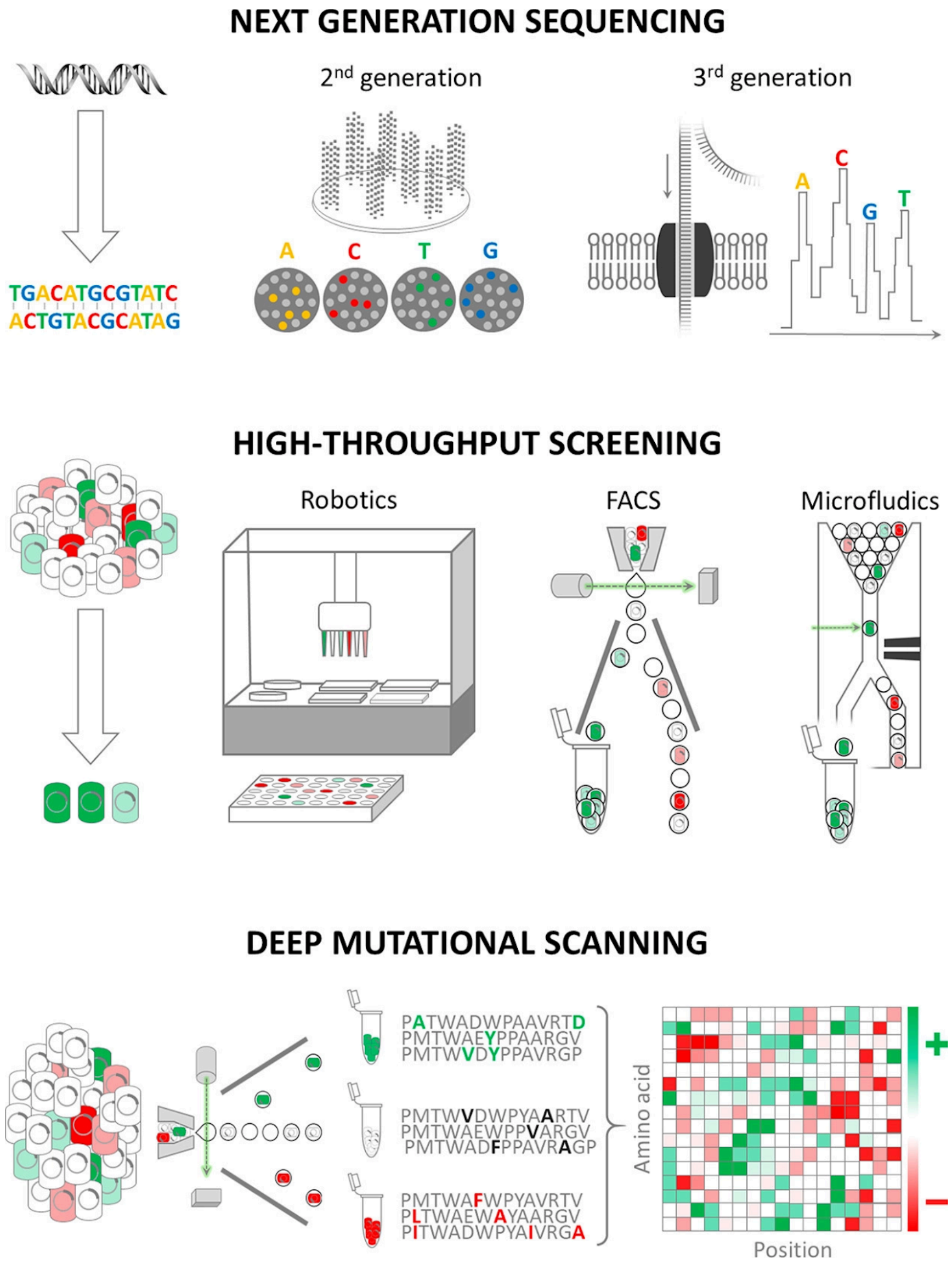

**Figure 36.** Schematic representation of the methods applicable for collection of robust and reliable data. (top) Next-generation sequencing (NGS) offers the high-throughput analysis of DNA/RNA sequences in the gigabase range per instrument. Second-generation instruments increased the throughput and accuracy by massive parallelization of short (100′ bp) DNA fragments reads after amplification. The third-generation (long-read) methods employ a single-molecule real-time sequencing of long DNA fragments (>1 Mbp). (middle) High-throughput screening (HTS) includes a wide range of different approaches: (i) liquid handling robotics with average throughput of 104 variants per day, (ii) fluorescence-activated cell sorting (FACS) enabling screening of up to 108 variants per day, and (iii) microfluidics with the production speed of up to 108 reaction droplets per day. (bottom) Deep mutational scanning (DMS) coupling high-throughput screening with next-generation sequencing offers a powerful strategy for comprehensively analyzing sequence–function relationships in enzymes. (Figure and Caption from reference[350] without change, under CC-BY License.[27])

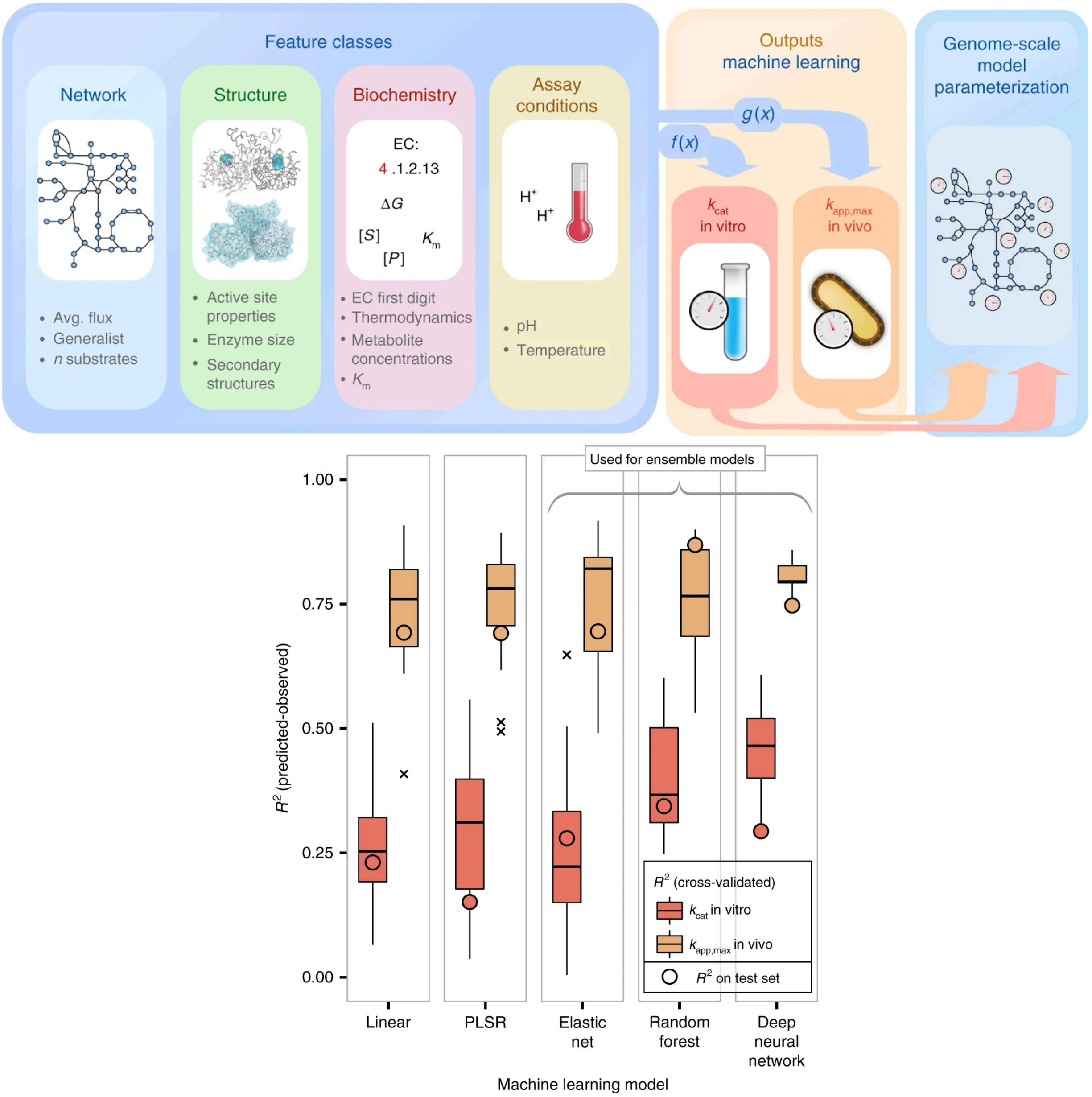

**Figure 37. Top**: Machine learning of catalytic turnover numbers for genome-scale metabolic model (GEM) parameterization. A feature set from diverse classes is curated and mapped to independently build machine learning (ML) models of both $k_{cat}$ *in vitro* (*f(x)*) and $k_{app,max}$ *in vivo* (*g(x)*). The inferred ML models are used to predict $k_{cat}$ *in vitro* or $k_{app,max}$ at the genome-scale to parameterize GEMs. **Bottom**:

Machine learning model performances for $k_{app,max}$ and $k_{cat}$ *in vitro*. Center lines show the median $R^2$ across five times repeated five-fold cross-validation (25 validations), except for the deep learning case, where the median for a single round of five-fold cross-validation (five validations) is shown. Box limits represent the 1st and 3rd quartiles, whiskers extend to values that lie within the 1.5x interquartile range, and the remaining points are shown as outliers (marked x). Circles show $R^2$ for a test set consisting of 20% of the available samples that were not used for hyperparameter optimization. This resulted in a training set of 172 observations of $k_{cat}$ *in vitro* and 106 observations of $k_{app,max}$. For the test set, 43 and 27 observations were used for $k_{cat}$ *in vitro* and $k_{app,max}$, respectively. (Figure and Caption from reference[377] without change, under Creative Commons Attribution 4.0 International License.[28])

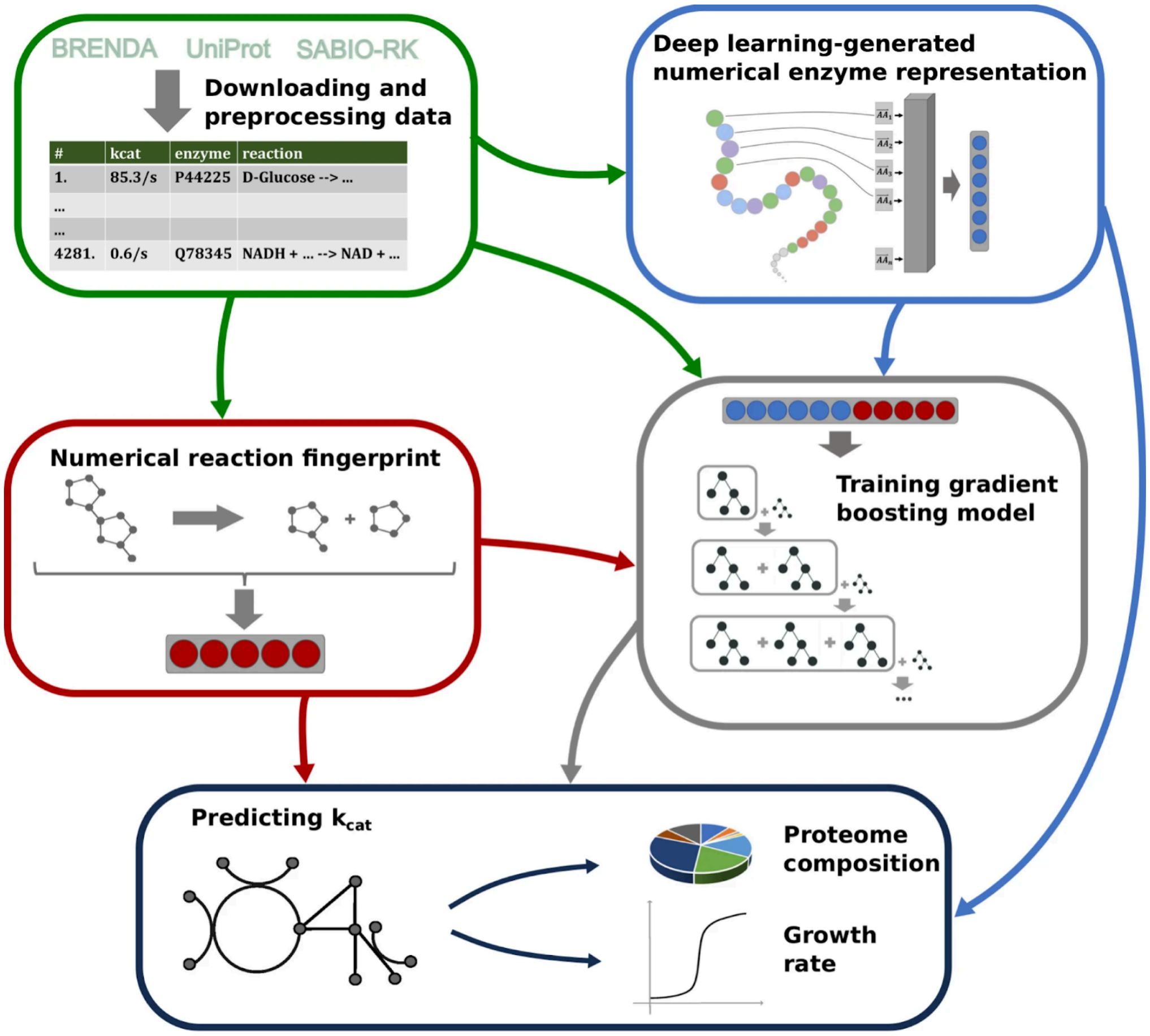

**Figure 38.** Experimentally measured $k_{cat}$ values are downloaded from three different databases. Enzyme information is represented with numerical vectors obtained from natural language processing (NLP) models that use the linear amino acid sequence as their input. Chemical reactions are represented using integer vectors. Concatenated enzyme-reaction representations are used to train a gradient boosting model to predict $k_{cat}$ . After training, the fitted model can be used to parameterize metabolic networks with $k_{cat}$ values. (Figure and Caption from reference[378] without change, under Creative Commons Attribution 4.0 International License.[28])

## 3.4 Small Molecule Drugs

Molecular crystals have a subtle property dependency on the weak forces exerted during their crystallization processes.[382] For example, it is systematically shown that the crystal structure of an organic small molecule is sensitive to its molecular structure, nucleation agents, nanoparticle or polymer additives, solvent, kinetics, as well as the intermolecular interactions involved in the nucleation and crystallization stages.[383–391] The crystal and molecular structures of organic small molecule drugs have a profound impact on their bioavailability and medical applications.

Recently, machine learning has made impressive achievements in drug discovery,[227,392–394] such as discovering a brand-new powerful antibiotic Halicin from the ZINC database mentioned earlier in **Table 1**.[395] A number of reviews highlight the previous progress in this field,[396–401] including a few dedicated to explainable AI for drug design.[402–404] A flowchart in **Figure 39** shows a deep learning algorithm that uses graph neural networks and sequence-based model for small molecule drug design.

With the onset of Covid19, and the search for effective small molecule antivirals, various screening methods have been employed toward the discovery of novel inhibitor molecules for the main protease (Mpro) of the SARS-CoV-2 virus.[405–407] These include for example a) using free-energy perturbation (FEP) calculations starting from a weak-hit drug molecule (Perampanel) for Mpro to design novel analogs with superior inhibition,[405,406] and b) using high throughput virtual screening to identify a competitive non-covalent inhibitor (Mcule-5948770040),[407] which was then verified to be effective via *in vitro* enzyme inhibition assays, as well as an X-ray crystal structure of a Mcule-5948770040-Mpro conjugate. Using the Mcule-5948770040 scaffold as a starting point, a structure−activity relationship (SAR) study was then performed on 19 synthesized novel derivatives of Mcule-5948770040, resulting in two derivatives that displayed enhanced inhibition.[408]

The utilization of AI in drug discovery for COVID-19 has gained considerable interest.[409] Generative AI has been employed to produce datasets[410] of potential drug molecules. Review articles [411,412] covering current research using artificial intelligence in *de novo* drug design and drug discovery for COVID-19 have appeared in the literature. Researchers at Oxford University and IBM have used generative AI to design and develop small-molecule inhibitors for not only Mpro, but also the spike protein receptor-binding domain (RBD) of SARS-CoV-2.[413]

Looking forward, machine and deep learning- based AI algorithms have significant potential [414–416] to accelerate drug discovery in target identification, discovery leading, preclinical optimization, clinical trial matching, personalized therapy, and continuous optimization.

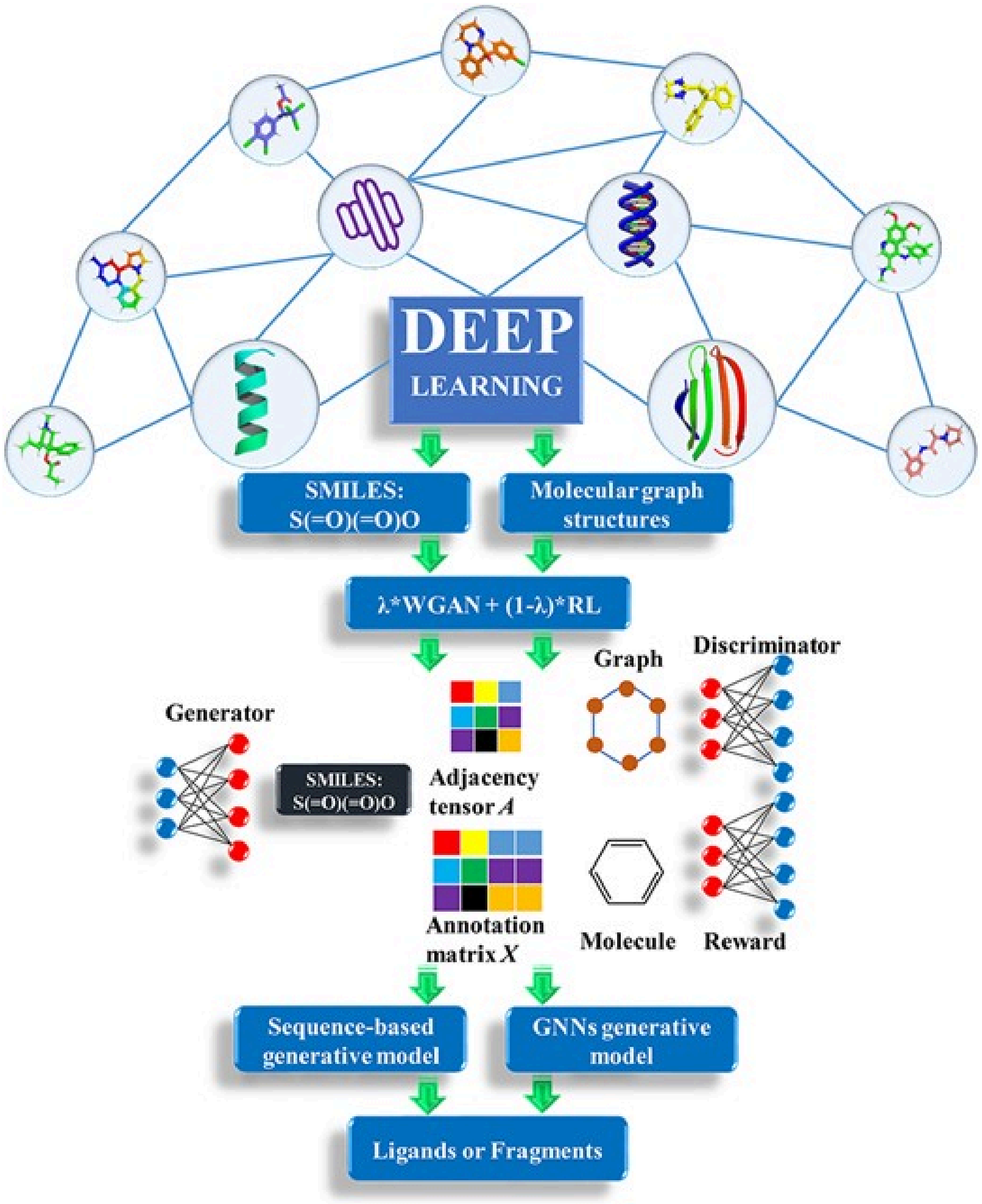

**Figure 39**. Flowchart of deep learning based on sequence-based model and graph neural networks model. (Figure and Caption from reference[392] without change, under Creative Commons Attribution 4.0 International License.[28])

# 4. Related Topics

## 4.1 Physics and AI

The relationship of Physics-based models and AI-driven methods may be as complicated as the one between Physics and AI itself.[417–419] Human scientists need to update our understanding on AI in a timely manner so that we are aware of its limitations, and at the same time, we are not restricted by our own limitations and imagination of what AI is capable of. We need to be aware that AI is not only capable of generating data-driven insights through the analysis of large datasets, it is also possible that AI can make profound discoveries on physics and chemistry that are hard if not possible for us to understand.

**Figure 40** is a multiscale hierarchy of various computational methods with their time and length scales. Each computational method has its own limitations in accuracy, resource requirements, and speed. Machine learning models can benefit greatly by incorporating insights from physics-based computational methods for interpretability, extrapolation, generalization, and efficient learning.[420,421] An example of physics based machine learning is given in **Figure 41**, where autonomous path-sampling algorithms are developed to understand molecular self-organization processes, such as ion assembly in water.

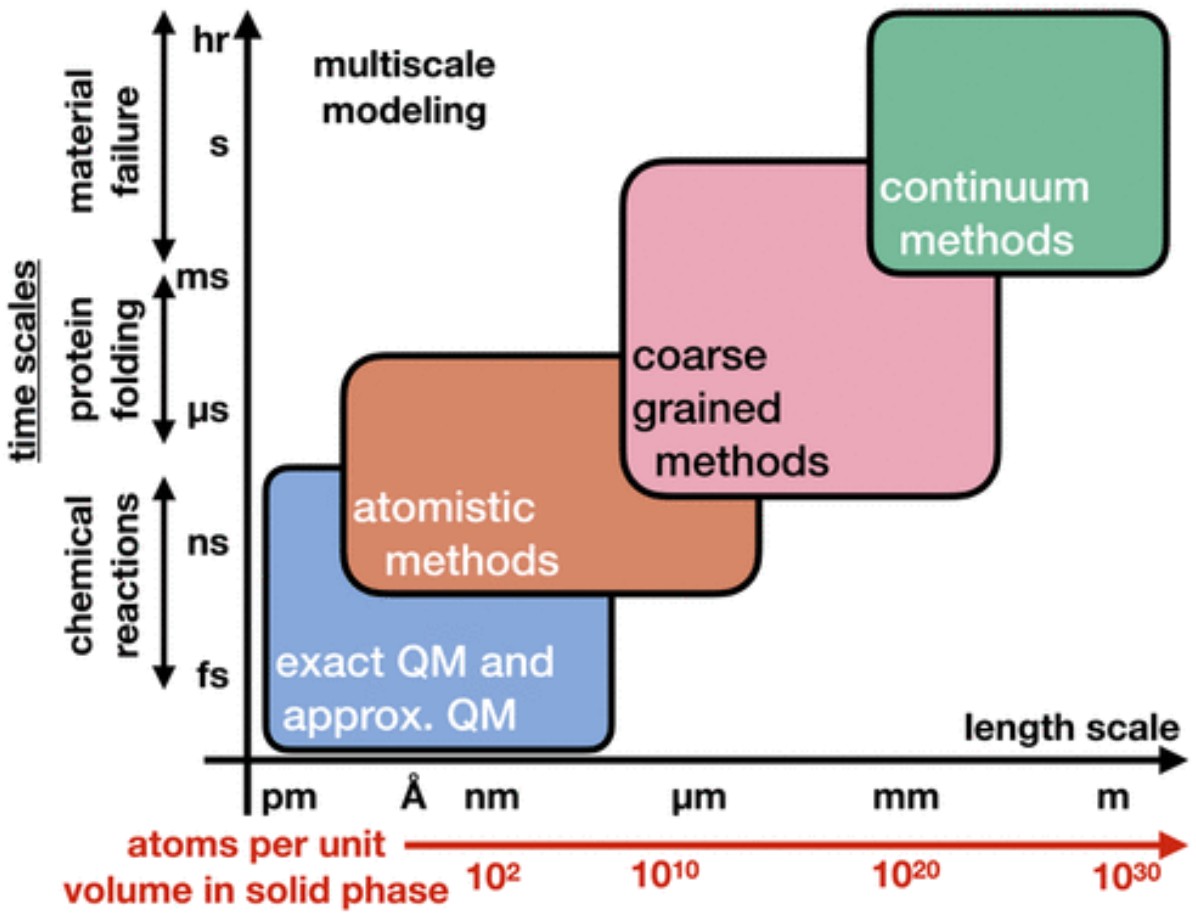

**Figure 40**. Hierarchy of computational methods and corresponding time and length scales. QM stands for Quantum Mechanics. (Figure and Caption from reference[422] without change, under under Creative Commons Attribution 4.0 International License.[28])

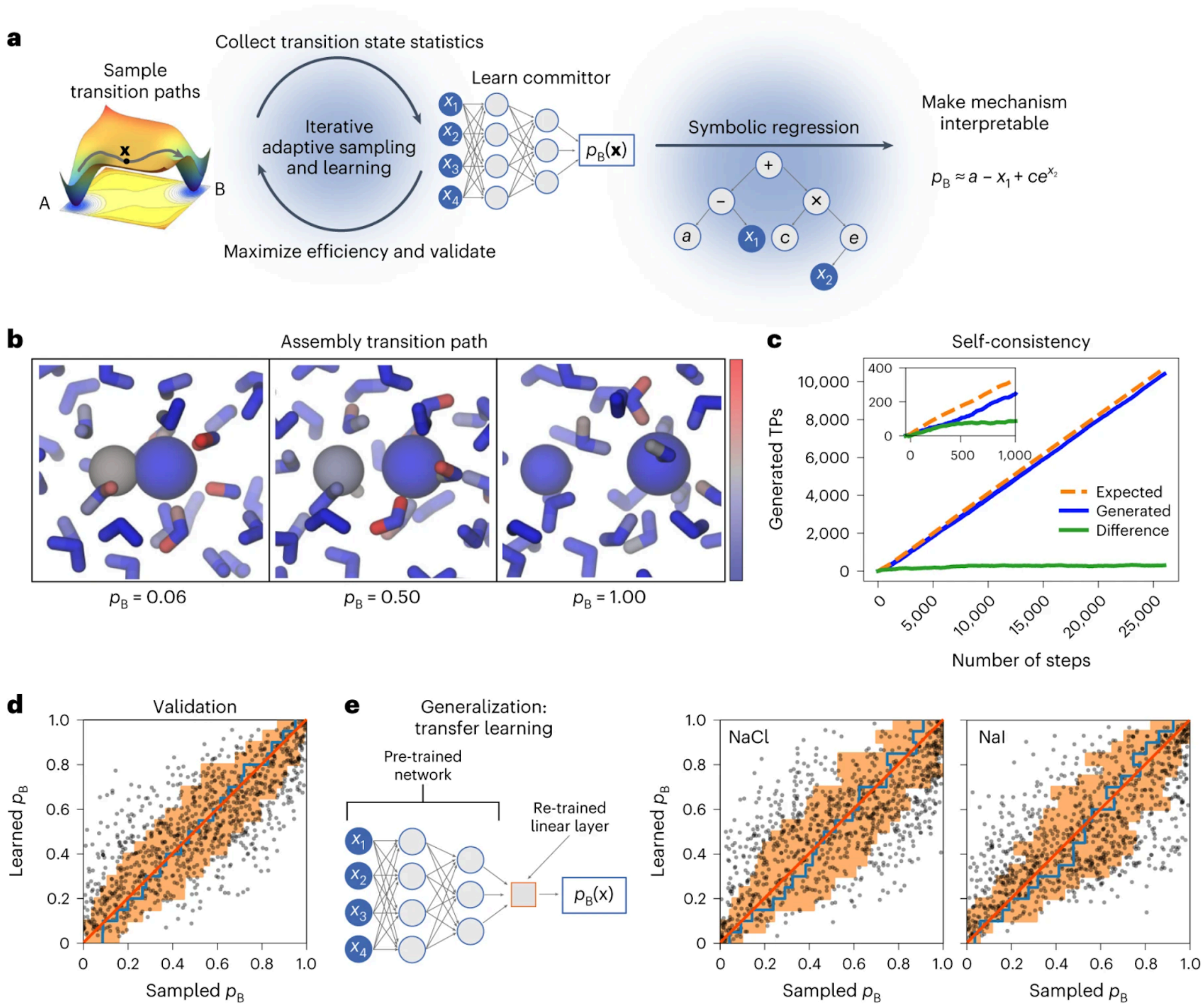

**Figure 41.** Learning the assembly mechanism of ions in water. a, Mechanism learning by path sampling. The method iterates between sampling transition paths from a configuration x between metastable states A and B (left), and learning the committor $p_{B(x)}$ (right). A neural network function of molecular features ($x1$ to $x4$) models the committor. The log predictor forming the last layer is not shown. At convergence, symbolic regression distills an interpretable expression that quantifies the molecular mechanism in terms of selected features ($x1, x2$) and numerical constants ($a, c$) connected by mathematical operations (here: +, −, ×, exp). b, Snapshots along a TP showing the formation of a LiCl ion pair (right to left) in an atomistic MD simulation. Water is shown as sticks, Li+ as a small sphere and Cl− as a large sphere. Atoms are colored according to their contribution to the reaction progress from low (blue) to high (red), as quantified by their contribution to the gradient of the reaction coordinate $q(\mathrm{x}|\mathrm{w})$. c, Self-consistency. Counts of the generated (blue line) and expected (orange dashed line) number of transition events. The green line shows the cumulative difference between the observed and expected counts. The inset shows a zoom-in on the first 1,000 iterations. d, Validation of the learned committor. Cross-correlation between the committor predicted by the trained network and the committor obtained by repeated sampling from molecular configurations on which the committor model was not trained. The average of the sampled committors (blue line) and their s.d. (orange shaded) were calculated in bins of the learned committor indicated by the vertical steps. For reference, the red line indicates the identity. e, Transferability of the learned committor. Representation of transfer learning, and cross-correlations between sampled committors for NaCl and NaI ion pairing and predictions of committor from a model trained on data for LiCl and adjusted by transfer learning using only 1,000 additional shooting outcomes each. Colors and s.d. (indicated by orange shading) are as in d. (Figure and Caption from reference[20] without change, under Creative Commons Attribution 4.0 International License.[28])

Other examples [423–427] include the estimation of interaction parameters of physical models from experimental measurements such as specular neutron reflectivity (NR) and small angle neutron scattering (SANS). In the past, physics-based models have been used for the interpretation of structures in thin films and extracted various interaction parameters from the specular neutron reflectivities of annealed thin films. In recent work,[423] multi-layer perceptron, an autoencoder, and a variational autoencoder were used to extract interaction parameters not only from neutron scattering length density profiles constructed using self-consistent field theory-based simulations, but also from a noisy *ad hoc* model constructed by subject matter experts. In particular, the variational autoencoder was shown to be the most promising tool when it comes to the reconstruction and extraction of parameters from an *ad hoc* neutron scattering length density profile of a thin film containing a symmetric di-block copolymer (poly(deuterated styrene-b-n-butyl methacrylate)). This work paves the way for automated analysis of specular neutron reflectivity from thin films using machine learning tools.

## 4.2 Model Explainability

Although deep learning and neural nets are powerful toolboxes for data science and AI, especially in computer vision and natural language processing, an explainable AI is important in trustworthiness, refinement, constraint identification, fairness evaluation, as well as confidence predictions.[220,428,429] Depending on the use case, there are times when deep learning interpretability is more or less crucial. But as AI begins to permeate critical infrastructure and decision making, explainability becomes essential to ensure these systems are trustworthy, fair and grounded. Interpretable AI can work with people by providing meaningful justifications for its behavior rather than just numerical confidence scores alone.[220,428,429] The future lies in balancing sophisticated machine learning with human intuitions about reasoning, fairness and ethics towards a future with human-AI collaboration and automated decision-making. Explainable deep learning models that understand why they predict what they predict will be key partners in tackling complex societal problems, with issues such as AI alignment, AI ethics, AI bias, and AI safety properly addressed.[220,428,429]

## 4.3 Autonomous Experiments

Some of the examples of autonomous experiments were already shown in the previous sections for additive manufacturing and flow chemistry. It is noted that AI/ML has significant potential to enable experimentation in automated experiment design, adaptive sampling, closed-loop control, distributed autonomy, virtual experimentation, and digital twins.[430] **Figure 42** illustrated a machine-vision control system, in which a feedback pipeline enables online extrusion parameter updates. Clearly, integration of AI/ML and robotics with a scientist in the loop can accelerate science and technology through the development and exploitation of closed-loop, autonomous experimentation systems.[431] For example, recent success has been demonstrated for adaptively driving X-ray diffraction via machine learning for autonomous phase identification.[432] Many other exciting examples such as scanning transmission electron microscopy,[433,434] scanning probe microscopy,[435–438] synthesis [439,440] are also abound. Thus, it is clear that the general concept of self-driving labs is readily becoming a reality[441] and this will shift our scientific paradigm from a traditional “make, measure, simulate” serial approach to an integrated self-driven approach for discovery and innovation. But realizing the promise of autonomous research requires data sharing, multidisciplinary collaboration, forethought about value alignment and job disruption. With progress in machine learning and policy changes facilitating access to resources, AI/ML can make experimentation smarter, faster, and more ambitious. This acceleration of discovery and innovation promises to address pressing challenges in fields like healthcare, transportation, and sustainability but also poses risks that must be managed. Overall, autonomous experimentation will depend on partnership between humans and AI to ask questions, set objectives, assess results, and ensure new capabilities are developed and applied responsibly.

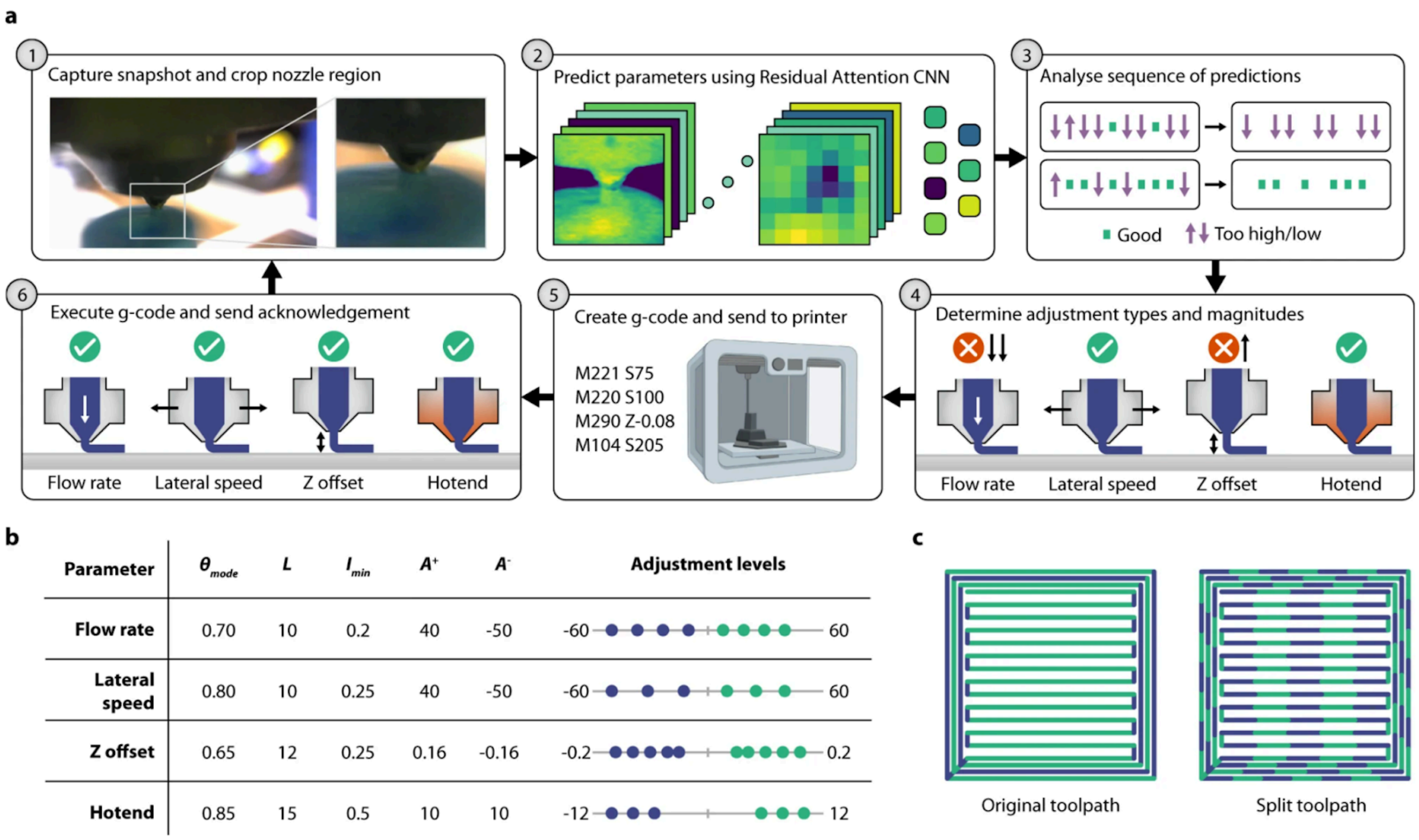

**Figure 42.** Machine vision control system pipeline and feedback parameters. a The six major steps in the feedback pipeline enable online parameter updates from images of the extrusion process. b Table containing θmode (mode threshold), L (sequence length), Imin (interpolation minimum), A+ (the largest increase), A− (largest decrease) for each printing parameter along with the possible levels of update amounts. c Simple example single layer geometry illustrating toolpath splitting into equal smaller segments. 1 mm lengths are used in the feedback process to enable rapid correction and reduce response time. (Figure and Caption from reference[442] without change, under Creative Commons Attribution 4.0 International License[28].)

## 4.4 Security, Ethics and Governance

To integrate physics-based, explainable AI with autonomous experimentation, it is essential to have a mechanism to guarantee the underlying quality, which may include but not limited to security, ethics,[443–445] and governance [446,447]. These issues can often broaden into the areas of infrastructure, social behaviors, philosophy and law. All parties of interest, including government, non-profit organization, academia, research institutions, and private companies, may need to take into account 5 ethics principles: 1). **transparency**, 2). **fairness**, 3). **doing no harm**, 4). **responsibility**, and 5). **privacy**.[443–445] A multidisciplinary approach, predesigned framework, and adaptable planning are required for effective governance and policy making for AI progressing, especially in sectors such as government, education, and healthcare.[446,447]

For AI models themselves, it is well known that a biased training dataset can almost guarantee a poor-quality output, or so called, garbage in and garbage out. To further demonstrate a small tip of the gigantic "iceberg" issues we discussed here, **Figure 43** highlighted the importance of cross-validation strategy and error balancing requirements for supervised learning, that can not be ignored or neglected in practical applications of AI and machine learning. For example, a biased model is based on oversimplified assumptions and tends to underfit the data, while a high variance model tends to pick up random noise and overfit. An optimal model complexity is required to minimize both bias and variance.

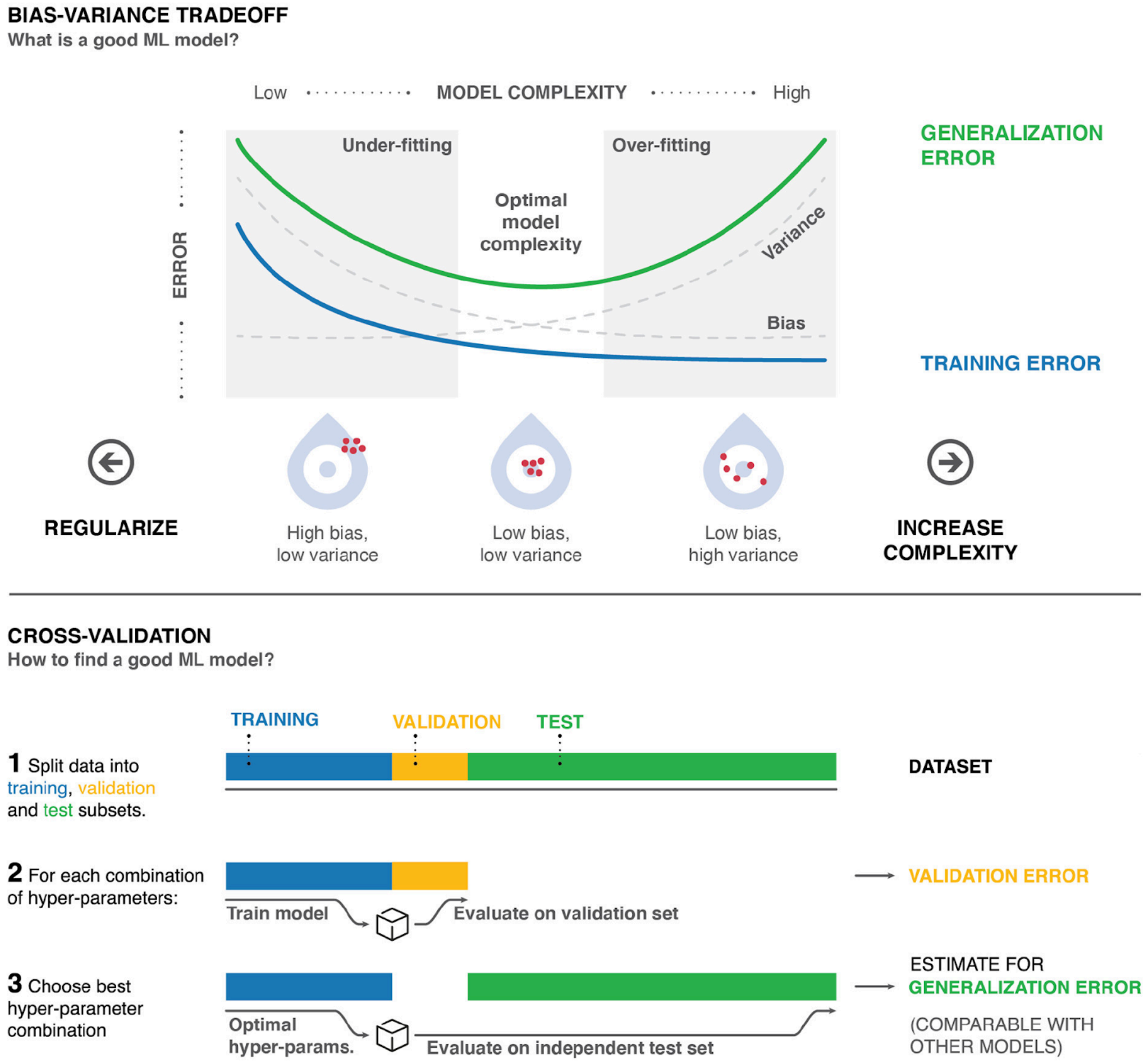

**Figure 43.** Supervised learning algorithms have to balance two sources of error during training: the bias and variance of the model. A highly biased model is based on flawed assumptions about the problem at hand (under-fitting). Conversely, a high variance causes a model to follow small variations in the data too closely, therefore making it susceptible to picking up random noise (overfitting). The optimal bias-variance trade-off minimizes the generalization error of the model, for example, how well it performs on unknown data. It can be estimated with cross-validation techniques. (Figure and Caption from reference[422] without change, under under Creative Commons Attribution 4.0 International License.[28])

# 5. Outlook

This article highlighted many fields which have been enhanced and will continue to be augmented by AI, which is a tool with vast potential to affect the science, engineering, and industrial domain as we know it. In the field of healthcare and personal medicine, AI will directly impact human intervention to prevent critical diseases and provide solutions to future diseases and pandemics. AI executes high-throughput screening to determine how large libraries of molecules may interact with drug targets or patients' genetic profiles, as well as incorporates data across disciplines to diagnose, design studies, interpret results, and propose personalized treatments that are most likely to yield safe, effective treatments. In the field of manufacturing, AI can explore vast design spaces to discover new materials or devices with desirable properties. AI both guides experiment design and predicts results to accelerate discoveries for battery, sensor, electronics, transportation, and renewable energy. **Figure 44** highlights some notable champion models for some of the manufacturing and healthcare applications discussed in this review. In summary, we expect Data Science, machine learning, and AI will revolutionize a wide range of application scenarios related to chemistry, engineering, manufacturing and health industries. The real promise lies in leveraging AI to connect advances across fields and achieve exponentially more through collaboration than any one area might alone.

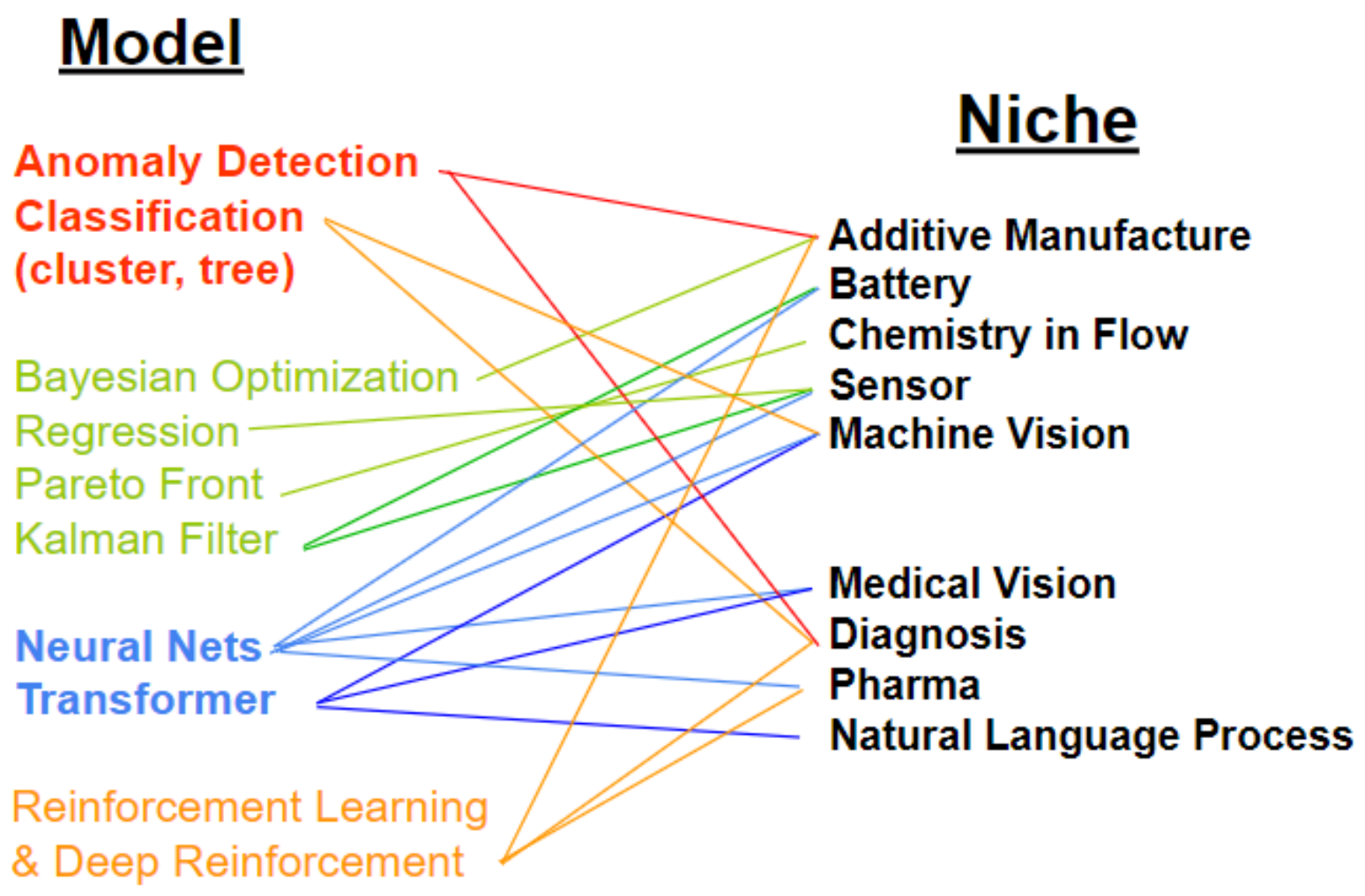

**Figure 44.** Some notable champion models for some of the manufacturing and healthcare applications discussed in this review

# Acknowledgment

This work was supported by Center for Nanophase Materials Sciences (CNMS), which is a US Department of Energy, Office of Science User Facility at Oak Ridge National Laboratory. A part of the first draft used Claude by Anthropic.

# Conflicts of Interest

The author declares no conflict of interest.

# Disclaimer

The findings and conclusions in this report are those of the author(s) and do not necessarily represent the official position of the National Institute for Occupational Safety and Health, Centers for Disease Control and Prevention.